\documentclass[a4paper,11pt,titlepage]{article}
\usepackage[english]{babel}
\makeatletter\AtBeginDocument{\let\@elt\relax}\makeatother{}
\usepackage[utf8x]{inputenc}
\usepackage[margin=1.5cm]{geometry}
\usepackage{graphicx,grffile} %% BEFORE SVG PACKAGE !!!
\usepackage{amsmath}
\usepackage{amsthm}
\usepackage{amsfonts}
\usepackage{mathrsfs}   % \mathscr{...}  lettre cursive majuscule
\usepackage{amssymb,yfonts,bbm}
\usepackage[bbgreekl]{mathbbol}	% to do letter with double bars use:  $\mathbb{ABCdef123}$
\usepackage[table,x11names]{xcolor}  				% Enable  textcolor
\usepackage{color,colortbl}
\usepackage{bm,mathtools}      				% Lettre grasses
\usepackage{booktabs}				
\usepackage{array}					
\usepackage{slashed}  				% to do letter with diagonal bar
\usepackage{chemarrow}				% for chemical arrows with auto size
\usepackage[]{hyperref}	% create link within the document (to biblio and equations)
\hypersetup{colorlinks=true,citecolor=blue,urlcolor=blue,linkcolor=blue, 
pdftitle={Spinor fields in GA},
pdfauthor={Gatien VERLEY}
}
\usepackage{tikz}					% to draw nice graph and sketch
\usepackage{accents}			% required for underbar \myubar command

\usepackage{verbatim}
\usepackage{listings}
\usepackage{enumitem}
\usepackage{url}
\usepackage{svg}
\usepackage{currfile}
\usepackage{pdflscape} 	% To switch to landscape pages
\usepackage[normalem]{ulem}  	% to draw  under letters : \dashuline{}, \dotuline, \uuline{}, \uwave{}
\usepackage[e]{esvect}   				% for nice vectors
\usepackage{accents} % to put symbol under letters 
\usepackage{paracol} % NE MARCHE PAS AVEC REVTEX
\usepackage{unicode-math-input} % Remplace le package ucs (incompatible avec hyperref qui donne alors une erreur de } manquant)
\usepackage{booktabs}

\usepackage{authblk}

\DeclareMathAlphabet{\mathpzc}{OT1}{pzc}{m}{it}

\allowdisplaybreaks

\makeatletter
\DeclareRobustCommand{\cev}[1]{%
  {\mathpalette\do@cev{#1}}%
}
\newcommand{\do@cev}[2]{%
  \vbox{\offinterlineskip
    \sbox\z@{$\m@th#1 x$}%
    \ialign{##\cr
      \hidewidth\reflectbox{$\m@th#1\vec{}\mkern4mu$}\hidewidth\cr
      \noalign{\kern-\ht\z@}
      $\m@th#1#2$\cr
    }%
  }%
}
\makeatother % fleche comme \vec mais en sens inverse

\renewcommand{\aa}{\mathbb{a}}
\newcommand{\bb}{\mathbb{b}}
\newcommand{\cc}{\mathbb{c}}

\newcommand{\ii}{\mathbb{i}}

\newcommand{\lbb}{\mathbb{l}}

\newcommand{\rr}{\mathbb{r}}

\newcommand{\0}{\mathbb{0}}
\newcommand{\1}{\mathbb{1}}
\newcommand{\2}{\mathbb{2}}
\newcommand{\3}{\mathbb{3}}
\newcommand{\4}{\mathbb{4}}
\newcommand{\5}{\mathbb{5}}
\newcommand{\6}{\mathbb{6}}
\newcommand{\7}{\mathbb{7}}
\newcommand{\8}{\mathbb{8}}
\newcommand{\e}{\bm{e}}
\newcommand{\B}{\bm{B}}

\renewcommand{\u}{{\bm{u}}}
\renewcommand{\v}{{\bm{v}}}

\newcommand{\Ac}{\mathcal{A}} % multivecteurs
\newcommand{\Bc}{\mathcal{B}} % multivecteurs
\newcommand{\Cc}{\mathcal{C}} % multivecteurs
\newcommand{\Kc}{\mathcal{K}} % multivecteurs
\newcommand{\Lc}{\mathcal{L}} % multivecteurs
\newcommand{\Mc}{\mathcal{M}} % multivecteurs
\newcommand{\Nc}{\mathcal{N}} % multivecteurs
\newcommand{\Oc}{\mathcal{O}} % multivecteurs
\newcommand{\Pc}{\mathcal{P}} % multivecteurs
\newcommand{\Qc}{\mathcal{Q}} % multivecteurs
\newcommand{\Rc}{\mathcal{R}} % multivecteurs
\newcommand{\Sc}{\mathcal{S}} % multivecteurs
\newcommand{\Vc}{\mathcal{V}} % multivecteurs
\newcommand{\Wc}{\mathcal{W}} % multivecteurs
\newcommand{\kbar}{\mathchar'26\mkern-9mu k}

\renewcommand{\L}{{\textsc{l}}} % grade index

\newcommand{\K}{{\textsc{k}}} % grade index
\newcommand{\J}{{\textsc{j}}} % grade index
\newcommand{\ps}{\bm{\mathrm{I}}}
\newcommand{\vide}{{\varnothing}}
\newcommand{\I}{{\textsc{i}}}
\newcommand{\II}{{\textsc{ii}}}
\newcommand{\III}{{\textsc{iii}}}
\newcommand{\IV}{{\textsc{iv}}}

\newcommand{\eee}{{\textsc{e}}}
\newcommand{\ooo}{{\textsc{o}}}
\newcommand{\ppp}{{\textsc{p}}}

\newcommand{\mf}{\mathsf{f}}
\newcommand{\muf}{\myubar{\mathsf{f}}}

\renewcommand{\P}{{\mathsf{P}}}

\newcommand{\fpartial}{{\partial \kern-0.87em \raisebox{-0.05em}{ $\bullet$} } } % Filled \partial for space gradient operator on manifolds
\newcommand\myubar[1]{\underaccent{\bar}{#1}} % For adjoint operator

\newcolumntype{L}{>{$}l<{$}} % math-mode version of "l" column type
\newcolumntype{C}{>{$}c<{$}} % math-mode version of "c" column type
\newcolumntype{R}{>{$}r<{$}} % math-mode version of "r" column type

\definecolor{LightBlue}{rgb}{195,77,79}

\newcommand{\rc}{\cellcolor{Red1!50}}
\newcommand{\gc}{\cellcolor{SpringGreen3!50}}
\newcommand{\bc}{\cellcolor{SteelBlue1!50}}

\begin{document}

\title{Fermions inside the geometric algebra of spacetime}

\author{Gatien Verley} 
\date{\textit{Universit\'e Paris-Saclay, CNRS/IN2P3, IJCLab, 91405 Orsay, France}\\[2ex]\today}

\maketitle
 %TO write file nb on the 1st page
%\begin{center}
	%\currfilename
%\end{center}
%\newpage

% Large figure (meme en double collone... grace au étoile après figure)
%\begin{figure*}
%\begin{center}
%%\parbox{19cm}{\hspace{-1cm}
%%\includegraphics[width= 6cm]{./Images/fig16.eps} \hspace{-0.4cm}
%%\includegraphics[width= 6cm]{./Images/fig15.eps} \hspace{-0.4cm}
%%\includegraphics[width= 6cm]{./Images/fig14.eps}}
%\parbox{18cm}{
%\caption{Ma legende... \label{fig15} } }
%\end{center}
%\end{figure*}

%% CITATION COMMAND WITH COMMENTS
%\cite{*[{prepended text}][{appended text}]key}

%\newpage

\abstract{
The (signed) basis versors of the geometric algebra of spacetime form the extraspecial groups of order 32. Ignoring signs and the unit element, we show that the remaining fifteen normalized basis versors can be written as, or factorized into, a unit space bivector (of three-dimensional space) and an isoversor. Isoversors are: the volume trivector of space, the pseudoscalar, and the basis time-axis vector. Our first result is that this central product structure, already known for extraspecial groups, also holds in the geometric algebra of spacetime and is crucial for exhibiting generators of continuous groups in the algebra. In a second main result, inspired by Hestenes' 1982 paper ``Spacetime structure of weak and electromagnetic interactions'', we uncover two pairs of special unitary groups of degree two and three (in various representations). A rich algebraic structure emerges, leading us to our third main result: the description \emph{inside} the geometric algebra of spacetime of thirty-two particles of matter and antimatter in a single generation. The sixteen matter particles, four leptons (left/right with up/down isospins) and four triplets of quarks (with three colors in addition to chirality and isospin), are gathered four by four into quadrispinors in two different ways: particles in a quadrispinor have either the same color or the same isospin and chirality. Thanks to this twofold distribution of the sixteen matter particles inside four quadrispinors, the same continuous group based on the isoversors produces the electroweak and strong gauge symmetry. However, chirality forbids the gauge mixing of left and right particles, which motivates the restriction to SU(2) $\times$ U(1) for the electroweak interaction. Finally, as a fourth main result, we thoroughly study the dynamics of quadrispinors by generalizing the conventional treatment of bispinors in geometric algebra.}

\newpage
\tableofcontents

\newpage

\section{Geometric algebras of spacetime}
\label{GAST}

Geometric Algebra (GA), sometimes called real Clifford algebra, is introduced for undergraduate students in Refs.~\cite{Book_Macdonald2010, Book_Macdonald2012}, for physicists in Ref.~\cite{Book_Doran2003} and for mathematicians in Ref.~\cite{Wilson2022vol}, among many other introductions, reviews and textbooks \cite{Hestenes1987vol, Book_Hestenes2004, Doran2005, Chisolm2012vol, Hitzer2007vol}. There is a plethora of notations and conventions in geometric algebra: we propose here some that are particularly convenient for the GA of spacetime\footnote{Actually, one must not use basis elements corresponding to spacetime planes (with one time and one space direction) as is commonly done: this prevents using two triplets of commuting elements as versor basis.}. Particle physics based on geometric algebra most often uses larger algebras than the geometric algebra of spacetime \cite{Trayling2001vol34, Castro2014vol92, McClellan2019vol29, McClellan2021vol31} or algebras based on four-dimensional metric space, but without Minkowski signatures \cite{McClellan2016vol27}. Our main objective is to show that it is possible to focus on the geometric algebra of spacetime to describe a complete generation of fermionic particles existing in extensions of the Standard Model of particle physics that include right-handed neutrinos \cite{Buchmueller2006vol}. To accomplish this task, we distinguish the algebraic construction from the additional assumptions required for its physical interpretation.

We start in the first Section with a thorough exploration of the geometric algebra of spacetime, i.e., the algebra built from the four-dimensional metric spaces with metric signature $(±,∓,∓,∓)$. As main results, we provide the central product structure of the algebra (which we have related to the extraspecial groups of order $2^{5}_{∓}$ thanks to Ref.~\cite{GAP4}), we review the most useful involutions, we present the different special unitary groups in various representations, and useful projectors existing within the spacetime algebra for signature $(+,-,-,-)$. 

In the second Section, we carefully connect the geometric algebra of spacetime and particle physics based on the common notion of spinor~\cite{Book_Doran2003, Doran2005, Steane2013vol}. Inside the geometric algebra of spacetime, a general multivector describes four spinors (spin-one-half particles) \cite{Hestenes1982vol12}, including their chirality and materiality (i.e., the fact of being matter or antimatter states) as intrinsic properties (i.e., defined inside the algebra). General multivectors of the geometric algebra of spacetime are coined quadrispinors as they naturally generalize Dirac bispinors. Many representations (including chiral and standard ones) of quadrispinors are provided, together with the Dirac operators, extending the results of Ref.~\cite{Book_Doran2003} (and in fact also Ref.~\cite{Doran2005}, which we found after recovering the same results) to quadrispinors. Next, thanks to the existence of a matter--antimatter exchange symmetry, two different types of quadrispinors exist. We call them matter-first and matter-last quadrispinors. We show that a pair of such quadrispinors can be transformed into a quadrispinor for matter particles only and one for antimatter particles only. For a generation of particles, this can be done in two different ways, leading to the di-doublet quadrispinors describing two electroweak doublets, and the polychrome quadrispinors describing one lepton and three quarks of different colors.

In the third Section, we begin constructing a single-generation fermion model based on geometric algebra. We consider the local and global gauge symmetries of the interacting wave equations for massless fermions. Indeed, mass operators naturally appear in the wave equation when switching to di-doublet or polychrome quadrispinors. These mass operators should be gauge-transformed as well, or dealt with by introducing a Higgs field and Yukawa couplings. For simplicity, we focus on massless fermions.  

We conclude with the many directions that remain to be explored to obtain a standard model of particle physics expressed within the geometric algebra of spacetime.

\subsection{Presentation of the extraspecial groups $2^{5}_{\pm}$ of versors}
\label{groupstructure}

\paragraph{Reference frame}
There are two spacetime algebras, denoted $\mathcal{G}(1,3)$ and $\mathcal{G}(3,1)$, according to the choice of signature. Spacetime algebra is the geometric algebra built using the following four generators $\e_{\mu}$ that are orthonormal vectors, with $\mu=0,1,2,3$. We call $\e_{0}$ the \emph{clock vector} and $\e_{i}$ (for $i=1,2,3$) the \emph{ruler vectors}. These names are chosen to emphasize that specifying a spacetime frame is a choice of observer with their observation tools (clock and ruler). Different choices are possible at the same position and time depending on the observer's motion when it passes through this spacetime position. With these vectors, we can present the two finite groups on which our two spacetime algebras are constructed. ``Presenting'' a group starts by giving generators as above, then taking their geometric product to construct group elements. In geometric algebra, an element that is a product of vectors is called a versor. The number of versors is infinite if one does not provide rules to relate them. Hence, a group presentation is~\cite{Naikvol}
\begin{equation}
	\left \langle \text{product of vectors} | \text{reduction rules} \right \rangle.
\end{equation}
If there are enough rules, then the group (of versors) is finite. Below, we present the two groups of 32 elements on which are based $\mathcal{G}(1,3)$ and $\mathcal{G}(3,1)$.

\paragraph{Finite group associated to $\mathcal{G}(1,3)$} The geometric products of our basis vectors are such that $\e_{0}^{2}=1$, $\e_{i}^{2}=-1$ (with $i=1,2,3$) and two different basis vectors anticommute $\e_{\mu}\e_{\nu}=-\e_{\nu}\e_{\mu}$ if $\mu\neq\nu$. This defines the Dirac algebra, often represented using the Dirac matrices. A presentation of the associated finite group is given by
\begin{equation}
	\left \langle \e_{0},\e_{1},\e_{2},\e_{3}  | 
	\e_{0}^{2}=\e_{1}^{4}=\e_{2}^{4}=\e_{3}^{4}=1,   
	\e_{0}\e_{i}=\e_{1}^2\e_{i}\e_{0},
	\e_{1}\e_{2}=\e_{1}^2\e_{2}\e_{1},
	\e_{1}\e_{3}=\e_{1}^2\e_{3}\e_{1}, 
	\e_{2}\e_{3}=\e_{1}^2\e_{3}\e_{2} \right\rangle, \label{presentation-group-minus}
\end{equation}
where $1$ is the unit element of the group. This group of basis versors is the extraspecial group of order $2^{5}=32$ and type ``$-$''  denoted $2^{5}_{-}$. The negative type indicates that there is an odd number of quaternion subgroups. The opposite choice of signature leads to a \emph{different} group, as we explain now.
% GAP ID = 50 for $2^{5}_{-}$ group extraspecial central inner product of D8 and Q8

\paragraph{Finite group associated to $\mathcal{G}(3,1)$} The geometric products of our basis vectors are now such that $\e_{0}^{2}=-1$, $\e_{i}^{2}=1$ (with $i=1,2,3$) and, here again, two different basis vectors anticommute $\e_{\mu}\e_{\nu}=-\e_{\nu}\e_{\mu}$ if $\mu\neq\nu$. A presentation of the associated finite group is given by
\begin{equation}
	\left \langle \e_{0},\e_{1},\e_{2},\e_{3}|
	\e_{0}^4=\e_{1}^2=\e_{2}^2=\e_{3}^2=1,
	\e_{0}\e_{i}=\e_{0}^2\e_{i}\e_{0},
	\e_{1}\e_{2}=\e_{0}^2\e_{2}\e_{1},
	\e_{1}\e_{3}=\e_{0}^2\e_{3}\e_{1}, 
	\e_{2}\e_{3}=\e_{0}^2\e_{3}\e_{2}
	\right \rangle
	\label{presentation-group-plus}
\end{equation}
where $1$ is the unit element of the group. This group of basis versors is the extraspecial group of order $2^{5}=32$ and type ``$+$'' denoted $2^{5}_{+}$. The positive type indicates that there is an even number of quaternion subgroups. 
% GAP ID = 49 for $2^{5}_{+}$ group extraspecial central inner product of D8 and D8, or central inner product of Q8 and Q8

\paragraph{Comments and terminology} The two extraspecial groups we have just presented are \emph{different}, and their corresponding geometric algebras therefore have different structures. This was recently emphasized in Refs.~\cite{McClellan2019vol29,McClellan2021vol31} although it is known in group theory. 
From the presentations of Eq.~\eqref{presentation-group-minus} and~\eqref{presentation-group-plus}, we have defined two groups of $32$ elements, hence with $16$ couples of opposite elements (when taking the $\mathbbm{Z}_{2}$ quotient). We name these 16 elements as follows: the \emph{reals} denoted $\alpha, \beta, \dots$, the \emph{space vectors} being linear combinations of $\e_{1}$, $\e_{2}$, $\e_{3}$, the \emph{time vectors} proportional to $\e_{0}$, the \emph{space bivectors} being linear combinations of $\e_{1}\e_{2}$, $\e_{2}\e_{3}$, $\e_{3}\e_{1}$, the \emph{mixed bivectors} combining $\e_{0}\e_{1}$, $\e_{0}\e_{2}$, $\e_{0}\e_{3}$, the \emph{mixed trivectors} combining $\e_{0}\e_{1}\e_{2}$, $\e_{0}\e_{2}\e_{3}$, $\e_{0}\e_{3}\e_{1}$, the \emph{cube trivector} proportional to $\e_{1}\e_{2}\e_{3}$ and finally the \emph{pseudoscalars} proportional to $\e_{0}\e_{1}\e_{2}\e_{3}$. Other names exist in the literature, but those we propose are the most explicit: the word 'mixed' indicating that the versor includes both space and time vectors. 

As generators of rotations in real space, the space bivectors $\B_{1} ≐ \e_{2}\e_{3}$, $\B_{2}≐\e_{3}\e_{1}$, $\B_{3}≐\e_{1}\e_{2}$ will play a prominent role in the following. Similarly will the \emph{isoversors} $ ℐ_{\1} ≐ \e_{1}\e_{2}\e_{3}$, $ ℐ_{\2} ≐ \e_{0}\e_{1}\e_{2}\e_{3}$ and $ ℐ_{3}≐\e_{0}$. These will define involutive isometries such as time reflection or duality. We differentiate the indices of space bivectors ($i=1,2,3$) and isoversors ($\aa=\1,\2,\3$) for convenience. Duality is defined by $(\dots)^{\star} ≐ (\dots)ℐ_{2}^{-1}$, i.e., multiplying on the right by the inverse of the pseudoscalar $\ps ≐ ℐ_{2}$. For instance, the dual bivectors are $\B_{i}^{\star} ≐ \B_{i}  ℐ_{\2}^{-1} = - \B_{i}  ℐ_{\2}$, namely $\B_{i}^{\star} = \e_{0} \e_{i} $.

\subsection{Linear space and algebra of multivectors}

Geometric algebras $\mathcal{G}(1,3)$ and $\mathcal{G}(3,1)$ include real linear spaces (or vector spaces) of dimension $2^{4}=16=1+4+6+4+1$ of basis elements given by the basis versors of Table \ref{structure}.  
An arbitrary element $\Mc$ of this linear space, with sixteen real components, is called a multivector. It is different from vectors $\bm x$, also called $\I$-vectors, that are linear combinations of the four unit vectors $\bm x = \sum_{\mu} x^{\mu} \e_{\mu}$ only. Here, the zeroth coordinate is $x^{0} = ct$, with $c$ the speed of light and $t$ time. By definition of a linear space, the addition of two multivectors yields a multivector whose components are the component-wise sum of the components of each multivector. The real numbers, usually denoted with Greek letters, are the null-grade multivectors and are used to define the external law of the linear space.

The geometric product is the law that promotes the linear space of multivectors to an algebra called geometric algebra. This product is bilinear, associative, and distributive over addition. The product of general multivector follows from the product of the basis versors using linearity and distributivity. Using the geometric product and the rules $\e_{0}^{2}=\eta_{0}$, $\e_{i}^{2}=\eta_{i}$, for a signature choice $\eta_{±}=±(-1,+1,+1,+1)$ associated to the extraspecial group $2_{±}^{5}$, and using the anticommutation of basis vectors, we build in Table~\ref{Cayley} the Cayley table of the groups. Doing so, we recognize that both extraspecial groups $2^{5}_{±}$ have a central product structure: each of them is the product of two subgroups that centralize each other.

\paragraph{Central product structure for $2^{5}_{-}$} This extraspecial group of negative type is the central product of the quaternion group with the dihedral group. This leads to
\begin{eqnarray}
\B_{i}\B_{j}&=& -\delta_{ij}+\sum_{k}\epsilon_{ijk}\B_{k}, \label{bivectors-minus}\\
 ℐ_{\aa} ℐ_{\bb}&=& \delta_{\aa\bb}(-1)^{\aa+1}+∑_{\cc}\zeta_{\aa\bb\cc} ℐ_{\cc}, \label{isoversors-minus} \\
	\B_{i}\times ℐ_{\aa} &=&  0 , \label{commutation-minus} 
\end{eqnarray}
where the cross product is the conventional half commutator of geometric algebra $\Mc \times \Oc ≐  (\Mc\Oc - \Oc\Mc)/2$ for two arbitrary multivectors $\Mc$ and $\Oc$. The totally antisymmetric tensor is $\epsilon_{ijk}$ with $\epsilon_{123}=1$ and $\epsilon_{213}=-1$ (and their cyclic permutations).  We introduce the tensor $\zeta_{\aa\bb\cc}$ antisymetric on its first two indices, non-null only if $\aa\bb\cc$ is a permutation of $\1\2\3$ and equal to $-1$ if $\aa < \bb$ and to $+1$ otherwise, i.e., 
\begin{eqnarray}
	\zeta_{\1\2\3}=\zeta_{\2\3\1}=\zeta_{\1\3\2}=-1, \\
	\zeta_{\2\1\3}=\zeta_{\3\2\1}=\zeta_{\3\1\2}=+1.
\end{eqnarray}
This tensor is the structure constant of the dihedral group of order $8$. By convention, we take $ ℐ_{\0} = \B_{0} = 1$. We remark that indices in the Latin alphabet run over space indices $1$, $2$, $3$, while indices in the Greek alphabet run over spacetime indices $0$, $1$, $2$, $3$. Then Eqs.~(\ref{bivectors-minus}--\ref{commutation-minus}) do not hold for $ ℐ_{\0}$ and $\B_{0}$.

\paragraph{Central product structure for $2^{5}_{+}$} Similarly, this extraspecial group of positive type is the central product of two quaternion groups. This leads to
\begin{eqnarray}
\B_{i}\B_{j}&=& -\delta_{ij}+∑_{k}\epsilon_{ijk}\B_{k}, \label{bivectors-plus}\\
 ℐ_{\aa} ℐ_{\bb}&=& -\delta_{\aa\bb}+∑_{\cc}\epsilon_{\aa\bb\cc} ℐ_{\cc}, \label{isoversors-plus} \\
	\B_{i}\times ℐ_{\aa} &=&  0. \label{commutation-plus} 
\end{eqnarray}

\begin{table}[t] \centering
	(a)\;\;\begin{tabular}{|c||c|c|c|c|}
		\hline
		   $ 1$ 			&\textcolor{red}{$\e_{1}\e_{2}\e_{3}$}	&\textcolor{red}{$\e_{0}\e_{1}\e_{2}\e_{3}$} &\textcolor{red}{$\e_{0}$}	 	\\ \hline \hline
		   \textcolor{blue}{$\e_{2}\e_{3}$}	&$-\e_{1}$ &$-\e_{0}\e_{1}$	&$\e_{0}\e_{2}\e_{3}$ 				\\ \hline
		   \textcolor{blue}{$\e_{3}\e_{1}$}	&$-\e_{2}$ &$-\e_{0}\e_{2}$	&$\e_{0}\e_{3}\e_{1}$ 				\\ \hline
		   \textcolor{blue}{$\e_{1}\e_{2}$}	&$-\e_{3}$ &$-\e_{0}\e_{3}$	&$\e_{0}\e_{1}\e_{2}$ 				\\ \hline
	\end{tabular} \hspace{0.2cm}
(b)\;\;\begin{tabular}{|c||c|c|c|c|}
		\hline 
		   $ 1$ 			&\textcolor{red}{$± \e_{0}^{\star}$}	&\textcolor{red}{$\ps$}	&\textcolor{red}{$\e_{0}$}	 	\\ \hline \hline
		   \textcolor{blue}{$\B_{1}$}	& $-\e_{1}$	&$-\B_{1}^{\star}$	&$± \e_{1}^{\star}$ 				\\ \hline
		   \textcolor{blue}{$\B_{2}$}	& $-\e_{2}$	&$-\B_{2}^{\star}$	&$± \e_{2}^{\star}$ 				\\ \hline
		   \textcolor{blue}{$\B_{3}$}	& $-\e_{3}$	&$-\B_{3}^{\star}$	&$± \e_{3}^{\star}$ 				\\ \hline
	\end{tabular} for versors of $2^{5}_{±}$ \\ ~ \\ ~ \\
(c)\;\;\begin{tabular}{|c||c|c|c|c|}
		\hline 
		   $ 1$ 			&\textcolor{red}{$ ℐ_{\1}$}		&\textcolor{red}{$ ℐ_{\2}$}	&\textcolor{red}{$ ℐ_{\3}$} 	\\ \hline \hline
		   \textcolor{blue}{$\B_{1}$}	&$ ℐ_{\1}\B_{1}$ 	&$ ℐ_{\2}\B_{1}$ 			&$ ℐ_{\3}\B_{1}$	\\ \hline
		   \textcolor{blue}{$\B_{2}$}	&$ ℐ_{\1}\B_{2}$ 	&$ ℐ_{\2}\B_{2}$ 			&$ ℐ_{\3}\B_{2}$	\\ \hline
		   \textcolor{blue}{$\B_{3}$}	&$ ℐ_{\1}\B_{3}$ 	&$ ℐ_{\2}\B_{3}$ 			&$ ℐ_{\3}\B_{3}$	\\ \hline
	\end{tabular} 	\hspace{0.2cm}
(d)\;\;	\begin{tabular}{|c||c|c|c|c|} \hline
		$\vide $ &\textcolor{red}{$\III$}&\textcolor{red}{$\IV$}& \textcolor{red}{$\I$}  \\ \hline \hline
		\textcolor{blue}{$\II$} &$\I$&$\II$&$\III$  \\ \hline
		\textcolor{blue}{$\II$} &$\I$&$\II$&$\III$  \\ \hline
		\textcolor{blue}{$\II$} &$\I$&$\II$&$\III$  \\ \hline
	\end{tabular} \qquad \qquad \qquad \quad
	\caption{Basis versors of spacetime algebras $\mathcal{G}(3,1)$ and $\mathcal{G}(1,3)$ in three different notations. 
	(a) Conventional notation in terms of the geometric product of the basis vectors generating the algebra; see the presentations of the extraspecial groups $2^{5}_{±}$. 
	(b) Dual notation in term of the pseudoscalar $\ps ≐ℐ_{2}$, the basis vectors and space bivectors with their duals. The relation between trivectors and vectors using duality depends on the signature: the sign $±$ is for the group $2^{5}_{±}$, e.g., $-\e_{0}^{\star} ≐ - \e_{0}\ps^{-1} = \e_{1}\e_{2}\e_{3}$ for $2^{5}_{-}$. 
	(c) Central notation with the space bivectors $\B_{i}$ and isoversors $ ℐ_{\aa}$. 
	(d) Grades in Latin numbers of each versor at the same location in other tables, from $\vide$ (real number) to $\IV$ (pseudoscalar). Versors at the same location in the tables (a), (b) and (c) are equal. \label{structure}}
\end{table}

\paragraph{Anticommutation and Cayley tables} Whatever the signature choice, we emphasize that the bivectors anticommute, and so do the isoversors:
\begin{eqnarray}
	\B_{i} \B_{j} &=& -\B_{j}\B_{i} \quad \text{if } i \neq j, \\
		 ℐ_{\aa}  ℐ_{\bb} &=& - ℐ_{\bb}  ℐ_{\aa} \quad \text{if } \aa \neq \bb. \label{anticommutation}
\end{eqnarray}
The Cayley table of the quaternion group for bivectors, the dihedral group of isoversors of $2^{5}_{-}$ and the quaternion group of isoversors of $2^{5}_{+}$ are given respectively in Table~(\ref{Cayley}a), (\ref{Cayley}b) and (\ref{Cayley}c). The multiplications with opposite group elements are not shown for simplicity.
\begin{table}[h]
	\centering
	(a)\;\;\begin{tabular}{|c||c|c|c|}
		\hline
		$2^{5}_{±}$ &$\B_{1}$	&$\B_{2}$	&$\B_{3}$	\\ \hline\hline
		$\B_{1}$	&$-1 $		&$\B_{3} $	&$-\B_{2} $	\\ \hline
		$\B_{2}$	&$-\B_{3} $	&$-1 $		&$\B_{1} $	\\ \hline
		$\B_{3}$	&$\B_{2} $	&$-\B_{1} $	&$-1 $	\\
		\hline
	\end{tabular} \hspace{0.2cm}
	(b)\;\;\begin{tabular}{|c||c|c|c|}
		\hline
		$2^{5}_{-}$&$ ℐ_{\1}$	&$ ℐ_{\2}$	&$ ℐ_{\3}$	\\ \hline\hline
		$ ℐ_{\1}$	&$1 $		&$- ℐ_{\3}$	&$- ℐ_{\2} $	\\ \hline
		$ ℐ_{\2}$	&$ ℐ_{\3}$&$-1 $		&$- ℐ_{\1} $	\\ \hline
		$ ℐ_{\3}$	&$ ℐ_{\2}$&$ ℐ_{\1}$&$1 $	\\
		\hline
	\end{tabular} \hspace{0.2cm}
	(c)\;\;\begin{tabular}{|c||c|c|c|}
	\hline
	$2^{5}_{+}$&$ ℐ_{\1}$	&$ ℐ_{\2}$	&$ ℐ_{\3}$	\\ \hline\hline
	$ ℐ_{\1}$	&$-1 $		&$ℐ_{\3}$		&$- ℐ_{\2} $	\\ \hline
	$ ℐ_{\2}$	&$-ℐ_{\3}$	&$-1 $		&$ℐ_{\1} $	\\ \hline
	$ ℐ_{\3}$	&$ ℐ_{\2}$	&$-ℐ_{\1}$	&$-1 $	\\
	\hline
	\end{tabular}
	\caption{(a) and (c), Cayley tables of the two quaternion groups in $2^{5}_{+}$. (a) and (b) Cayley tables of the quaternion and dihedral groups in $2^{5}_{-}$. Each group has eight elements: $1$, $±\B_{1}$, $±\B_{2}$, $±\B_{3}$ or $1$, $± ℐ_{\1}$, $± ℐ_{\2}$, $± ℐ_{\3}$ \label{Cayley}}
\end{table}

\paragraph{Notations} Table~\ref{structure} summarizes the different notations to write basis versors of $\mathcal{G}(3,1)$ and $\mathcal{G}(1,3)$ forming the group elements of $2^{5}_{+}$ and $2^{5}_{-}$: group elements at the same location in the three tables are identical. In the \emph{conventional notation} of Table \ref{structure}(a), one writes versors as a geometric product of basis vectors. This notation is useful for building the spacetime algebra, as we just did in Section \ref{groupstructure}, and for grasping the geometrical meaning of each versor (as subspaces such as lines, planes, or volumes). To shorten notation, one often encounters in geometric algebra the \emph{dual notation} of Table \ref{structure}(b). It uses duality to relate supplementary subspaces and their associated versors, e.g., $\e_{0}$ and $\e_{0}^{\star}=±\e_{1}\e_{2}\e_{3}$ for $2^{5}_{±}$. 
The central product structure guides us to the most convenient notation, given in Table \ref{structure}(c), that we call the \emph{central notation}.
In practice, this means that any basis versor writes as a space bivector, an isoversor, or a product of the two, whatever their order, given Eqs.~\eqref{commutation-minus} and \eqref{commutation-plus}. As a drawback of the central notation, grades are not immediately visible, but the commutation of space bivectors and isoversors greatly simplifies calculations. 

Given Eqs.~(\ref{bivectors-minus}--\ref{commutation-minus}) and Eqs.~(\ref{bivectors-plus}--\ref{commutation-plus}), all calculations in the framework of geometric algebra are equivalent to calculations with hypercomplex numbers. Readers not familiar with geometric algebra can take it for granted as an abstract algebra, and most of the remaining part of this article is accessible to them. Nonetheless, some concepts and terminology of geometric algebra are useful to introduce. For instance, grade projection is used to extract components of a given grade in a multivector; see Table (\Ref{structure}d). We denote by $(\dots)_{\vide}$ the projection of a multivector to zero grade $\vide$ corresponding to real numbers. More generally, we denote by $(\dots)_{\K}$ the projection to grade $\K$, i.e., the projection that keeps only terms of this grade in a multivector. Uppercase letters and Roman numerals are used for grades to distinguish them from indices of other types. This allows us to remove parentheses for the grade projection $(\Mc)_{\K} = \Mc_{\K}$. By convention, other projections (such as even/odd grade projection) are denoted using subscripts.

As mentioned, we use different fonts for indices labeling bivectors and isoversors.  By convention, the same letter with subscripts of different fonts, such as $\bbmu$ and $μ$, are different objects, e.g., $\Rc_{\bbmu}  ≠ \Rc_{\mu}$. 
Space refers to the three-dimensional ordinary space in which we define position. Spacetime is the four-dimensional space in which we define events occurring at a given position and time, or, for short, at a given spacetime position. Finally, we use boldface letters to denote blades in geometric algebra. A blade is the outer product of several vectors. For instance, a space bivector can always be written as a $\II-$blade (the exterior product of two vectors), but a spacetime bivector may not. For instance, one cannot find a $\II$-blade that is equal to the sum of two $\II$-blades in supplemental bivector subspaces of spacetime algebra. We remark that spacetime vectors are always $\I-$blades and spacetime trivectors are always $\III-$blades. Our notation, using different fonts for different elements of the algebra, as summarized in Table~\ref{NotationTable}, helps distinguish them, improving the readability of calculations. 

\begin{table} \centering
	\begin{tabular}{|L|l|}
			\hline
			\text{Notation} & Description \\
			\hline
			α,β,γ,\dots  & Real numbers \\
			a,b,c,\dots & Real numbers (usually components) \\
			\bm a, \bm b , \bm c, \dots & Spacetime vectors \\
			\vec a, \vec b, \vec c, \dots & Space vectors (in 3D space)\\
			\bm A, \bm B, \bm C, \dots & Space bivectors ($\II$-blades of 3D space) \\
			A, B, C, \dots & Spacetime bivectors (blades or sum of blades) \\
			\Ac, \Bc, \Cc, \dots & Multivectors \\
			\Ac_{\K}, \Bc_{\K}, \Cc_{\K}, \dots & Homogenous multivectors of grade $\K$ \\ 
			\nabla = \e_{0}∂_{0}+\vec \nabla & Spacetime gradient with $\vec \nabla = \e_{1}∂_{1}+\e_{2}∂_{2}+\e_{3}∂_{3}$ \\
			\mf, \mathsf{g},\dots & Operators acting on multivector	\\ 
			\text{Superscripts} & Involutions and powers\\
			\text{Subscript} & Projections and components\\
			\hline
			i,j,k, \dots & index in $1,2,3$ \\
			μ,ν,ζ, \dots & index in $0,1,2,3$ \\
			a,b,c, \dots & index in $0,1,2,3,4,5,6,7$ \\
			\aa,\bb,\cc, \dots & index in $\1,\2,\3$ or $\0,\1,\2,\3,\4,\5,\6,\7$ \\
			\I,\J,\K, \dots & Grades in roman numbering $\vide,\I,\II,\III,\IV$ \\ \hline
	\end{tabular}
	\caption{Notations used for specific elements inside spacetime algebra. For involutions, see Section~\ref{SecInvolutions}. Notation for operators is standard, although we also use Greek letters for historical reasons (Pauli, Dirac, and Gell-Mann operators). \label{NotationTable}}
\end{table}

\subsection{Involutions of spacetime algebra}
\label{SecInvolutions}

In matrix algebra, transposition, complex conjugation, and their combination (Hermitian conjugation) lead to the norm of a complex vector. Similarly, in geometric algebra, several involutions, some of which have geometric interpretations, yield the norm of a multivector. The composition of reversion and space- or time-reflection (according to the signature convention) yields a Hestenesian conjugation, first introduced by David Hestenes in Refs.~\cite{Hestenes1996vol, Hestenes1981vol}. It seems appropriate to name this involution in honor of David Hestenes given his pioneer role in geometric algebra. Then, Hermitian and Hestenesian conjugation apply, respectively, to complex vectors and to real multivectors to define their respective norms.

\paragraph{Reversion}  In geometric algebra, the reversion changes the order of the vectors appearing in a versor. It does not have an obvious geometric meaning: it just changes the signs of some basis versors. The reverse of a space bivector changes its sign 
\begin{equation}
	\forall i, \; \B_{i}^\mathsf{r}=-\B_{i}, 
\end{equation}
and for the isoversor we have
\begin{eqnarray}
	 ℐ_{\1}^\mathsf{r} &=& - ℐ_{\1}, \\
	 ℐ_{\2}^\mathsf{r} &=& + ℐ_{\2}, \\
	 ℐ_{\3}^\mathsf{r} &=& + ℐ_{\3},
\end{eqnarray}
for both signature conventions.

\paragraph{Space reflection } 
\label{space-reflection}
Space reflection is an involution that changes the sign of the space vectors $\e_{i}$, $i=1,2,3$, whatever the signature choice. It is defined by 
\begin{equation}
	\Mc^\mathsf{s}≐  ∓ ℐ_{\3}\Mc ℐ_{\3} = ∓ \e_{0}\Mc\e_{0}, \label{SpaceReflection}
\end{equation}
for $2^{5}_{±}$ for any grades versors in $\Mc$. With this definition $\e_{i}^\mathsf{s} = ∓ \e_{0}\e_{i}\e_{0} =  ± (\e_{0})^{2}\e_{i} = -\e_{i}$ while $\e_{0}^{\mathsf{s}}=\e_{0}$. 
Given that $ ∓ \e_{0}^{2}= 1$ and $ \e_{0} \times \B_{i} = 0$,
the action of space reflection on space bivectors does not depend on the signature convention
\begin{equation}
	\forall i, \; \B_{i}^\mathsf{s}= \B_{i}.
\end{equation}
Similarly, the action of space reflection on isoversors is independent of the signature choice:
\begin{eqnarray}
	 ℐ_{\1}^\mathsf{s} &=& - ℐ_{\1}, \\
	 ℐ_{\2}^\mathsf{s} &=& - ℐ_{\2}, \\
	 ℐ_{\3}^\mathsf{s} &=& + ℐ_{\3}. 
\end{eqnarray}
In geometric algebra, a reflection with respect to a hyperplane (a space with dimension three for spacetime algebra) is written as a conjugacy action based on the vector orthogonal to the hyperplane. It is the case in Eq.~\eqref{SpaceReflection} since the $\I$-vector $\e_{0}=  ℐ_{\3}$ is orthogonal to the space hyperplane $ ℐ_{1} = \e_{1}\e_{2}\e_{3}$ corresponding to the three-dimensional space. We emphasize that conjugation by a multivector does not change the grade; hence, space reflection commutes with grade projection. The space reflection also preserves the geometric product of the spacetime algebra, $( \Mc\Pc)^{\mathsf{s}} = \Mc^{\mathsf{s}}\Pc^{\mathsf{s}}$. Real numbers are preserved by space reflection: $1^\mathsf{s}=1$.

\paragraph{Time reflection } Time reflection is an involution that changes the sign of the time vector $\e_{0}$, whatever the signature choice. It is defined by 
\begin{equation}
\Mc_{\K}^\mathsf{t}≐  ∓(-1)^{\K} ℐ_{\3}\Mc _{\K}ℐ_{\3} = ∓(-1)^{\K} \e_{0}\Mc_{\K}\e_{0} \label{TimeReflection}
\end{equation}
for $2^{5}_{±}$ and for a versor $\Mc_{\K}$ of grade $\K$. With this definition $\e_{i}^\mathsf{t} = ± \e_{0}\e_{i}\e_{0} =  ∓ (\e_{0})^{2}\e_{i} = \e_{i}$ while $\e_{0}^{\mathsf{t}}=± (\e_{0})^{2}\e_{0}= -\e_{0}$.
The action of time reflection on space bivectors does not depend on the signature convention
\begin{equation}
	\forall i, \; \B_{i}^\mathsf{t}= \B_{i}.
\end{equation}
Similarly, the action of time reflection on isoversors is independent of the signature choice:
\begin{eqnarray}
	 ℐ_{\1}^\mathsf{t} &=& + ℐ_{\1}, \\
	 ℐ_{\2}^\mathsf{t} &=& - ℐ_{\2}, \\
	 ℐ_{\3}^\mathsf{t} &=& - ℐ_{\3}. 
\end{eqnarray}
Time reflection preserves the grade and the geometric product of the spacetime algebra, $( \Mc\Pc)^{\mathsf{t}} = \Mc^{\mathsf{t}}\Pc^{\mathsf{t}}$. Real numbers are preserved by time reflection: $1^\mathsf{t}=1$.

\paragraph{Hestenes conjugation} For $2^{5}_{-}$ and $\mathcal{G}(1,3)$ [respectively for $2^{5}_{+}$ and $\mathcal{G}(3,1)$],  the Hestenesian conjugate of $\Mc$ is the multivector $\Mc^{\dag}$ that is obtained from $\Mc$ by a combination of reversion and space reflection (respectively time reflection). Hence, we have
\begin{eqnarray}
	\Mc^{\dag} &≐& \Mc^\mathsf{rs} = \Mc^\mathsf{sr} \quad \text{ for } 2^{5}_{-}, \\
	\Mc^{\dag} &≐& \Mc^\mathsf{rt} = \Mc^\mathsf{tr} \quad \text{ for } 2^{5}_{+} .
\end{eqnarray} 
Hestenesian conjugation always changes the sign of the space bivectors and of the pseudoscalar; we say that they are antihestenian. Hestenian conjugation on $ℐ_{1}$ and $ℐ_{3}$ depends on the choice of signature  
\begin{equation}
	\forall i, \; \B_{i}^{\dag}= -\B_{i}.
\end{equation}
and
\begin{eqnarray}
	 ℐ_{\1}^{\dag}&=& ∓ ℐ_{\1}, \\
	 ℐ_{\2}^{\dag}&=& -ℐ_{\2}, \\
	 ℐ_{\3}^{\dag}&=& ∓ ℐ_{\3},
\end{eqnarray} 
for the groups $2^{5}_{±}$. Using Hestenesian conjugation, we define  the scalar product and the norm of multivectors as 
\begin{eqnarray}
	\Mc * \Vc &=& (\Mc \Vc^{\dag})_{\vide}, \label{scalprod} \\
	|\Mc| &≐& (\Mc \Mc^{\dag})_{\vide}^{1/2} \geq 0. \label{norm}
\end{eqnarray}  
Since reversion, space reflection, and time reflection commute with grade projection, Hestenian conjugation commutes with grade projection. Let's emphasize that the scalar product of Eq.~\eqref{scalprod} is symmetric.

\paragraph{Adjoint and self-adjoint operators} Given the scalar product of Eq.~\eqref{scalprod} and an operator $\mf$, we define its adjoint operator $\muf$ as the operator such that for any multivectors $\Mc$ and $\Vc$
\begin{equation}
	\mf(\Mc)*\Vc = \Mc * \muf(\Vc) \qquad \text{or} \qquad \Mc*\mf(\Vc) = \muf(\Mc) * \Vc,
\end{equation}
the two definitions being equivalent by symmetry of the scalar product. Self-adjoint operators verify $\muf=\mf$, and anti-adjoint $\muf=-\mf$.

\paragraph{Grade involution as spacetime reflection } The composition of space and time reflection is the grade involution of spacetime algebra, i.e., the involution that changes the sign of the odd grade components of a multivector
\begin{equation}
	\Mc^\mathsf{st}=\Mc^\mathsf{ts}=\sum_{\K}(-1)^{\K}\Mc_{\K}≐\Mc^\mathsf{o}
\end{equation}
where, as usual, we use superscripts for involutions. Here, $\mathsf{o}$ denotes \emph{odd}.

\begin{table}\centering
	\begin{tabular}{|C|C|l|} \hline
		\tilde{\Mc} & \Mc^\mathsf{r}		& Reversion of multivector $\Mc$ \\
		σ_{i} & ℐ_{\2}\B_{i} 				& Relative vectors basis 	\\
		γ_{μ} & \e_{μ} 						& Basis spacetime vectors  \\
		\hat{σ}_{i} & \B_{i}(\dots)\B_{3}& Pauli operators 			\\
		\hat{γ}_{μ} & γ_{μ}   				& Dirac operators 			\\
		\langle \dots\rangle_{k}& (\dots)_\K & Projection to grade $k=\K$ \\
		\Mc_{k} ⋅ \Nc_{l} = \langle \Mc_{k}\Nc_{l} \rangle_{|k-l|}& \Mc_{\K} ⋅ \Nc_{\L}=\left(\Mc_{\K} \Nc_{\L}\right)_{\K-\L} & Inner product convention \\
		 \hline
	\end{tabular}
	\caption{Comparison of the notations in Ref.~\cite{Doran2005} and in the present work. The use of the left contraction as the inner product is promoted in Ref.~\cite{Dorst2002vol}.}
\end{table}

%%%%%%%%%%%%%%%%%%%%%%%%%%%%%%%%%%%%%%%%%%%%%%%%%%%%%%%%%%%%%%%%%%%%%

\subsection{Bivector and Isoversor SU(2)}

We define below two different representations of the special unitary group of dimension two, one based on bivectors and another based on operators using isoversors that will turn out to be involved in the definition of Dirac operators. We focus on the definitions in this Section, intending to justify and physically ground them in Section~\ref{DefSpinorAndDyn}.

\paragraph{Bivector SU(2) as $2$ representation} 
A representation of the special unitary group SU(2) in matrix algebra is based on the exponential of the Pauli matrices multiplied by the complex number $\imath$
 \begin{equation}
	  \sigma^{1} = \begin{bmatrix}
		0& 1  \\
		1& 0 \\
	\end{bmatrix}, \quad
	\sigma^{2} =\begin{bmatrix}
		0 & -ι  \\
		ι &0 \\
	\end{bmatrix}, \quad
	\sigma^{3} =\begin{bmatrix}
		1& 0  \\
		0&-1 \\
	\end{bmatrix}. \label{PauliMatrices}
  \end{equation}
Those matrices act on Weyl's spinors defined as two-component complex vectors. Their equivalent in geometric algebra (see Section \ref{WeylSpinors} for a thorough explanation) are the so-called space spinors that are the sum of a real number and a space bivector, such as
\begin{equation}
	  \Sc = \phi_{0} + \phi_{1} \B_{1} + \phi_{2} \B_{2} + \phi_{3} \B_{3}.
\end{equation}
The Pauli operators (analog of Pauli matrices when acting on space spinors) are written in geometric algebra
\begin{equation}
σ^{i} ≐ \B_{i}(\dots)\B_{3}. \label{Pauli-op-GA}
\end{equation}

\begin{figure}[t] \centering
	\includegraphics[width=0.47\textwidth]{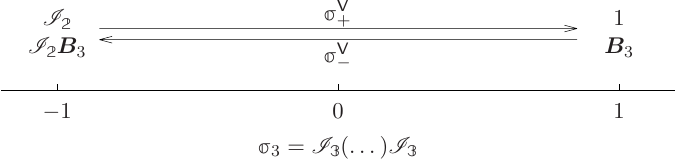} 
	\includegraphics[width=0.47\textwidth]{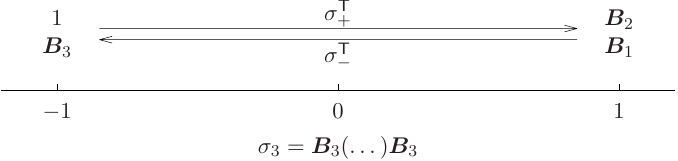}
	\caption{Weight diagrams of isoversor (left) and bivector (right) SU(2) representations. \label{SU2bivectorANDisoversor}}
\end{figure}

The multiplication by the unit complex number $\imath $ is given $(\dots)\B_{3}$. The down spin states are for $1$ and $\B_{3}$, while the up spin states for $\B_{1}$ and $\B_{2}$, see Figure \ref{SU2bivectorANDisoversor}.

The spin raising and lowering operators are $σ_{±}^\mathsf{T}$ defined as
(and using circular permutations of subscripts for the other operators)
\begin{eqnarray}
	σ^\mathsf{T}_{±}(\Qc) &=&\frac{-\B_{1}\Qc\B_{3}±\B_{2} \Qc}{2}, \label{AllMixingBivector1}  \\
	σ^\mathsf{U}_{±}(\Qc) &=&\frac{-\B_{3}\Qc\B_{2}±\B_{1} \Qc}{2}, \label{AllMixingBivector2} \\
	σ^\mathsf{V}_{±}(\Qc) &=&\frac{-\B_{2}\Qc\B_{1}±\B_{3} \Qc}{2}. \label{AllMixingBivector3}
\end{eqnarray}
All these raising operators transform $1$ into a basis bivector (and vice versa for the lowering operators). All the ladder operators square to the null operator as required. The action of these operators is illustrated in the diagram of Figure \ref{WeightsFundamental} for comparison with the ladder operators for the special unitary group of dimension 3. 
Cartan operators are obtained from the commutator of the raising and lowering operators:
\begin{eqnarray}
	\left[σ^\mathsf{T}_{+},σ^\mathsf{T}_{-}\right] &=& \B_{3}(...)\B_{3}, \label{CommutatorsSU2bivector1} \\
	\left[σ^\mathsf{U}_{+},σ^\mathsf{U}_{-}\right] &=& \B_{2}(...)\B_{2}, \\
	\left[σ^\mathsf{V}_{+},σ^\mathsf{V}_{-}\right] &=& \B_{1}(...)\B_{1}.  \label{CommutatorsSU2bivector3}
\end{eqnarray}

\paragraph{Isoversor SU(2) as $\2$ representation}
Another representation of the special unitary group of dimension two exists based on the Dirac operators (written here in standard representation)
\begin{equation}
	γ^{5}\,\!' 	= \begin{bmatrix} 0 & \1 \\ \1 & 0 \end{bmatrix}, \quad 
	\imath γ^{0}\,\!'γ^{5}\,\!' = \begin{bmatrix} 0 & \imath\1 \\ -\imath\1 & 0  \end{bmatrix}, \quad
	γ^{0}\,\!' 	= \begin{bmatrix} \1 & 0 \\ 0 & -\1 \end{bmatrix}. 
\end{equation}
This representation is $(-1)^{i+1}\sigma^{i}\otimes \1$ for $i=1,2,3$ (no sum), i.e., the tensor product of the Pauli matrices with the $2×2$ identity matrix denoted $\1$. Alternatively, in geometric algebra and when acting on spacetime spinors (i.e., even multivectors), the corresponding operators are written
\begin{eqnarray}
	\bbsigma^{1} &≐& ℐ_{\2}(\dots)\B_{3}, \label{ssigmas-def-1}\\
	\bbsigma^{2} &≐& -ℐ_{\1}(\dots)ℐ_{\3},\\
	\bbsigma^{3} &≐& ℐ_{\3}(\dots)ℐ_{\3}, \label{ssigmas-def-3}
\end{eqnarray}
although this is much less trivial than in matrix algebra.
We have the usual composition rule (up to a global sign)
\begin{equation}
	\bbsigma^{i}\circ \bbsigma^{j}(\dots)= δ^{ij}-\epsilon^{ijk}\bbsigma^{k}(\dots)\B_{3}
\end{equation}
with $\B_{3}$ playing the role of the complex number like for the $\B_{i}(\dots)\B_{3}$ operators. The generators $\bbsigma^{i}$ act on spinors defined by
\begin{equation}
	\Kc = k_{0}+k_{3}\B_{3}+ (\kbar_{0} + \kbar_{3}\B_{3})ℐ_{\2}\B_{3} \qquad \Leftrightarrow \qquad k = \begin{pmatrix} k_{0} + \imath k_{3} \\ \kbar_{0} +\imath\kbar_{3} \end{pmatrix}
\end{equation}
with the bar letters being for the lower spin components, cf. Figure \ref{SU2bivectorANDisoversor}. The lower and upper real components are $k_{0}$ and $\kbar_{0}$ respectively; the lower and upper imaginary components are $k_{3}$ and $\kbar_{3}$ respectively.
With this convention, the raising and lowering operators are
\begin{equation}
	\bbsigma^\mathsf{V}_{±}(\dots)=\frac{1}{2}\left[ ± \bbsigma^{1}(\dots)\B_{3}+\bbsigma^{2}(\dots) \right] .
\end{equation}
This, including circular permutations, write
\begin{eqnarray}
	\bbsigma^\mathsf{V}_{±}(\dots)&=& \frac{1}{2}\left[ ∓ℐ_{\2}(\dots)-ℐ_{\1}(\dots)ℐ_{\3} \right], \\
	\bbsigma^\mathsf{U}_{±}(\dots)&=& \frac{1}{2}\left[ ∓ℐ_{\3}(\dots)-ℐ_{\2}(\dots)ℐ_{\1} \right], \\
	\bbsigma^\mathsf{T}_{±}(\dots)&=& \frac{1}{2}\left[ ∓ℐ_{\1}(\dots)-ℐ_{\3}(\dots)ℐ_{\2} \right]. 
\end{eqnarray}
These ladder operators produce the diagram in Figure \ref{WeightsFundamentalIsoversorBis}. It is much less symmetric than the one in Figure \ref{WeightsFundamentalBis}. The diagram similarity arises from the arrows (associated with ladder operators of types $\mathsf{T,U,V}$) sharing the same positions on the two diagrams. The differences arise from the orientation of these arrows and from a possible phase of $-1$.
Cartan operators are obtained from the commutator of the raising and lowering operators:
\begin{eqnarray}
	\left[\bbsigma^\mathsf{T}_{+},\bbsigma^\mathsf{T}_{-}\right] &=& ℐ_{\1}(...)ℐ_{\1}, \label{CommutatorsSU2isoversor1}\\
	\left[\bbsigma^\mathsf{U}_{+},\bbsigma^\mathsf{U}_{-}\right] &=& ℐ_{\2}(...)ℐ_{\2}, \\
	\left[\bbsigma^\mathsf{V}_{+},\bbsigma^\mathsf{V}_{-}\right] &=& ℐ_{\3}(...)ℐ_{\3}.  \label{CommutatorsSU2isoversor3}
\end{eqnarray}

Finally, let's remark that the commuting operators of Eq.~\eqref{Pauli-op-GA} and Eqs.~(\ref{ssigmas-def-1}--\ref{ssigmas-def-3}), associated with $\1 \otimes σ^{i}$  $(-1)^{i+1}\sigma^{i}\otimes \1$, respectively, produce in geometric algebra the expected structure for the complexified Lie algebra of the Lorentz group
\begin{equation}
	\mathfrak{so}(1,3)_{\mathbb C}\cong
\mathfrak{sl}(2,\mathbb C)\oplus\mathfrak{sl}(2,\mathbb C).
\end{equation}

\subsection{Isoversor extended SU(3) of $\mathcal{G}(1,3)$}
\label{isoversorSU3}

We provide below, for the geometric algebra $\mathcal{G}(1,3)$, the generators of the special unitary group involving the isoversors $ℐ_{\aa}$ and one bivector only (we took $ \ B _ {3} $ arbitrarily here). The latter play the role of the complex number for the group. It can be placed on the left or right of the multivector on which the generators of the group act (we chose a right action to agree with the convention justified below, although a left action is preferable for switching to raising and lowering operators and massive gauge bosons). As usual, the unitary operators follow from exponentiation of a linear combination of the generators multiplied by the complex number of the group.

\subsubsection{Fundamental and conjugate representations}
\label{FundIsoversorSU3}

We define the fundamental (respectively conjugate) representation of the isoversor extended SU(3) continuous symmetry group in spacetime algebra from its generators $λ_{\aa}$ for $\aa=\1,\dots,\8$ in the left (respectively $\bar{λ}_{\aa}$ for the right) part of Table~\ref{IsoversorGroup}. We denote this four-dimensional representation $\3+\1$ (respectively $\bar{\3}+\1$). The structure constants $f_{abc}$, antisymmetric when permuting any two indices, follow from the commutators
\begin{equation}
	\boxed{[λ_{\aa},λ_{\bb}] = \sum_{\cc} f_{\aa\bb\cc}\,2 λ_{\cc}(\dots) \B_{3}}. \label{StructureCte}
\end{equation}
The non-zero structure constants are (modulo index permutations)
\begin{align}
	f_{\1\2\3}&=1, \qquad \qquad f_{\4\5\8} = f_{\6\7\8} = \frac{\sqrt{3}}{2}, \label{struct1}\\
	f_{\1\4\7}&=-f_{\1\5\6}=f_{\2\4\6} = f_{\2\5\7} = f_{\3\4\5} = - f_{\3\6\7} = \frac{1}{2}, \label{struct3} 
\end{align}
like for the Gell-Mann matrices \cite{GellMann1962vol125}, as reproduced in appendix~\ref{GuellMann}. 
The Hermitian nature of these matrices is translated in GA by the fact that $λ$ operators are self-adjoint. The traceless nature of these matrices means in geometric algebra that $\sum_{\bbmu}ℐ_{\bbmu}*λ_{\aa}(ℐ_{\bbmu})=0$.

We remark that the following relations hold between these operators: %\textcolor{blue}{OK with Bacry's book}
\begin{eqnarray}
	2\mathsf{T}_{\0}	&=&  λ_{\8} =  \frac{2}{\sqrt{3}}(\mathsf{U}_{\3}-\mathsf{V}_{\3})  , \label{T0opfirst} \\
	2\mathsf{U}_{\0}	&=&  -\frac{1}{2}(λ_{\3}\sqrt{3}+λ_{\8}), \\
	2\mathsf{V}_{\0}	&=& \frac{1}{2}(λ_{\3}\sqrt{3}-λ_{\8}). 
\end{eqnarray}
In particular, we find the zeroth and third components of $(\mathsf{T},\mathsf{U},\mathsf{V})$ satisfy
\begin{eqnarray}
	\mathsf{T}_{\3}+\mathsf{U}_{\3}+\mathsf{V}_{\3} &=& 0, \label{SumT3} \\
	\mathsf{T}_{\0}+\mathsf{U}_{\0}+\mathsf{V}_{\0} &=& 0. \label{SumTO}
\end{eqnarray}
Eqs.~(\ref{T0opfirst}--\ref{SumTO}) hold for the four continuous groups defined in Tables~\ref{IsoversorGroup} and \ref{BivectorGroup}, i.e., for operators with or without overbars, whatever the subscript type. 
Charge $\mathsf{Q}$, hypocharge $\mathsf{X}$ et hypercharge $\mathsf{Y}$ operators (each associated to one of the $T,U,V$ axis) are
\begin{eqnarray}
	-\mathsf{Q} &=& \frac{2\mathsf{U}_{\0}}{\sqrt{3}}=\mathsf{V}_{\3}-\frac{\mathsf{X}}{2}=-\mathsf{T}_{\3}-\frac{\mathsf{Y}}{2}, \label{chargeSU3isoversor} \\
	\mathsf{X} &=& \frac{2\mathsf{V}_{\0}}{\sqrt{3}}= \mathsf{T}_{\3}-\frac{\mathsf{Y}}{2}=-\mathsf{U}_{\3}+\frac{\mathsf{Q}}{2}, \\
	\mathsf{Y} &=& \frac{2\mathsf{T}_{\0}}{\sqrt{3}}= \mathsf{U}_{\3}+\frac{\mathsf{Q}}{2}=-\mathsf{V}_{\3}-\frac{\mathsf{X}}{2} = \frac{λ_{\8}}{\sqrt{3}}, \label{hypercharge}
\end{eqnarray}
Common eigenstates of operators $\mathsf{T}_{\3}$ and $\mathsf{Y}$ are the isoversors $ℐ_{\bbmu}$ as shown on Figure~\ref{WeightsFundamental}. We emphasize that 
\begin{equation}
	\forall \aa \in \{\1,\2,\dots,\8\}, \; λ_{\aa}(ℐ_{\1})=0,
\end{equation}
as shown on Table.~\ref{SU3images} that also provides the images of the isoversors $ℐ_{\bbmu}$ by the $λ_{\aa}$ operators.

\begin{figure} \centering
	\includegraphics[height=7cm]{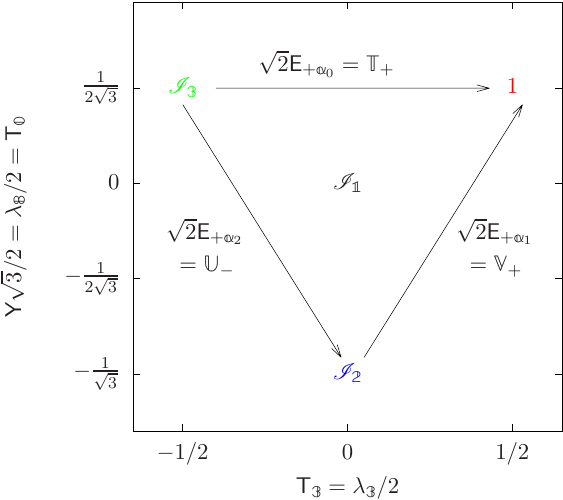}\includegraphics[height=7cm]{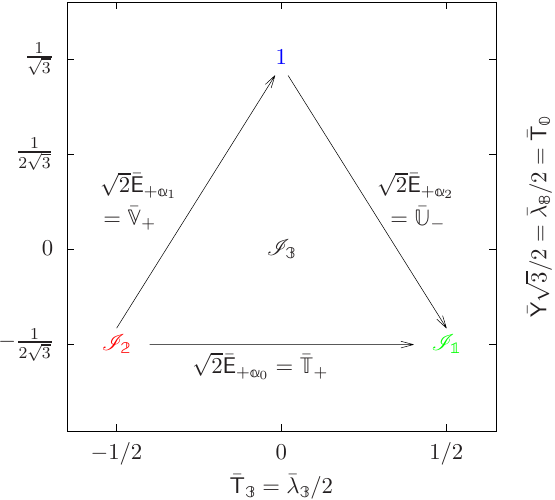}
	\caption{Weight diagrams of the fundamental (left) and conjugate (right) representations $\3\oplus 1$ and $\bar{\3} \oplus 1$ respectively, for the four common eigenstates of operators $\mathsf{T}_{\3}$ (eigenvalue in abscissa) and $\mathsf{Y}$ (eigenvalue in ordinate). Arrows are associated with raising operators $\mathsf{E}_{+\bbalpha_{n}}$. Operator $\mathsf{E}_{-\bbalpha_{n}}$ goes in opposite direction. 
	\label{WeightsFundamentalIsoversor}
	} \vspace{1em}\hspace{-1cm}
	\includegraphics[height=7cm]{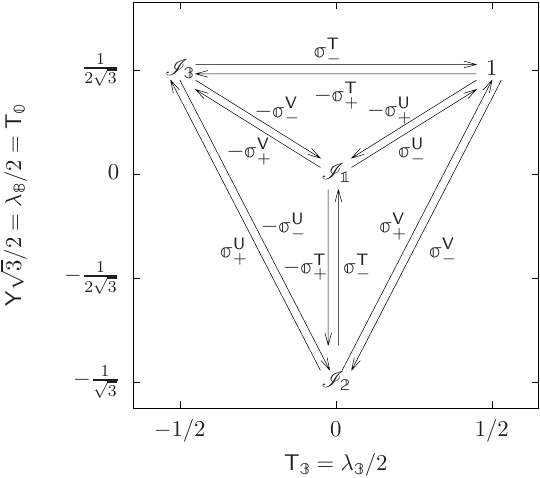}
	\caption{ Weight diagrams of the three different representations for isoversor SU(2) for the four common eigenstates of operators $\mathsf{T}_{\3}$ (eigenvalue in abscissa) and $\mathsf{Y}\sqrt{3}/2$ (eigenvalue in ordinate) that are not Cartan operators according to Eqs.~(\ref{CommutatorsSU2isoversor1}--\ref{CommutatorsSU2isoversor3}). \label{WeightsFundamentalIsoversorBis}}
\end{figure}

\begin{table} \centering
{\renewcommand{\arraystretch}{1.3}
	\begin{tabular}{l||cccc}
					&	$1$				&	$ℐ_{\1}$	&	$ℐ_{\2}$	&	$ℐ_{\3}$  \\ \hline \hline
		$λ_{\1}$	&	$-ℐ_{\3}\B_{3}$	&	$0$	&	$0$	&	$\B_{3}$              \\
		$λ_{\2}$	&	$ ℐ_{\3}$			&	$0$	&	$0$	&	$1$                    \\ \hline
		$λ_{\3}$	&	$(+1)$				&	$0$	&	$0$	&	$(-1)ℐ_{\3}$              \\ \hline
		$λ_{\4}$	&	$ℐ_{\2}$			&	$0$	&	$1$	&	$0$                 \\
		$λ_{\5}$	&	$ℐ_{\2}\B_{3}$	&	$0$	&	$-\B_{3}$	&	$0$                 \\
		$λ_{\6}$	&	$0$				&	$0$	&	$-ℐ_{\3}\B_{3}$	&	$ℐ_{\2}\B_{3}$                 \\
		$λ_{\7}$	&	$0$				&	$0$	&	$-ℐ_{\3}$	&	$-ℐ_{\2}$                 \\ \hline
		$λ_{\8}$	&	($1/\sqrt{3})$	&	$0$	&	$(-2/\sqrt{3})ℐ_{\2}$	&	$(1/\sqrt{3})ℐ_{\3}$                 \\ \hline
	\end{tabular} \hspace{1cm}
	\begin{tabular}{l||cccc}
				&	$1$				&	$ℐ_{\1}$	&	$ℐ_{\2}$	&	$ℐ_{\3}$  \\ \hline \hline
	$\bar{λ}_{\1}$	&	$0$	&	$-ℐ_{\2}\B_{3}$	&	$ℐ_{\1}\B_{3}$	&	$0$              \\
	$\bar{λ}_{\2}$	&	$0$	&	$ℐ_{\2}$	&	$ℐ_{\1}$	&	$0$                    \\ \hline
	$\bar{λ}_{\3}$	&	$0$				&	$(+1)ℐ_{\1}$	&	$(-1)ℐ_{\2}$	&	$0$              \\ \hline
	$\bar{λ}_{\4}$	&	$ℐ_{\2}$			&	$0$	&	$1$	&	$0$                 \\
	$\bar{λ}_{\5}$	&	$ℐ_{\2}\B_{3}$	&	$0$	&	$-\B_{3}$	&	$0$                 \\
	$\bar{λ}_{\6}$	&	$-ℐ_{\1}\B_{3}$				&	$\B_{3}$	&	$0$	&	$0$                 \\
	$\bar{λ}_{\7}$	&	$ℐ_{\1}$				&	$1$	&	$0$	&	$0$                 \\ \hline
	$\bar{λ}_{\8}$	&	($2/\sqrt{3})$	&	$(-1/\sqrt{3})ℐ_{\1}$	&	$(-1/\sqrt{3})ℐ_{\2}$	&	$0$                 \\ \hline
\end{tabular}
}
	\caption{Images by $λ_{\aa}$ (Left) and $\bar{λ}_{\aa}$ (Right) for $\aa \in \{\1,\dots,\8\}$ of $1$ and the three isoversors $ℐ_{\bbmu}$ for $\bbmu = \1,\2,\3$. \label{SU3images}
	}
\end{table}

In Section \ref{DefSpinorAndDyn}, we define charge conjugation of a general spacetime multivector. This definition is crucial to introduce the conjugate representations of the special unitary group for isoversors. Indeed, the interaction operators, when acting on the charge-conjugate partner of each particle, are written using operators in the conjugate representation.
We will see that charge conjugation exchanges the role of isoversor $1$ and $ℐ_{\2}$ and those of $ℐ_{\3}$ and $ℐ_{\1}$; therefore, this must be so when going from the fundamental to the conjugate representations of SU(3). 
As required, the commutators of operators in the conjugate representation yield the same structure constants as in the fundamental representation; see Eqs.~(\ref{StructureCte}--\ref{struct3}).

\subsubsection{Ladder operators and eigenstates}

The raising and lowering operators of isoversor SU(3) are written
\begin{eqnarray}
	\mathsf{E}_{(±1,0)}(\dots)&=&\frac{∓λ_{\1}(\dots)\B_{3}+λ_{\2}(\dots)}{2\sqrt{2}} = \mathsf{E}_{±\bbalpha_{0}}(\dots) = \frac{\mathbb{T}_{±}(\dots)}{\sqrt{2}}, \label{Tpm}  \\
		\mathsf{E}_{(±1/2,±\sqrt{3}/2)}(\dots)&=&\frac{λ_{\4}(\dots)±λ_{\5}(\dots)\B_{3}}{2\sqrt{2}} = \mathsf{E}_{±\bbalpha_{1}}(\dots)= \frac{\mathbb{V}_{±}(\dots)}{\sqrt{2}},\\
			\mathsf{E}_{(∓1/2,±\sqrt{3}/2)}(\dots)&=&\frac{±λ_{\6}(\dots)\B_{3}-λ_{\7}(\dots)}{2\sqrt{2}} =\mathsf{E}_{∓\bbalpha_{2}}(\dots)= \frac{\mathbb{U}_{±}(\dots)}{\sqrt{2}}, \label{Vpm}
\end{eqnarray}
where we have introduced the weight vectors 
\begin{align}
	+\bbalpha_{0} &= (+1,0), 					&-\bbalpha_{0} &= (-1,0), 					\\
	+\bbalpha_{1} &= (+1/2,+\sqrt{3}/2), 	&-\bbalpha_{1} &= (-1/2,-\sqrt{3}/2), 	\\
	+\bbalpha_{2} &= (+1/2,-\sqrt{3}/2), 	&-\bbalpha_{2} &= (-1/2,+\sqrt{3}/2).
\end{align}
The conjugate raising and lowering operators are similarly defined 
\begin{eqnarray}
	\bar{\mathsf{E}}_{(±1,0)}(\dots)&=&\frac{∓\bar{λ}_{\1}(\dots)\B_{3}+\bar{λ}_{\2}(\dots)}{2\sqrt{2}} = \bar{\mathsf{E}}_{±\bbalpha_{0}}(\dots) = \frac{\bar{\mathbb{T}}_{±}(\dots)}{\sqrt{2}}, \label{Tpmbar}  \\
	\bar{\mathsf{E}}_{(±1/2,±\sqrt{3}/2)}(\dots)&=&\frac{\bar{λ}_{\4}(\dots)±\bar{λ}_{\5}(\dots)\B_{3}}{2\sqrt{2}} =\bar{ \mathsf{E}}_{±\bbalpha_{1}}(\dots)= \frac{\bar{\mathbb{V}}_{±}(\dots)}{\sqrt{2}},\\
	\bar{\mathsf{E}}_{(∓1/2,±\sqrt{3}/2)}(\dots)&=&\frac{∓\bar{λ}_{\6}(\dots)\B_{3}+\bar{λ}_{\7}(\dots)}{2\sqrt{2}} =\bar{\mathsf{E}}_{∓\bbalpha_{2}}(\dots)= \frac{\bar{\mathbb{U}}_{±}(\dots)}{\sqrt{2}}, \label{Vpm}
\end{eqnarray}
adding an overbar (although there is a sign change for the last equation). The action of these operators is summarized in Fig.~\ref{WeightsFundamental}.

\subsubsection{Projectors of isoversor SU(3)}
\label{ProjIsoversor}
Let's define the linear operators
\begin{eqnarray}
	\mathsf{K}_{\0}(\dots) &≐& -ℐ_{\1} (\dots)ℐ_{\1}+ℐ_{\2} (\dots)ℐ_{\2}-ℐ_{\3} (\dots)ℐ_{\3}, \\
	\mathsf{K}_{\1}(\dots) &≐& -ℐ_{\1} (\dots)ℐ_{\1}-ℐ_{\2} (\dots)ℐ_{\2}+ℐ_{\3} (\dots)ℐ_{\3}, \label{K11}\\
	\mathsf{K}_{\2}(\dots) &≐& +ℐ_{\1} (\dots)ℐ_{\1}+ℐ_{\2} (\dots)ℐ_{\2}+ℐ_{\3} (\dots)ℐ_{\3}, \\
	\mathsf{K}_{\3}(\dots) &≐& +ℐ_{\1} (\dots)ℐ_{\1}-ℐ_{\2} (\dots)ℐ_{\2}-ℐ_{\3} (\dots)ℐ_{\3}. \label{K33}
\end{eqnarray}
The eigenstates of these operators are the isoversors $ℐ_{\bbmu}$ for $\bbmu=\0,\1,\2,\3$, with eigenvalues are $-3$ or $+1$ according to
\begin{eqnarray}
	\mathsf{K}_{\bbmu}(ℐ_{\bbnu}) = \left \{ \begin{array}{l} (-3) ℐ_{\bbmu}  \quad \text{ if } \bbmu = \bbnu, \\ (+1)ℐ_{\bbnu} \quad \text{ if } \bbmu ≠ \bbnu.  \end{array}
	 \right.
\end{eqnarray}
From these operators, we define the four projectors
\begin{equation}
	\mathsf{P}_{\bbmu}(\Qc) ≐ \frac{\Qc - \mathsf{K}_{\bbmu} (\Qc)}{4} \label{isoversorPROJ}
\end{equation}
that extracts (from the quadrispinor $\Qc$) the term in $ℐ_{\bbmu}$ and its associated space spinor:
\begin{equation}
	\mathsf{P}_{\bbmu} (ℐ_{\bbnu}) = δ_{\bbnu,\bbmu} ℐ_{\bbmu}.
\end{equation}
We could also introduce the supplemental projector (such that $\mathsf{P}_{\bbmu}+ \mathsf{P}^{\perp}_{\bbmu} = \mathsf{Id}$) that removes the spinorial components attached to $ℐ_{\bbmu}$
\begin{equation}
	\mathsf{P}^{\perp}_{\bbmu}(\Qc) ≐ \frac{3\Qc + \mathsf{K}_{\bbmu} (\Qc)}{4}   .
\end{equation}
Let's finally remark that 
\begin{eqnarray}
	ℐ_{\1}\mathsf{K}_{\2}(\dots) &=& \mathsf{K}_{\3}(ℐ_{\1} \dots), \label{pregauge1} \\
	ℐ_{\2}\mathsf{K}_{\3}(\dots) &=& \mathsf{K}_{\1}(ℐ_{\2} \dots), \label{pregauge} \\
	ℐ_{\3}\mathsf{K}_{\1}(\dots) &=& \mathsf{K}_{\2}(ℐ_{\3} \dots). \label{pregauge3} 
\end{eqnarray}

\subsection{Bivector extended SU(3) of $\mathcal{G}(1,3)$}

We provide below, for the geometric algebra $\mathcal{G}(1,3)$, the special unitary group that involves the bivectors $\B_{i}$ and the pseudoscalar $ℐ_{\2}$ only. The latter serves as the complex number in the group. It can be placed on the left or right of the multivector on which the group's generators act (we chose a right action arbitrarily, although a left action is preferable for switching to raising and lowering operators).

\subsubsection{Fundamental and conjugate representations}
\label{FundBivectorSU3}

We define the fundamental (respectively conjugate) representation of the bivector SU(3) continuous symmetry group in spacetime algebra from its generators $λ_{a}$ for $a=1,\dots,8$ in the left (respectively $\bar{λ}_{a}$ for the right) part of Table~\ref{BivectorGroup}. We denote this four-dimensional representation $3\oplus 1$ (respectively $\bar{3}\oplus 1$).
For both, the structure constants are the same as those of Eqs.~(\ref{struct1}--\ref{struct3}): 
\begin{equation}
	\boxed{[λ_{a},λ_{b}] = \sum_{c} f_{abc}\,2 λ_{c}(\dots)ℐ_{\2}}. \label{structureSU3bivector}
\end{equation}
These operators are self adjoint and traceless $\sum_{μ} \B_{μ}*λ(\B_{μ}) =0$. In Ref.~\cite{Hestenes1982vol12}, David Hestenes introduces very similar generators of SU(3). In this reference, the generators act only on spacetime bivector fields, such as $\sum_{i}(a_{i}+ℐ_{\2}b_{i})B_{i}$. Moreover, the operator $λ_{8}$ is defined with the correct eigenvalues, but produces incorrect structure constants. More precisely, one does not have $λ_{8}=2(\mathsf{U}_{3}-\mathsf{V}_{3})$ as required for special unitary group of degree three.
For the operator defined in Table~\ref{BivectorGroup}, analogous relations or similar definitions to those in Eqs.~(\ref{SumT3}--\ref{hypercharge}) also hold straightforwardly.

Let's note that the square of each generator yields a projector, as in matrix algebra. However, the matrix representations of generators of the special unitary groups act on complex vectors of given dimension (at least three in the fundamental representation $(1,0)≐3$ and its conjugate representation $(0,1)≐\bar 3$, higher specific dimensions for the other representation e.g., $(1,1)$). On the other hand, the generators given in Table~\ref{BivectorGroup} can act on any multivector. They can even act on space spinors, the GA equivalent of two-component complex vectors of matrix algebra (Weyl's spinors).

\begin{figure} \centering
	\includegraphics[height=7cm]{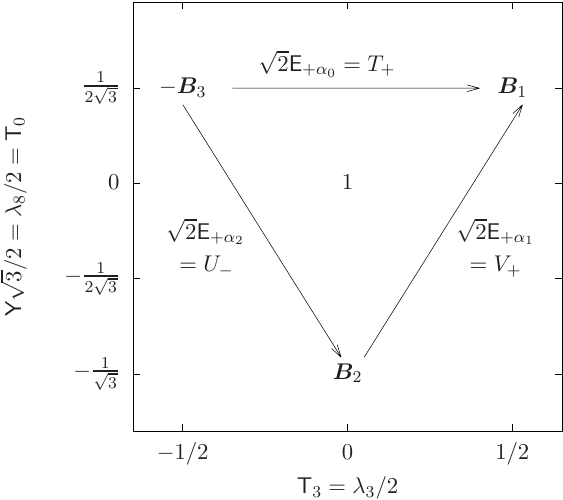}
	\includegraphics[height=7cm]{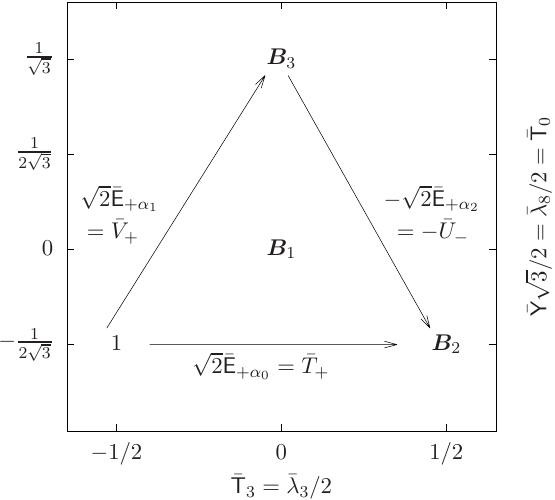}
	\caption{Weight diagrams of the fundamental (left) and conjugate (right) representations $3\oplus 1$ and $\bar 3 \oplus 1$ respectively, for the four common eigenstates of operators $\mathsf{T}_{3}$ (eigenvalue in abscissa) and $\mathsf{Y}\sqrt{3}/2$ (eigenvalue in ordinate). Arrows are associated with raising operators $\mathsf{E}_{+\alpha_{n}}$, operator $\mathsf{E}_{-\alpha_{n}}$ goes in the opposite direction. 
	\label{WeightsFundamental}
	} \vspace{0.5em}\hspace{-1cm}
	\includegraphics[height=7cm]{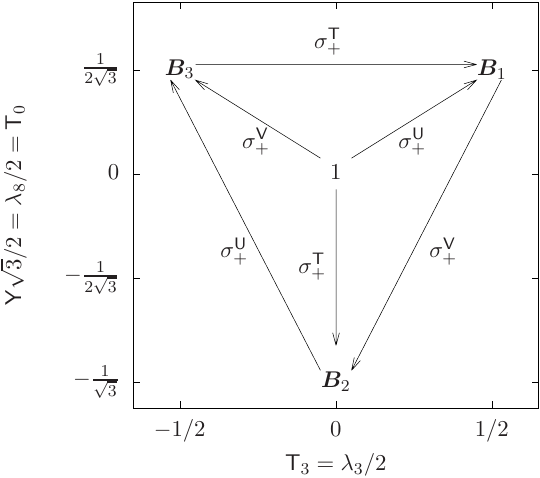}
	\caption{Weight diagrams of the three different representations for bivector SU(2) for the four common eigenstates of operators $\mathsf{T}_{3}$ (eigenvalue in abscissa) and $\mathsf{Y}\sqrt{3}/2$ (eigenvalue in ordinate). 
	For each arrow with a raising operator $\sigma^\mathsf{T,U,V}_{+}$, the opposite arrow (not shown) exists for the lowering operator $\sigma^\mathsf{T,U,V}_{-}$.
	\label{WeightsFundamentalBis}
	}
\end{figure}

\begin{table}
\centering
{\renewcommand{\arraystretch}{1.3}
\begin{tabular}{L||CCCC} 
			&	1	&	\B_{1}			&	\B_{2}		& \B_{3} \\ \hline \hline
	λ_{1}	&	0	&	-\B_{3}			&	0				&	-\B_{1}	\\ 
	λ_{2}	&	0	&	-ℐ_{\2}\B_{3}	&	0				&	ℐ_{\2}\B_{1}	\\ \hline
	λ_{3}	&	0	&	\B_{1}			&	0				&	-\B_{3}	\\ \hline
	λ_{4}	&	0	&	\B_{2}			&	\B_{1}		&	0	\\ 
	λ_{5}	&	0	&	ℐ_{\2}\B_{2}	&-ℐ_{\2}\B_{1}	&	0	\\ 
	λ_{6}	&	0	&	0					&	-\B_{3}		&	-\B_{2}	\\ 
	λ_{7}	&	0	&	0					&ℐ_{\2}\B_{3}	&	-ℐ_{\2}\B_{2}	\\ \hline
	λ_{8}	&	0	&	(1/\sqrt{3})\B_{1}	&	(-2/\sqrt{3})\B_{2}	&	(1/\sqrt{3})\B_{3}	\\ \hline
\end{tabular}
\begin{tabular}{L||CCCC} 
			&	1	&	\B_{1}			&	\B_{2}		& \B_{3} \\ \hline \hline
	\bar{λ}_{1}	&	-\B_{2}			&	0	&	-1			& 	0				\\ 
	\bar{λ}_{2}	&	ℐ_{\2}\B_{2}	&	0	&	-ℐ_{\2}	& 	0				\\ \hline
	\bar{λ}_{3}	&	-1					&	0	&	\B_{2}			&	0				\\ \hline
	\bar{λ}_{4}	&	-\B_{3}			&	0	&	0					&	-1		\\ 
	\bar{λ}_{5}	&	ℐ_{\2}\B_{3}	&	0	&	0					& -ℐ_{\2}\\ 
	\bar{λ}_{6}	&	0					&	0	&	-\B_{3}			& 	-\B_{2}		\\ 
	\bar{λ}_{7}	&	0					&	0	&	ℐ_{\2}\B_{3}	& -ℐ_{\2}\B_{2}	\\ \hline
	\bar{λ}_{8}	&	(-1/\sqrt{3})	&	0	&	(-1/\sqrt{3})\B_{2}	&	(2/\sqrt{3})\B_{3}	\\ \hline
\end{tabular}
}
	\caption{Images by $λ_{a}$ (Left) and $\bar{λ}_{a}$ (Right) for $a \in \{1,\dots,8\}$ of $1$ and the three bivectors $\B_{\mu}$ for $\mu = 1,2,3$. \label{SU3imagesbivector}}
\end{table}

\subsubsection{Ladder operators and eigenstates}

The raising and lowering operators of bivector SU(3) writes
\begin{eqnarray}
	\mathsf{E}_{(±1,0)}(\dots)&=&\frac{λ_{1}(\dots)±λ_{2}(\dots)ℐ_{\2}}{2\sqrt{2}} = \mathsf{E}_{±α_{0}}(\dots) = \frac{\mathsf{T}_{±}}{\sqrt{2}},   \\
		\mathsf{E}_{(±1/2,±\sqrt{3}/2)}(\dots)&=&\frac{λ_{4}(\dots)±λ_{5}(\dots)ℐ_{\2}}{2\sqrt{2}} = \mathsf{E}_{±α_{1}}(\dots) = \frac{\mathsf{V}_{±}}{\sqrt{2}},\\
			\mathsf{E}_{(∓1/2,±\sqrt{3}/2)}(\dots)&=&\frac{λ_{6}(\dots)±λ_{7}(\dots)ℐ_{\2}}{2\sqrt{2}} = \mathsf{E}_{∓α_{2}}(\dots) = \frac{\mathsf{U}_{±}}{\sqrt{2}},
\end{eqnarray}
for the same weight vectors as for isoversor SU(3)
\begin{align}
	+α_{0} &= (+1,0), 					&-α_{0} &= (-1,0), 					\\
	+α_{1} &= (+1/2,+\sqrt{3}/2), 	&-α_{1} &= (-1/2,-\sqrt{3}/2), 	\\
	+α_{2} &= (+1/2,-\sqrt{3}/2), 	&-α_{2} &= (-1/2,+\sqrt{3}/2).
\end{align}
We use different fonts for these weight vectors to indicate which group the raising and lowering operators belong to. As before, the conjugate raising and lowering operators are similarly defined 
\begin{eqnarray}
	\bar{\mathsf{E}}_{(±1,0)}(\dots)&=&\frac{\bar{λ}_{1}(\dots)∓\bar{λ}_{2}(\dots)ℐ_{\2}}{2\sqrt{2}} = \bar{\mathsf{E}}_{±α_{0}}(\dots) = \frac{\bar{\mathsf{T}}_{±}}{\sqrt{2}},   \\
		\bar{\mathsf{E}}_{(±1/2,±\sqrt{3}/2)}(\dots)&=&\frac{-\bar{λ}_{4}(\dots)∓\bar{λ}_{5}(\dots)ℐ_{\2}}{2\sqrt{2}} = \bar{\mathsf{E}}_{±α_{1}}(\dots) = \frac{\bar{\mathsf{V}}_{±}}{\sqrt{2}},\\
			\bar{\mathsf{E}}_{(∓1/2,±\sqrt{3}/2)}(\dots)&=&\frac{\bar{λ}_{6}(\dots)±\bar{λ}_{7}(\dots)ℐ_{\2}}{2\sqrt{2}} = \bar{\mathsf{E}}_{∓α_{2}}(\dots) = \frac{\bar{\mathsf{U}}_{±}}{\sqrt{2}}.
\end{eqnarray}

\subsubsection{Projectors of bivector SU(3)}
\label{ProjBivector}

Let's define the linear operators 
\begin{eqnarray}
	\mathsf{K}_{0} &≐& +\B_{1}(\dots)\B_{1}+\B_{2}(\dots)\B_{2}+\B_{3}(\dots)\B_{3}, \\
	\mathsf{K}_{1} &≐& +\B_{1}(\dots)\B_{1}-\B_{2}(\dots)\B_{2}-\B_{3}(\dots)\B_{3}, \\
	\mathsf{K}_{2} &≐& -\B_{1}(\dots)\B_{1}+\B_{2}(\dots)\B_{2}-\B_{3}(\dots)\B_{3}, \\
	\mathsf{K}_{3} &≐& -\B_{1}(\dots)\B_{1}-\B_{2}(\dots)\B_{2}+\B_{3}(\dots)\B_{3}. 
\end{eqnarray}
The eigenstates of these operators are $\B_{\mu}$ for $\mu=0,1,2,3$, with eigenvalues are $-3$ or $+1$ according to
\begin{eqnarray}
	\mathsf{K}_{μ}(\B_{ν}) = \left \{ \begin{array}{l} (-3) \B_{\mu}  \quad \text{ if } \mu = \nu, \\ (+1)\B_{\nu} \quad \text{ if } \mu ≠ \nu.  \end{array}
	 \right.
\end{eqnarray}
From these operators, we define the four projectors and supplemental projectors
\begin{eqnarray}
	P_{μ}(\Qc) &=& \frac{\Qc-\mathsf{K}_{μ}(\Qc)}{4},  \\
	P^{\perp}_{μ}(\dots) &=& \frac{3\Qc +\mathsf{K}_{μ}(\Qc)}{4},  
\end{eqnarray}
that respectively keeps only the terms proportional to $\B_{μ}$ or remove them all respectively. This suggests that two different and complementary four-fold decompositions of spacetime algebra are possible: one with space spinors in product of isoversors that we use in the following, and one in which the role of bivectors and isoversors is exchanged.

\subsection{Tensor product of representations}

\begin{figure} \centering
	\includegraphics{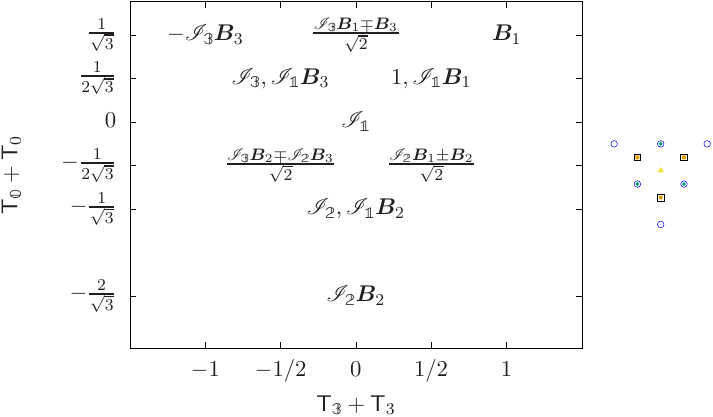}
	\caption{Eigenstates of the tensor product representation $(3\oplus 1)\otimes (\3\oplus \1) = 6 \oplus \bar{3} \oplus 3 \oplus 3 \oplus 1$ as shown on the right weight diagram. Full blue dots of representation $\bar 3$ are for the linear combination with the lower sign in the left plot, while the blue circles of representation $6$ use the three linear combinations with the upper sign in the left plot. Similarly, there are two representations $3$: the filled and the open squares, respectively. One is associated with versors before the comma and the other with those after the comma. The $(\bar 3\oplus 1) \otimes (\bar{\3}\oplus \1)$ has the weight diagram with opposite weights associated to different eigenstates (of the conjugate Cartan operators).
	\label{4*44}}
	\vspace{1em}

	\includegraphics{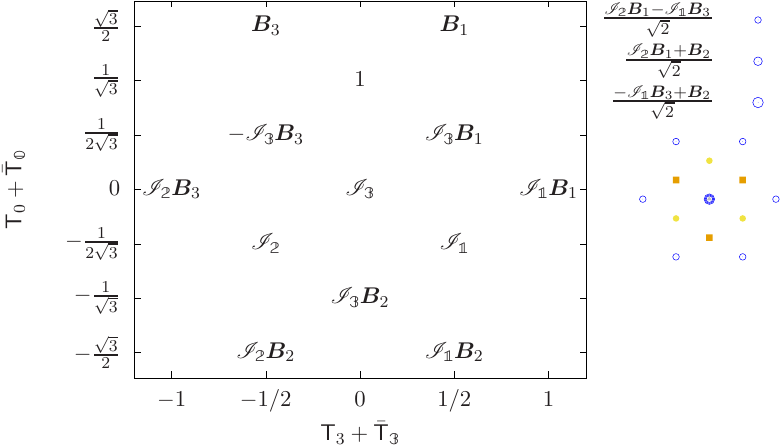}
	\caption{Eigenstates of the tensor product representation $(3\oplus 1)\otimes (\bar{\3}\oplus \1) = 9 \oplus \bar{3} \oplus 3 \oplus 1$ as shown on the right weight diagram. Open blue dots of representation $ 9$ include the three weights at the origin, with eigenstates given in the legend. Similarly, the representation $\bar{\3}$ is given by the yellow dots and representation $3$ by the orange filled squares. The singlet is the grey triangle; it can be associated with a $\bar{\3}$ to recover the four-dimensional representation $\bar{\3}\oplus \1$ generated by the isoversors. The $(\bar{3}+1) \otimes (\3+\1)$ has the same weight diagram, since it is symmetric with respect to the origin, but also has different eigenstates (of the conjugate Cartan operators).
	\label{4*44b}
	}
\end{figure}

The ladder operators of bivector and isoversor SU(3) (extended) groups commute in any representation (fundamental or conjugate). The Cartan generators, chosen as $(\mathsf{T}_{0},\mathsf{T}_{3})$ and $(\mathsf{T}_{\0},\mathsf{T}_{\3})$ also commute. Therefore, it is possible to define tensor product representations such as 
\begin{equation}
	(3\oplus 1) \otimes (\3\oplus \1), \quad 
	(3\oplus 1) \otimes (\bar {\3} \oplus \1), \quad 
	(\bar 3\oplus 1) \otimes (\3\oplus \1), \quad
	(\bar 3\oplus 1) \otimes (\3\oplus \1).
\end{equation}
For the former two, the eigenstates and weight diagrams are shown in Fig.~\ref{4*44} and \ref{4*44b}, respectively. The ladder operators are $\mathsf{E}_{±α} ≐ \mathsf{E}_{±α}+\mathsf{E}_{±\bbalpha}$ and $\mathsf{E}_{±α}+\bar{\mathsf{E}}_{±\bbalpha}$ respectively. Those operators yield the expected eigenstates on the weight diagram, although the state normalization is not correctly preserved in all cases upon application of the ladder operators. In the same line, we find a representation $9$ in the tensor product representation $(3\oplus 1)\otimes (\bar{\3}\oplus \1)$. The three states with null weights cannot be combined linearly to create a singlet state (i.e., any linear combination of these three states has a non-null image by at least one of the ladder operators). Indeed, to get commuting ladder operators for the isoversor and bivector SU(3), we introduced a phase factor in Eqs.~(\ref{Tpm}--\ref{Vpm}) as compared to their definitions in matrix algebra. This phase affects the tensor product representation: we do not find exactly the $3 \otimes \bar{ \3} = 8 \oplus 1$ representation. The two continuous groups (bivector and isoversor) are not independent because the complex number of one is an element of the other and vice versa. This mutual dependence prevents the construction of the usual tensor product representation $3 \otimes \bar{\3} $.

\subsection{Representation changes for operators} 
\label{RepChoice}

We discuss below (for completeness) the notion of representation change, although this is rather straightforward to translate into geometric algebra. Illustrations based on the continuous groups defined in the previous Section are provided.

\paragraph{Commutation relations in the new representation} 
Let's assume that we have defined an operator $λ_{\aa}$ in a given representation. We can define it in another representation using unitary operators $\mathsf{U}_{±}$, i.e., operators such that $\mathsf{U}_{+}\mathsf{U}_{-}=\mathsf{U}_{-}\mathsf{U}_{+} = \mathsf{Id}$, e.g., operators defined in Eq.~\eqref{SwitchRepOp}.
If the operator $\mathsf{U}_{+}$ is used to produce an image of the state multivector, then the operators acting on the new multivectors (i.e., in the new representation) are 
\begin{equation}
	λ_{\aa}' ≐ \mathsf{U}_{+} \circ λ_{\aa} \circ \mathsf{U}_{-}.
	\label{RepChange}
\end{equation} 
If the operator belongs to a continuous group with specific structure constants, it is immediate to check that the commutation relations for the $λ_{\aa}'$ are those of the $λ_{\aa}$
\begin{eqnarray}
	[λ_{\aa}',λ_{\bb}'] &=& \mathsf{U}_{+} \circ λ_{\aa} \circ \mathsf{U}_{-} \circ \mathsf{U}_{+} \circ λ_{\bb} \circ \mathsf{U}_{-} - \mathsf{U}_{+} \circ λ_{\bb} \circ \mathsf{U}_{-} \circ \mathsf{U}_{+} \circ λ_{\aa} \circ \mathsf{U}_{-} \\
	&=& \mathsf{U}_{+} \circ (λ_{\aa}  \circ λ_{\bb}  - λ_{\bb} \circ   λ_{\aa}) \circ \mathsf{U}_{-} \\
	&=& \mathsf{U}_{+} \circ  [λ_{\aa},λ_{\bb}]\circ \mathsf{U}_{-} \\
	&=& \sum_{c} f_{\aa\bb\cc} λ_{\cc}  '
\end{eqnarray}
Below we provide four different representations of each SU(3) group that are remarkable because they act on different sub-parts of spacetime algebra: Each one of the isospinors $ℐ_{\bbmu}$ or space bivectors $\B_{μ}$ is alternatively in the kernel of all the operators in each representation.

\paragraph{Four different isoversor SU(3)}
\label{FourSU3}

From the considerations of the previous paragraph, infinitely many representations exist for the special unitary groups. However, by cyclicity, four of them are simpler to define and interesting enough because their generators have different kernels, which prevent them from mixing the same parts of the spacetime algebra $\mathcal{G}(1,3)$. 

Let's define for simplicity
\begin{equation}
	λ_{\aa}^{(\0)}(\dots) = ℐ_{\1}λ_{\aa}(ℐ_{\1}\dots) \quad \Rightarrow \quad λ_{\aa}(\dots) = ℐ_{\1}λ^{(\0)}_{\aa}(ℐ_{\1}\dots).
\end{equation}
Then, the four isoversor SU(3) are given by
\begin{eqnarray}
	λ_{\aa}^{(\bbnu)} (\dots) = ℐ_{\bbnu}λ^{(0)}_{\aa}(ℐ_{\bbnu}^{†}\dots) \quad \text{ for } \bbnu =\0,\1,\2,\3.
\end{eqnarray}
By construction, we have $\forall \aa\in \{\0,\dots,\7\}, λ_{\aa}^{(\bbnu)}(ℐ_{\bbnu})=0$, i.e., all these operators have different kernels.

\paragraph{Four different bivector SU(3)}
\label{FourSU3}

Similarly, the generators of bivector SU(3) given in Table \ref{BivectorGroup} can be used to define a total of four special unitary groups and their conjugates. The four isoversor SU(3) are given by
\begin{eqnarray}
	λ_{a}^{(ν)} (\dots) = \B_{ν}λ_{a}(\B_{ν}^{†}\dots) \quad \text{ for } ν =0,1,2,3.
\end{eqnarray}
By construction, as before, we have $\forall a \in \{0,\dots,7\}, λ_{a}^{(ν)}(\B_{ν})=0$, i.e., all these operators have different kernels which prevent them from mixing the same parts of the spacetime algebra $\mathcal{G}(1,3)$. 

\paragraph{Chiral, Standard and other representations}
 In the next Section, we define the chiral and standard representations of the Dirac operators acting on an arbitrary multivector of the geometric algebra of spacetime. In addition to this representation change, generated by operators $\mathsf{U}_{±}$, two others are remarkable:
\begin{eqnarray}
	  \mathsf{U}_{±}(\Mc) &=& \frac{\Mc ± ℐ_{\1} \Mc ℐ_{\3}\B_{3}}{\sqrt{2}}, \\
	  \mathsf{V}_{±}(\Mc) &=& \frac{\Mc ± ℐ_{\3} \Mc ℐ_{\2}}{\sqrt{2}}, \\
	  \mathsf{W}_{±}(\Mc) &=& \frac{\Mc ± ℐ_{\2} \Mc ℐ_{\1}}{\sqrt{2}}.
 \end{eqnarray}
The operators in the three above equations commute. Hence, we can compose them to define up to eight different representations. These eight representation changes are particularly interesting (at least as much as standard and chiral representations) because they transform the basis isoversors $ℐ_{\bbmu}$ into multivectors involving $3_{±}=(1±ℐ_{\2}\B_{3})$, $\1_{±} = (1±ℐ_{\1})/2$ and $\3_{±}=(1±ℐ_{\3})/2$ respectively. Those multivectors all have unit squares and are orthogonal projectors since $3_{+}3_{-}= 3_{-}3_{+} = 0$, $\3_{+}\3_{-} = \3_{-}\3_{+} = 0$, etc. 

\subsection{Partial conclusion}

In this Section, we have introduced the extraspecial groups $2^{5}_{±}$ that exist in the two spacetime algebras of different signatures. We have introduced our notations and terminology with simplicity in mind: the central notation eases calculations thanks to the commutation of bivectors and isoversors; the names of different objects in the algebra are chosen to geometrically or physically illustrate their meaning. We have also defined several involutions that lead to a multivector norm in spacetime algebra. Continuous groups, with a central role in particle physics, were finally introduced in different representations, along with different projectors.

\begin{landscape}
	\begin{table}
	\begin{tabular}{c|c}
	Fundamental representation (Isoversor) & Conjugate representation (Isoversor)  \\ \hline
			\begin{minipage}{0.68\textwidth}
\begin{eqnarray}
% FONDAMENTAL ISOVERSOR
	2\mathsf{T}_{\0} &=& \frac{1}{2\sqrt{3}}\left[ ℐ_{\1}(\dots)ℐ_{\1} + ℐ_{\2}(\dots)ℐ_{\2} + 2 ℐ_{\3}(\dots)ℐ_{\3} \right] = λ_{\8}, \label{T0op} \\
	2\mathsf{T}_{\1}&=& \frac{1}{2}\left[ ℐ_{\1}(\dots)ℐ_{\2} - ℐ_{\2}(\dots)ℐ_{\1} \right]\B_{3} = λ_{\1} \\
	2\mathsf{T}_{\2}&=& \frac{1}{2}\left[ ℐ_{\3}(\dots) + (\dots)ℐ_{\3}\right] = λ_{\2}\\
	2\mathsf{T}_{\3}&=& \frac{1}{2}\left[ ℐ_{\1}(\dots)ℐ_{\1} - ℐ_{\2}(\dots)ℐ_{\2} \right] = λ_{\3} \\
&& \nonumber \\
	2\mathsf{U}_{\0} &=&  \frac{1}{2\sqrt{3}}\left[-2 ℐ_{\1}(\dots)ℐ_{\1} + ℐ_{\2}(\dots)ℐ_{\2} -  ℐ_{\3}(\dots)ℐ_{\3} \right]  \\
	2\mathsf{U}_{\1}&=& \frac{1}{2}\left[ ℐ_{\3}(\dots)ℐ_{\2} + ℐ_{\2}(\dots)ℐ_{\3} \right]\B_{3} = λ_{\6} \\
	2\mathsf{U}_{\2}&=& \frac{1}{2}\left[ ℐ_{\1}(\dots) - (\dots)ℐ_{\1}\right] = λ_{\7}\\
	2\mathsf{U}_{\3}&=& \frac{1}{2}\left[ ℐ_{\2}(\dots)ℐ_{\2} + ℐ_{\3}(\dots)ℐ_{\3} \right] = \frac{1}{2}(-λ_{\3}+λ_{\8}\sqrt{3}), \\
&& \nonumber \\
	 2\mathsf{V}_{\0} &=&  \frac{1}{2\sqrt{3}}\left[ ℐ_{\1}(\dots)ℐ_{\1} -2 ℐ_{\2}(\dots)ℐ_{\2} - ℐ_{\3}(\dots)ℐ_{\3} \right] \\
	 2\mathsf{V}_{\1}&=& \frac{1}{2}\left[ ℐ_{\3}(\dots)ℐ_{\1} - ℐ_{\1}(\dots)ℐ_{\3} \right] = λ_{\4} \\
	-2\mathsf{V}_{\2}&=& \frac{1}{2}\left[ ℐ_{\2}(\dots)\B_{3} + (\dots)ℐ_{\2}\B_{3} \right] = λ_{\5}\\
	-2\mathsf{V}_{\3}&=& \frac{1}{2}\left[ ℐ_{\1}(\dots)ℐ_{\1} + ℐ_{\3}(\dots)ℐ_{\3} \right] = \frac{1}{2}(λ_{\3}+λ_{\8}\sqrt{3}) \label{V3op} 
\end{eqnarray}
			\end{minipage} &
			\begin{minipage}{0.7\textwidth}
\begin{eqnarray}
% CONJUGATE ISOVERSOR
	2\bar{\mathsf{T}}_{\0} &=&  \frac{1}{2\sqrt{3}}\left[ ℐ_{\1}(\dots)ℐ_{\1} - ℐ_{\2}(\dots)ℐ_{\2} + 2 ℐ_{\3}(\dots)ℐ_{\3} \right]	= \bar{λ}_{\8}  \\
	2\bar{\mathsf{T}}_{\1}&=& \frac{1}{2}\left[ -ℐ_{\1}(\dots)ℐ_{\2} - ℐ_{\2}(\dots)ℐ_{\1} \right]\B_{3} = \bar{λ}_{\1} \\
	2\bar{\mathsf{T}}_{\2}&=& \frac{1}{2}\left[ ℐ_{\3}(\dots) - (\dots)ℐ_{\3}\right] = \bar{λ}_{\2}\\
	2\bar{\mathsf{T}}_{\3}&=& \frac{1}{2}\left[ ℐ_{\1}(\dots)ℐ_{\1} + ℐ_{\2}(\dots)ℐ_{\2} \right] = \bar{λ}_{\3} \\
&& \nonumber \\
	2\bar{\mathsf{U}}_{\0} &=& -\frac{1}{2\sqrt{3}}\left[2 ℐ_{\1}(\dots)ℐ_{\1} + ℐ_{\2}(\dots)ℐ_{\2} +  ℐ_{\3}(\dots)ℐ_{\3} \right]  \\
	2\bar{\mathsf{U}}_{\1}&=& \frac{1}{2}\left[ -ℐ_{\3}(\dots)ℐ_{\2} + ℐ_{\2}(\dots)ℐ_{\3} \right]\B_{3} = \bar{λ}_{\6} \\
	2\bar{\mathsf{U}}_{\2}&=& \frac{1}{2}\left[ ℐ_{\1}(\dots) + (\dots)ℐ_{\1}\right] = \bar{λ}_{\7}\\
	2\bar{\mathsf{U}}_{\3}&=& \frac{1}{2}\left[- ℐ_{\2}(\dots)ℐ_{\2} + ℐ_{\3}(\dots)ℐ_{\3} \right] = \frac{1}{2}(-\bar{λ}_{\3}+\bar{λ}_{\8}\sqrt{3}), \\
&& \nonumber \\
	 2\bar{\mathsf{V}}_{\0} &=&   \frac{1}{2\sqrt{3}}\left[ ℐ_{\1}(\dots)ℐ_{\1} +2 ℐ_{\2}(\dots)ℐ_{\2} - ℐ_{\3}(\dots)ℐ_{\3} \right]  \\
	 2\bar{\mathsf{V}}_{\1}&=& \frac{1}{2}\left[ ℐ_{\3}(\dots)ℐ_{\1} - ℐ_{\1}(\dots)ℐ_{\3} \right] = \bar{λ}_{\4} \\
	-2\bar{\mathsf{V}}_{\2}&=& \frac{1}{2}\left[ ℐ_{\2}(\dots)\B_{3} + (\dots)ℐ_{\2}\B_{3} \right] = \bar{λ}_{\5}\\
	-2\bar{\mathsf{V}}_{\3}&=& \frac{1}{2}\left[ ℐ_{\1}(\dots)ℐ_{\1} + ℐ_{\3}(\dots)ℐ_{\3} \right] = \frac{1}{2}(\bar{λ}_{\3}+\bar{λ}_{\8}\sqrt{3}) 
\end{eqnarray}
			\end{minipage} 
	\end{tabular}
	\caption{Operators defining the special unitary groups of degree $3$ acting on the isoversor part of the geometric algebra of spacetime. Nonetheless, one bivector say $\B_{3}$ (acting as complex number) is also involved in these symmetry groups. Remark that $\B_{3}$ appears whenever $ℐ_{\2}$ is present once in each term.  \label{IsoversorGroup}}
	\end{table}
	
	\begin{table}[t]
	\begin{tabular}{c|c}
	Fundamental representation (Bivector) & Conjugate representation (Bivector)  \\ \hline
			\begin{minipage}{0.68\textwidth}
\begin{eqnarray}
	% FONDAMENTAL BIVECTOR
	\\
	2\mathsf{T}_{0}&=& \frac{1}{2\sqrt{3}}\left[ -\B_{3}(\dots)\B_{3} - \B_{1}(\dots)\B_{1} +  2\B_{2}(\dots)\B_{2}  \right] = λ_{8} \nonumber \\ 
	2\mathsf{T}_{1}&=&\frac{1}{2}\left[ \B_{3}(\dots)\B_{1} +  \B_{1}(\dots)\B_{3}  \right] = λ_{1} \\
	2\mathsf{T}_{2}&=& \frac{1}{2}\left[ \B_{2}(\dots) -  (\dots)\B_{2}  \right]ℐ_{\2} =  λ_{2} \\
	2\mathsf{T}_{3}&=& \frac{1}{2}\left[ \B_{3}(\dots)\B_{3} -  \B_{1}(\dots)\B_{1}  \right] = λ_{3} \\
	&~& \nonumber \\ 
	2\mathsf{U}_{0}&=& -\frac{1}{2}(λ_{3}\sqrt{3}+λ_{8})\\
	2\mathsf{U}_{1}&=&\frac{1}{2}\left[ \B_{3}(\dots)\B_{2} +  \B_{2}(\dots)\B_{3}  \right] = λ_{6} \\
	2\mathsf{U}_{2}&=& \frac{1}{2}\left[ \B_{1}(\dots) -  (\dots)\B_{1}  \right]ℐ_{\2} =  λ_{7} \\
	2\mathsf{U}_{3}&=& \frac{1}{2}\left[ \B_{2}(\dots)\B_{2} -  \B_{3}(\dots)\B_{3}  \right]  = \frac{1}{2}(-λ_{3}+λ_{8}\sqrt{3}) \\
	&~& \nonumber \\
	2\mathsf{V}_{0}&=& \frac{1}{2}(λ_{3}\sqrt{3}-λ_{8}) \\
	2\mathsf{V}_{1}&=&\frac{1}{2}\left[ -\B_{2}(\dots)\B_{1} -  \B_{1}(\dots)\B_{2}  \right] = λ_{4} \\
	-2\mathsf{V}_{2}&=& \frac{1}{2}\left[ \B_{3}(\dots) -  (\dots)\B_{3}  \right]ℐ_{2}=  λ_{5} \\
	-2\mathsf{V}_{3}&=& \frac{1}{2}\left[ -\B_{1}(\dots)\B_{1} +  \B_{2}(\dots)\B_{2}  \right]= \frac{1}{2}(λ_{3}+λ_{8}\sqrt{3}) 
\end{eqnarray}
			\end{minipage} &
			\begin{minipage}{0.7\textwidth}
\begin{eqnarray}
% CONJUGATE BIVECTOR 
\nonumber \\
	2\bar{\mathsf{T}}_{0}&=& \frac{1}{2\sqrt{3}}\left[ -\B_{3}(\dots)\B_{3} + \B_{1}(\dots)\B_{1} +  2\B_{2}(\dots)\B_{2}  \right] =  \bar{λ}_{8}  \\  
	2\bar{\mathsf{T}}_{1}&=& \frac{1}{2}\left[ \B_{1}(\dots)\B_{3}  -\B_{3}(\dots)\B_{1} \right] = \bar{λ}_{1} \\
	2\bar{\mathsf{T}}_{2}&=& \frac{1}{2}\left[ \B_{2}(\dots) +  (\dots)\B_{2}  \right]ℐ_{\2} =  \bar{λ}_{2} \\
	2\bar{\mathsf{T}}_{3}&=& \frac{1}{2}\left[ \B_{3}(\dots)\B_{3} +  \B_{1}(\dots)\B_{1}  \right] = \bar{λ}_{3} \\
	&~& \nonumber \\ 
	2\bar{\mathsf{U}}_{0}&=& -\frac{1}{2}(\bar{λ}_{3}\sqrt{3}+\bar{λ}_{8})\\
	2\bar{\mathsf{U}}_{1}&=&\frac{1}{2}\left[ \B_{3}(\dots)\B_{2} +  \B_{2}(\dots)\B_{3}  \right] = \bar{λ}_{6} \\
	2\bar{\mathsf{U}}_{2}&=& \frac{1}{2}\left[ \B_{1}(\dots) -  (\dots)\B_{1}  \right]ℐ_{\2} =  \bar{λ}_{7} \\
	2\bar{\mathsf{U}}_{3}&=& \frac{1}{2}\left[ \B_{2}(\dots)\B_{2} -  \B_{3}(\dots)\B_{3}  \right]  = \frac{1}{2}(-\bar{λ}_{3}+\bar{λ}_{8}\sqrt{3}) \\
	&~& \nonumber \\
	2\bar{\mathsf{V}}_{0}&=& \frac{1}{2}(\bar{λ}_{3}\sqrt{3}-\bar{λ}_{8}) \\
	2\bar{\mathsf{V}}_{1}&=&\frac{1}{2}\left[ \B_{2}(\dots)\B_{1} -  \B_{1}(\dots)\B_{2}  \right] = \bar{λ}_{4} \\
	-2\bar{\mathsf{V}}_{2}&=& \frac{1}{2}\left[ \B_{3}(\dots) +  (\dots)\B_{3}  \right]ℐ_{\2}=  \bar{λ}_{5} \\
	-2\bar{\mathsf{V}}_{3}&=& \frac{1}{2}\left[ \B_{1}(\dots)\B_{1} +  \B_{2}(\dots)\B_{2}  \right]= \frac{1}{2}(\bar{λ}_{3}+\bar{λ}_{8}\sqrt{3}) 
\end{eqnarray}
			\end{minipage} 
	\end{tabular}
	\caption{Operators defining the special unitary groups of degree $3$ acting on the bivector part of the geometric algebra of spacetime. Nonetheless, the isoversor $ℐ_{\2}$ (acting as complex number) is also involved in these symmetry groups.
	Remark that $ℐ_{\2}$ appears whenever at most one bivector is present in each term.
	\label{BivectorGroup}}
	\end{table}
\end{landscape}

%\newpage

%%%%%%%%%%%%%%%%%%%%%%%%%%%%%%%%%%%%%%%%%%%%%%%%%%%%%%%%%%%%%%%%%%%%%%
\section{Definition and dynamics of quadrispinor fields}
\label{DefSpinorAndDyn}

In the left column below, I review spinor theory based on matrix algebra \cite{Book_Maggiore2005, Steane2013vol}. In the right column, I revisit and develop spinor theory based on geometric algebra~\cite{Book_Doran2003, Francis2005vol317, Book_Hestenes2015}. First, I introduce Weyl's spinors with left and right chiralities, respectively, in the $(1/2,0)$ and $(0,1/2)$ representations of the Lorentz group. Second, I define Dirac bispinors in the $(1/2,0) \oplus (0,1/2)$ representation. For those bispinors, I precisely define the standard and chiral representations, which are not widely discussed in geometric algebra \cite{Doran2005}. I also show that a general multivector field in the spacetime algebra is a quadrispinor field since it is associated with a pair of (Dirac) bispinor fields. Third, I build the quadrispinor of homogeneous materiality whose space spinor components are all associated with matter particles (or antimatter ones) from the quadrispinor of homogeneous mass that includes two couples of matter-antimatter particles.
As done when defining the special unitary groups, I focus on the signature convention $\e_{0}^{2}=1$ and $\e_{i}^{2}=-1$ associated with the extraspecial group $2^{5}_{-}$ since it corresponds to the Dirac Algebra (of gamma matrices). Identical symbols may appear in the left and right columns. They usually have different definitions.

\subsection{Weyl or Space spinors and their transformations}
\label{WeylSpinors}
% one can begin a section for both columns with
% \begin{paracol}{2}[\section{Part I}]

\begin{paracol}{2}[] %%%%%%%%%%%%%%%%%%%%%%%%%%%%%%%

% MATRIX
\paragraph{Weyl spinor and scalar product} Weyl spinors in matrix algebra are vectors with two complex components. Let's take $ s,r \in \mathbb{C}^{2}$:
\begin{equation}
	s  ≐ \begin{pmatrix} a \\ b \end{pmatrix} =  \begin{pmatrix} \phi_{0} -ι \phi_{3} \\ \phi_{2}-ι \phi_{1} \end{pmatrix}, \quad r ≐ \begin{pmatrix} c \\ d \end{pmatrix} = \begin{pmatrix} ψ_{0} -ι ψ_{3} \\ ψ_{2}-ι ψ_{1} \end{pmatrix} . \label{weylspinor}
\end{equation}
The scalar product of these Weyl spinors is 
\begin{equation}
	s^{†} r = \begin{pmatrix} a^{*} & b^{*} 	\end{pmatrix}  \begin{pmatrix} c \\ d \end{pmatrix} = a^{*} c + b^{*} d,	
\end{equation}
with $†$ denoting the Hermitian conjugation. The mapping between matrix and geometric algebra is justified based on the fact that the same spin direction is obtained with Weyl's spinors $s$ or space spinors $\Sc$.

%GEOMETRIC
\switchcolumn\paragraph{Space spinor and scalar product} 
Space spinors in geometric algebra are the sum of a real number and a space bivector. Let's take two space spinors
  \begin{eqnarray}
	  \Sc &≐& \phi_{0} + \phi_{1} \B_{1} + \phi_{2} \B_{2} + \phi_{3} \B_{3}, \label{spacespinor} \\
	 \Rc &≐& ψ_{0} + ψ_{1}\B_{1}+ ψ_{2}\B_{2}+ ψ_{3}\B_{3}. \nonumber
 \end{eqnarray}
Their scalar product writes \cite{Book_Doran2003}
\begin{eqnarray}
	\langle \Sc|\Rc\rangle &≐& (\Sc^\mathsf{r} \Rc)_{\vide} - (\Sc^\mathsf{r} \Rc\B_{3})_{\vide}\B_{3}, \label{scalprodSpinor} \\ 
	\langle \Sc|\Rc\rangle  &=& ψ_{0}\phi_{0}+ψ_{1}\phi_{1}+ψ_{2}\phi_{2}+ψ_{3}\phi_{3} \nonumber \\ && + \B_{3}(-ψ_{0}\phi_{3}+ψ_{1}\phi_{2}-ψ_{2}\phi_{1}+ψ_{3}\phi_{0}), \nonumber
\end{eqnarray}
keeping only the scalar term and the term in $\B_{3}$ in the geometric product of $\Sc^\mathsf{r}$ with $\Rc$. We remark that the above scalar product amounts to extracting the scalar and $\B_{3}$ components of the geometric product of a reverse spinor with another spinor. Hence, one can usefully introduce the projectors
\begin{equation}
	(\dots)_{μ} ≐ (\dots \B_{μ}^\mathsf{r})_{\vide}\B_{μ},
\end{equation}
leading to $\langle \Sc|\Rc\rangle = (\Sc\Rc)_{0}+(\Sc\Rc)_{3} $.
\end{paracol}

\begin{paracol}{2}[] %%%%%%%%%%%%%%%%%%%%%%%%%%%%%%%

\paragraph{Charge conjugate Weyl spinor} Starting from the weyl spinor $s$, we define another spinor 
\begin{equation}
	\tilde s ≐ \begin{pmatrix} b^{*} \\ -a^{*} \end{pmatrix} = \begin{pmatrix}  \phi_{2}+ι \phi_{1}\\  -\phi_{0} -ι \phi_{3}  \end{pmatrix} 
\end{equation}
where $*$ is the complex conjugation. The two spinors are orthogonal: $s^{\dag} \tilde s = 0$. The tilde operation defined above is not involutive ($\tilde{\tilde{s}}=-s$). Then, we define another involution with an upperscript $\mathrm{c}$ called charge conjugation, and that is an involution ($(s^\mathrm{c})^\mathrm{c} = s$). Assuming the spinor of Eq.~\eqref{weylspinor} is a left spinor
\begin{equation}
	s_{\textsc{l}}  ≐ \begin{pmatrix} a \\ b \end{pmatrix} \quad \underset{\mathrm{c}}{\longrightarrow} \quad s_{\textsc{l}}^\mathrm{c} = \begin{pmatrix} b^{*} \\ -a^{*} \end{pmatrix} = \tilde s_{\textsc{l}}  ,
\end{equation}
the associated right spinor and its charge conjugate are 
\begin{equation}
 s_{\textsc{r}} ≐ \begin{pmatrix} c \\ d \end{pmatrix} \quad \underset{\mathrm{c}}{\longrightarrow} \quad s^\mathrm{c}_{\textsc{r}} ≐ \begin{pmatrix} -d^{*} \\ c^{*} \end{pmatrix} = - \tilde s_{\textsc{r}},
\end{equation}
see Ref.~\cite{Book_Maggiore2005} (page 28). We emphasize that $s^\mathrm{c}_{\textsc{l}}$ is a right spinor, while $s^\mathrm{c}_{\textsc{r}}$ is a left spinor (as the notation does not show). With these definitions, $(s_{χ}^\mathrm{c})^\mathrm{c}= s_{χ}$ for $χ=\textsc{l},\textsc{r}$. 
This charge conjugation involution requires an extrinsic notion of (left and right) chirality. Then, the charge conjugation of a right spinor is defined as the opposite of the one acting on a left spinor, and then ``$\mathrm{c}$'' is involutive, i.e., $\mathrm{c}^{2}=1$. We say that the difference between $\tilde s$ and $s^\mathrm{c}$ is extrinsic because one has to provide the chirality of the Weyl spinor when giving its vector in $\mathbb{C}^{2}$.

%GEOMETRIC
\switchcolumn\paragraph{Charge conjugate space spinor}
We use the tilde notation for the right action of $-\B_{2}$, i.e.,
\begin{eqnarray}
	\tilde \Sc &≐& - \Sc \B_{2}, \quad \text{ although } \tilde \Sc ≠ \Sc^\mathrm{c}. %\\
\end{eqnarray}
The tilde operation does not change the chirality (behavior under pure boost) 
%, or versatility) 
as compared to the definition of Eq.~\eqref{chargeconjugation} of charge conjugation 
and denoted with a superscript~c. 
Remaining at the level of space spinors, we define the charge conjugation by
\begin{eqnarray}
\Sc_\textsc{l}^\mathrm{c} &≐& -\Sc_\textsc{l} \B_{2} = \tilde \Sc_\textsc{l} . \label{C-left} \\
\Sc_\textsc{r}^\mathrm{c} &≐& \Sc_\textsc{r} \B_{2} = - \tilde \Sc_\textsc{r}, \label{C-right}
\end{eqnarray}
Like in matrix algebra, the notion of left or right chirality is extrinsic when using space spinors (only). It will become intrinsic at the level of even (and odd) multivectors that represent the Dirac bispinors in the matricial framework. For now, we indicate the extrinsic spinor's chirality as a subscript only when needed.

\end{paracol}

\begin{paracol}{2}[] %%%%%%%%%%%%%%%%%%%%%%%%%%%%%%%

 % MATRIX
\paragraph{Pauli matrices} 
We take $σ^{0}=\1$ the $2×2$ identity matrix. The three Pauli matrices $\sigma^{i}$ are defined in Eq.~\eqref{PauliMatrices}.
 %\begin{equation}
	  %\sigma^{1} = \begin{bmatrix}
		%0& 1  \\
		%1& 0 \\
	%\end{bmatrix}, \quad
	%\sigma^{2} =\begin{bmatrix}
		%0 & -ι  \\
		%ι &0 \\
	%\end{bmatrix}, \quad
	%\sigma^{3} =\begin{bmatrix}
		%1& 0  \\
		%0&-1 \\
	%\end{bmatrix}.
  %\end{equation}
These matrices satisfy the following algebra:
\begin{equation}
	σ^{i}σ^{j}=δ_{ij}\1+ι ε_{ijk}σ^{k}.
\end{equation} 
Hence the matrices $ι σ^{j}$ are such that
\begin{equation}
	(ι σ^{i})(ισ^{j})=-δ_{ij}\1+ ε_{ijk}(ι σ^{k}) \label{PauliMatrixProduct}
\end{equation} 
that is isomorphic to the geometric product of space bivectors.

%GEOMETRIC
\switchcolumn\paragraph{Pauli operators} 
With the correspondence defined on Eqs.~\eqref{spacespinor} of the left and right columns, one can directly check that the $i$th Pauli operator acting on a space spinor $\Sc$ is given by $\B_{i} \Sc \B_{3}$. Similarly, the multiplication by the complex number $ι$ is realized by $(\dots)\B_{3}$, i.e., by right multiplying by bivector $\B_{3}$. Therefore, the matrix $-ι σ^{i}$ acting on a Weyl spinor $s$ is equivalent to the left multiplication by $\B_{i}$ of the space spinor $\Sc$. The correspondence is summarized as
\begin{eqnarray}
	-ι σ^{i} s \quad &\leftrightarrow& \quad \B_{i} \Sc, \\
	σ^{i} s \quad &\leftrightarrow& \quad \B_{i} \Sc \B_{3}, \label{Pauli-op} \\
		ι \quad &\leftrightarrow& \quad (\dots)\B_{3}. \label{i-mult}
\end{eqnarray}
As required, we remark that $(-ι σ_{i})^{2}=-1$ and $\B_{i}^{2} = -1$, and more generally that
\begin{equation}
	\B_{i}\B_{j}=-δ_{ij}+ ε_{ijk}\B_{k}, \label{BivectorProduct}
\end{equation}
in agreement with Eq.~\eqref{PauliMatrixProduct} of the left column. In the following, we justify the correspondence between Weyl's spinors and space spinors of Eqs.~\eqref{spacespinor} by showing that their associated spin vector points in the same space direction. 

\end{paracol}

\begin{paracol}{2}[] %%%%%%%%%%%%%%%%%%%%%%%%%%%%%%%

%MATRIX
\paragraph{Spin matrix} Following Ref.~\cite{Steane2013vol}, we define the spin matrix $\hat S$ as the matrix that has eigenvalues $1$ associated to eigenvector $s$ and eigenvalue $-1$ associated to eigenvector $\tilde s$. Then, $\sigma^{3}$ is the diagonalized form of the spin matrix $\hat S$. The unitary matrix ($\hat V \hat V^{\dag} = \1$)
\begin{equation}
	  \hat V = \frac{1}{|s|} \begin{bmatrix} a & -b^{*} \\ b & a^{*} \end{bmatrix}, \, \text{with }  |s| = \sqrt{ s^{\dag} s  } = \sqrt{|a|^{2}+|b|^{2}},
\end{equation}
changes from $\sigma^{3}$ (in the eigenstate basis) to the spin matrix
\begin{equation}
	\hat  S ≐ \hat V \sigma^{3} \hat V^{\dag} = \frac{1}{|s|^{2}}\begin{bmatrix}|a|^{2}-|b|^{2} & 2ab^{*} \\ 2a^{*}b & |b|^{2}-|a|^{2} \end{bmatrix}. \label{eigen-equation}
\end{equation}
The hat denoting normalization, such that $S = |s|^{2} \hat S$ or $V= |s|\hat V$.
Interestingly, the spin matrix %$\hat S ≐  S / |s|^{2} $
follows also from
\begin{eqnarray}
	\hat S &=& \sum_{i} \frac{s^{\dagger} \sigma^{i} s }{|s|^{2}} \sigma^{i},  \\ 
	\hat S &=& \frac{ab^{*}+a^{*}b}{|a|^{2}+|b|^{2}} \sigma^{1} + ι\frac{ab^{*}-a^{*}b}{|a|^{2}+|b|^{2}} \sigma^{2} + \frac{|a|^{2}-|b|^{2}}{|a|^{2}+|b|^{2}} \sigma^{3}. \nonumber \\  \label{spinmatrix}
\end{eqnarray}
Hence, the spin matrix provides a spin direction in space when the Pauli matrices are interpreted as the three spatial directions.

We remark that the spin matrix is related to the spin vector by
\begin{eqnarray}
	2 ss^{†} &=& |s|^{2} (\1 + \hat S) = |s|^{2}(\hat V\hat V^{†}+\hat Vσ^{3} \hat V^{†}) \nonumber \\ && \\
	2 ss^{†} &=& 2\begin{bmatrix} 
	|a|^{2}& a b^{*} \\ a^{*}b & |b|^{2}\\
	\end{bmatrix}
\end{eqnarray}
The above operator is an order-2 spinor, i.e., written in matrix form, but of rank 1 given its null determinant.

%GEOMETRIC
\switchcolumn\paragraph{Spin bivector} We normalize the space spinor 
\begin{equation}
	\hat\Sc ≐ \Sc/ |\Sc|, \, \text{with } |\Sc|^{2} = \Sc * \Sc = \sum_{\mu} \phi_{\mu}^{2}.
\end{equation}
By analogy with the matrix algebra, we introduce the (normalized) spin bivector as 
\begin{equation}
	\hat {\bm S} = \hat \Sc \B_{3} \hat \Sc^{\mathsf{r}} , \label{SpinBivectorDef}
\end{equation}
corresponding to Eq.~\eqref{eigen-equation} in the left column. 
Computing the components of $\hat{\bm{S}}$ leads to 
\begin{eqnarray}
	\hat{\bm S} &=& 2\frac{\phi_{0}\phi_{2}+\phi_{1}\phi_{3}}{|\Sc|^{2}}\B_{1}+2\frac{\phi_{2}\phi_{3}-\phi_{0}\phi_{1}}{|\Sc|^{2}}\B_{2} \nonumber\\&& +\frac{\phi_{0}^{2}-\phi_{1}^{2}-\phi_{2}^{2}+\phi_{3}^{2}}{|\Sc|^{2}}\B_{3} \label{spinbivector}
\end{eqnarray}
Identifying the left column Eq.~\eqref{spinmatrix} with the right column Eq.~\eqref{spinbivector} leads to the following correspondence between the matrix and geometric frameworks:
\begin{eqnarray}
	a&=&\phi_{0}-ι \phi_{3}, \\
	b&=&\phi_{2}-ι \phi_{1}. \label{spin-vecteur-rotor}
\end{eqnarray}
We remark that in this case $|s| = |\Sc|^{2} $, with the left-hand side of the equality for $|s|$ defined in the left column, and the right-hand side for $|\Sc|^{2}$ defined in the right column.

The equivalent of an order 2 spinor matrix in geometric algebra is the space spinor $2\underline{\Sc} ≐ \Sc \Sc^\mathsf{r} + \Sc \B_{3} \Sc^{\mathsf{r}}$ that is of ``rank 1'' since the grade $\vide$ component has the same norm as the grade $\II$ component. Notice that if $\Sc$ transforms under left action (as any spinor), a second-order spinor $\underline{\Sc}$ transforms under conjugacy action.

\end{paracol}

\begin{paracol}{2}[]	
% MATRIX
\paragraph{Spacetime spin vector}
	One can see the real term \\ $(s^{\dagger} \sigma^{i} s )/|s|^{2}$ as the $i$th component of matrix $\hat S$ on the Pauli matrix basis $\sigma^{i}$. Then, the spacetime spin vector has components $\mathsf{s}^{\mu} = s^{\dag} \sigma^{\mu} s$. More explicitly, it writes
\begin{equation}
	(\mathsf{s}^{\mu})_{μ=0,\dots,3} = \begin{pmatrix} |a|^{2}+|b|^{2} \\  ab^{*}+ a^{*}b \\ ι (ab^{*}- a^{*}b) \\ |a|^{2}-|b^{2}| \end{pmatrix} =  \begin{pmatrix} s^{\dag} s \\ s^{\dag} \sigma^{1}s \\ s^{\dag} \sigma^{2}s \\ s^{\dag} \sigma^{3}s \end{pmatrix} \label{spin-vecteur}
\end{equation}
in terms of the contravariant coordinates. It is a null spacetime vector $∑_{\mu}\mathsf{s}^{\mu} \mathsf{s}_{\mu} = 0$.
For a right spinor $\tilde s $, we find
\begin{equation}
	(\mathsf{s}_{μ})_{μ=0,\dots,3} = \begin{pmatrix} |a|^{2}+|b|^{2} \\  -(ab^{*}+ a^{*}b) \\ -ι (ab^{*}- a^{*}b) \\ -(|a|^{2}-|b^{2}|) \end{pmatrix} =  \begin{pmatrix} \tilde s^{\dag} \tilde s \\ \tilde s^{\dag} \sigma^{1}\tilde s \\ \tilde s^{\dag} \sigma^{2}\tilde s \\ \tilde s^{\dag} \sigma^{3}\tilde s \end{pmatrix}. \label{spin-vecteur}
\end{equation}
We use a covariant subscript notation, because the contravariant upperscript quadrivector $\mathsf{s}^{μ}$ has the opposite sign for the space part of its components ($η_{μν}=\mathrm{diag}(1,-1,-1,-1)_{μν}$, then taking down the index changes the sign of the spatial part, $\mathsf{s}_{μ}=Σ_{ν}η_{μν}\mathsf{s}^{ν}$).

% GEOMETRIC
\switchcolumn 

\paragraph{Spacetime spin vector} As in the matrix framework, the spacetime spin vector is 
\begin{eqnarray}
	\bm s &=& |\Sc|^{2} \e_{0} + 2(\phi_{0}\phi_{2}+\phi_{1}\phi_{3}) \e_{1}  +2(\phi_{2}\phi_{3}-\phi_{0}\phi_{1})\e_{2} \nonumber \\&& +(\phi_{0}^{2}-\phi_{1}^{2}-\phi_{2}^{2}+\phi_{3}^{2}) \e_{3} \\
	\bm s &=& |\Sc|^{2} \e_{0} + \Sc \e_{3} \Sc^{\mathsf{r}}   \label{spinor2nullvector}
\end{eqnarray}
It is a null vector $\bm s^{2} = 0$. We can use a spacetime split to write $\bm s =  |\Sc|^{2} \e_{0} + \vec s$ keeping in mind that $|\Sc|^{2} = |\vec s|$ for $\vec s$ the space vector part of $\bm s$. For the conjugated spinor $\Sc^\mathrm{c}= ±\Sc \B_{2}$ obtained by charge conjugation of a left spinor ($-$ case) or a right ($+$ case), we find
\begin{eqnarray}
	|\Sc^\mathrm{c}|^{2} \e_{0} + \Sc^\mathrm{c} \e_{3} {\Sc^\mathrm{c}}^{\mathsf{r}} 
	&=& |\Sc|^{2} \e_{0} + \Sc \B_{2} \e_{3} \B_{2}^{\mathsf{r}} \Sc^{\mathsf{r}} \label{spinvecchargeconj1} \\
	&=& |\Sc|^{2} \e_{0} - \Sc \e_{3}  \Sc^{\mathsf{r}} \\
	&=& |\Sc|^{2} \e_{0} - \vec s \\
	&=& \bm s^\mathsf{s}  \label{spinvecchargeconj4}
\end{eqnarray}
that is indeed the space reversal of $\bm s$. In the end, charge-conjugated spinors describe parity-related vectors of spacetime (vectors related by space reflection).
Even though we have to specify a spacetime split, we use a frame-free description of vectors, explaining why we don't need the equivalent of covariant and contravariant indices to distinguish between forms and vectors; this subtlety arises for space vectors (or forms) when decomposed onto coordinates using a specific basis.

\end{paracol}

\begin{paracol}{2}[]
\paragraph{Transformation of spinors} A spinor can be transformed using an arbitrary Lorentz transformation
  \begin{equation}
	  \Lambda = \begin{bmatrix} \Lambda_{11} & \Lambda_{12} \\ \Lambda_{21} & \Lambda_{22} \end{bmatrix} \quad \text{with } \Lambda_{11}\Lambda_{22}-\Lambda_{12}\Lambda_{21} =1  
  \end{equation}
where the components are complex numbers. Fixing the determinant removes two real degrees of freedom, leaving the six real degrees of freedom of an arbitrary Lorentz transformation. Noticing that 
  \begin{equation}
	  \Lambda^{T} \epsilon \Lambda = \epsilon
\qquad \text{ when } \qquad
	  \epsilon = \begin{bmatrix} 0 & 1 \\ -1 & 0 \end{bmatrix},
  \end{equation}
one finds the following Lorentz-invariant quantity made with two spinors $s$ and $w$:
  \begin{equation}
	   s^{T} \Lambda^{T}\epsilon \Lambda w =  s^{T} \epsilon w .
  \end{equation}
The matrix $\epsilon$ is the spinor Minkowski metric. 

An infinitesimal Lorentz transformation ($θ_{k} \ll 1$ and $η_{k} \ll 1$ for $k=1,2,3$) writes
\begin{equation}
	Λ_{θ,η} = 1+  ι  \frac{\bm {θ}}{2} ⋅ \bm{σ}  + \frac{\bm {η}}{2} ⋅ \bm{σ} 
\end{equation}
when acting on a left Weyl spinor and
\begin{equation}
	Λ_{θ,-η} = 1 + ι  \frac{\bm {θ}}{2} ⋅ \bm{σ}  - \frac{\bm {η}}{2} ⋅ \bm{σ} 
\end{equation}
when acting on a right Weyl spinor.

\switchcolumn\paragraph{Transformation of spinors} The Lorentz operator $\mathsf{L}_{θ,±η}$ (in chiral representation) or the Lorentz rotor $\hat \Lc$ satisfying $\hat \Lc^\mathsf{r} \hat \Lc = 1$ (standard representation) allows one to Lorentz transform the space spinor $\Sc$ into $\mathsf{L}_{θ,±η}(\Sc)$ or $\hat \Lc \Sc$ respectively. One can find the following Lorentz-invariant quantity made with two space spinors $\Sc$ and $\Wc$
  \begin{equation}
	   (\hat\Lc\Sc)^\mathsf{r}\hat\Lc \Wc = \Sc^\mathsf{r} \hat\Lc^\mathsf{r}\hat\Lc \Wc = \Sc^{\mathsf{r}} \Wc.
  \end{equation}
Hence there is no need for a metric ($\epsilon$ in the left column) when using space spinors of geometric algebra. We see here that the reverse plays the role of matrix transposition.

An infinitesimal Lorentz transformation ($θ_{k} \ll 1$ and $η_{k} \ll 1$ for $k=1,2,3$) writes
\begin{equation}
	\mathsf{L}_{θ,η}(\dots) = 1 -  \sum_{k}\frac{θ_{k}}{2}\B_{k}(\dots) + \sum_{k}\frac{η_{k}}{2}\B_{k} (\dots)\B_{3} \label{LorentzLeft}
\end{equation}
when acting on a left space spinor and
\begin{equation}
	\mathsf{L}_{θ,-η}(\dots) = 1 -  \sum_{k}\frac{θ_{k}}{2}\B_{k}(\dots) - \sum_{k}\frac{η_{k}}{2}\B_{k} (\dots)\B_{3} \label{LorentzRight}
\end{equation}
when acting on a right space spinor.

\end{paracol}

\begin{paracol}{2}[] %%%%%%%%%%%%%%%%%%%%%%%%%%%%%%%
% MATRIX
\paragraph{Four Weyl spinors} Based on Weyl's spinor $s$ one can define three other spinors $s^{*}$, $\epsilon s$ and $\epsilon s^{*}$ using the matrix $\epsilon$ and complex conjugation: 
  \begin{eqnarray}
	s &=&  \begin{pmatrix} a \\ b \end{pmatrix} =  \begin{pmatrix} \phi_{0} -ι \phi_{3} \\ \phi_{2}-ι \phi_{1} \end{pmatrix},  \\
	s^{*} &=& \begin{pmatrix} a^{*} \\ b^{*} \end{pmatrix} =  \begin{pmatrix} \phi_{0} +ι \phi_{3} \\ \phi_{2}+ι \phi_{1} \end{pmatrix}, \\
	\epsilon s &=& \begin{pmatrix} b \\ -a \end{pmatrix} =  \begin{pmatrix}  \phi_{2}-ι \phi_{1} \\ -\phi_{0} +ι \phi_{3}  \end{pmatrix}, \\
	\tilde s &=& \epsilon | s^{*} \rangle = 
	\begin{bmatrix} 0 & 1 \\ -1 & 0 \end{bmatrix} 
	\begin{pmatrix} a^{*} & b^{*} \end{pmatrix} 
	= \begin{pmatrix} b^{*} \\ -a^{*} \end{pmatrix}, \nonumber \\
  \end{eqnarray}
the latter being the right spinor associated to the left spinor $s$. 

%GEOMETRIC
\switchcolumn\paragraph{Four space spinors} Based on the space spinor $\Sc_{\textsc{l}}$, one can write the equivalent of the three other spinors of matrix algebra, respectively $\Sc_{\textsc{l}}^{*}$, $\Sc_{\textsc{l}}^{\epsilon}$ and $\tilde \Sc_{\textsc{l}}$ defined by:
  \begin{align}
  \Sc_{\textsc{l}} &= \phi_{0} + \phi_{1} \B_{1} + \phi_{2} \B_{2} + \phi_{3} \B_{3} \\
	\Sc_{\textsc{l}}^{*} &= \phi_{0} - \phi_{1} \B_{1} + \phi_{2} \B_{2} - \phi_{3} \B_{3}   \\
	\Sc_{\textsc{l}}^{\epsilon} &= \phi_{2} - \phi_{3} \B_{1} - \phi_{0} \B_{2} + \phi_{1} \B_{3} \\
	\tilde \Sc_{\textsc{l}} &= \phi_{2} + \phi_{3} \B_{1} - \phi_{0} \B_{2} - \phi_{1} \B_{3}   
  \end{align}
If we assume that $ \Sc_{\textsc{l}} $ is a left (space) spinor, then the corresponding right (space) spinor is $\tilde \Sc_{\textsc{l}} = \Sc_{\textsc{l}}^{\epsilon*}= -\Sc_{\textsc{l}}\B_{2}$ as aforementioned. We see, for instance, that the complex conjugation $*$ reverses the sign of the space bivector with odd subscript.
From this correspondence $s \leftrightarrow \Sc_{\textsc{l}}$ follows the other space spinors: 
\begin{eqnarray}
	s^{*} &\leftrightarrow& \Sc_{\textsc{l}}^{*} = -\B_{2}\Sc_{\textsc{l}}\B_{2} \label{CCinGA} \\
	\epsilon s &\leftrightarrow& \Sc_{\textsc{l}}^{\epsilon} = -\B_{2} \Sc_{\textsc{l}} \\
	\tilde s &\leftrightarrow& \tilde \Sc_{\textsc{l}} = - \Sc_{\textsc{l}} \B_{2}
\end{eqnarray}
The right multiplication by $\B_{2}$ realizes both the exchange of the up and down components and the complex conjugation.
The three equations above are compatible given that
\begin{equation}
\tilde s = εs^{*} \leftrightarrow -\B_{2}(-\B_{2}\Sc_{\textsc{l}})\B_{2} = - \Sc_{\textsc{l}} \B_{2}.
\end{equation} 
The scalar product defined in Eq.~\eqref{scalprodSpinor} gives $\langle \Sc_{\textsc{l}} | \tilde \Sc_{\textsc{l}} \rangle=0$ as required.
\end{paracol}

\begin{figure} \centering
	\includegraphics[width=8cm]{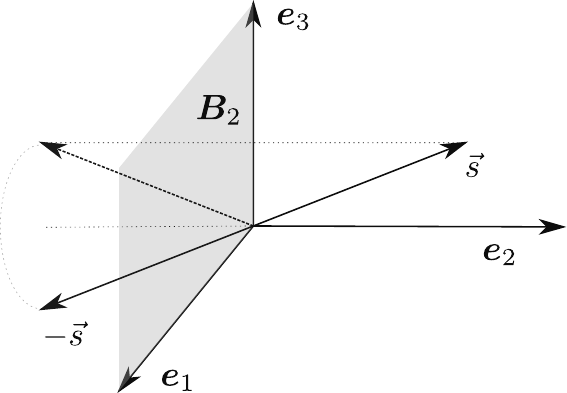}
	\caption{Reflection about plane $\B_{2}$ and $π$ rotation around axis $\e_{2}$ of the spin vector $\vec s$. \label{SpinDir}}
\end{figure}

\begin{paracol}{2}[] %%%%%%%%%%%%%%%%%%%%%%%%%%%%%%%

%MATRIX
\paragraph{Interpretation of charge conjugation} The flagpole direction of the spinor $\vec s$ is $s^{\dag} \bm \sigma s$ where $\bm \sigma = (\sigma^{1},\sigma^{2},\sigma^{3})$. The complex conjugation on the flagpole changes the sign of its second component since $\sigma^{2}$ is the only Pauli matrix with complex component $\sigma^{2*}=-\sigma^{2}$. Then, the complex conjugation is just the reflection with respect to the plane $\e_{3}\e_{1}$ that changes the sign in the second direction $\e_{2} \rightarrow -\e_{2}$.

Now the action of charge conjugation for the spin vector $\vec s$ is: First, a complex conjugation that is a reflection with respect to the plane $\B_{2}$, second, the action of $\epsilon$ that is a rotation of angle $\pi$ in the plane $\B_{2} = \e_{3}\e_{1}$. All in all, charge conjugation changes the sign of the spin vector as illustrated in Fig.~\ref{SpinDir}. However, this operation is different from the parity operation.

%GEOMETRIC
\switchcolumn\paragraph{Interpretation of tilde conjugation} The complex conjugation changes the sign of $\B_{1}= \e_{2}\e_{3}$ and $\B_{3}=\e_{1}\e_{2}$, but not $\B_{2}=\e_{3}\e_{1}$ since it changes the sign of $\e_{2}$ only (by analogy with matrix algebra). This is compatible with Eq.~\eqref{CCinGA} that encodes the reflection with respect to the plane $\B_{2}$ of an even multivector~\cite{Book_Macdonald2010}. Eqs.~(\ref{spinvecchargeconj1}--\ref{spinvecchargeconj4}) show that charge conjugation on a spinor also switches the spin direction in geometric algebra.

\end{paracol}

\begin{paracol}{2}[] %%%%%%%%%%%%%%%%%%%%%%%%%%%%%%%

% MATRIX 
\paragraph{Wave equation for Weyl's spinors}
From Ref.~\cite{Book_Maggiore2005} (page 57), the Weyl--Pauli equations for a pair of Weyl's spinors, a left one $s_\textsc{l}$ and a right one $s_\textsc{r}$, are written as
\begin{eqnarray}
	\imath\left( σ^{0}∂_{0}- ∑_{i}σ^{i}∂_{i} \right) s_\textsc{l} &=& m s_\textsc{r}, \label{Pauli1}\\
	\imath\left( σ^{0}∂_{0}+ ∑_{i}σ^{i}∂_{i} \right) s_\textsc{r} &=& m s_\textsc{l}, \label{Pauli2}
\end{eqnarray}
for (anti-)particles of mass $m$.

\switchcolumn\paragraph{Wave equation for space spinors}

The wave equations for a pair of chiral space spinors, a left one  $\Mc_{\0}$ and a right one $\Mc_{\2}$, are written 
\begin{eqnarray}
	∂_{0}\Mc_{\0}\B_{3} - ∑_{i}\B_{i} ∂_{i}(\Mc_{\0}\B_{3})\B_{3} &=& m \Mc_{\2} \nonumber \\ &&\\
	∂_{0}\Mc_{\2}\B_{3} + ∑_{i}\B_{i} ∂_{i}(\Mc_{\2}\B_{3})\B_{3} &=& m \Mc_{\0} \nonumber \\
	\end{eqnarray}
from direct correspondence with Weyl--Pauli equations using Eqs.~(\ref{Pauli-op}--\ref{i-mult}) and for (anti-)particles of mass $m$.

\end{paracol}

\subsection{Quadrispinor of inhomogeneous materiality and associated wave equations}

A quadrispinor is a generic multivector of spacetime algebra that describes four fermionic particles or antiparticles. We say that it has inhomogeneous materiality when the quadrispinor includes particle--antiparticle partners. Then, a unique equation in spacetime algebra describes the evolution of two particle--antiparticle partners.

\subsubsection{Standard and chiral representations}
\label{StdChiralRep}

We introduce below the standard and chiral representations for quadrispinors. In geometric algebra, the standard representation of Dirac bispinors is usually assumed, although this representation choice is often implicit. Given the central role of chirality for the electroweak interaction, it is fundamental to expose clearly the differences between the two representations, emphasizing the intrinsic chirality of each spinorial component of a quadrispinor. On this solid ground, the chiral nature of the electroweak interaction follows naturally, as shown in Section \ref{InternalDOF}.

\begin{paracol}{2}[]
% MATRIX
\paragraph{Dirac Bispinors} A bispinor is a four-component vector $ψ\in \mathbb{C}^{4}$ that has a left Weyl spinor, say $\phi_{\textsc{l}}$ for its two first components, and a right Weyl spinor, say $χ_{\textsc{r}}$ for its two last components:
\begin{equation}
	ψ ≐  \begin{pmatrix} \phi_{\textsc{l}} \\ χ_{\textsc{r}} \end{pmatrix}. \label{diracspinor}
\end{equation} 
We use the conventions of Maggiore \cite{Book_Maggiore2005} (page 32). 
% and not Steane conventions!
For the above bispinor, the chirality operator is 
\begin{equation}
	γ^{5}= \begin{bmatrix} -\1 & 0 \\ 0 & \1 \end{bmatrix}.
\end{equation}
The eigenvalues of the chirality operator are $-1$ for the left spinors and $+1$ for right spinors
\begin{eqnarray}
	γ^{5} \begin{pmatrix} \phi_{\textsc{l}} \\ 0   \end{pmatrix} &=& - \begin{pmatrix} \phi_{\textsc{l}} \\ 0   \end{pmatrix} ,\\
	γ^{5} \begin{pmatrix} 0 \\ χ_{\textsc{r}}  \end{pmatrix} &=&  \begin{pmatrix}  0 \\ χ_{\textsc{r}}   \end{pmatrix} .
\end{eqnarray}
The above bispinors are in chiral representation (see below). The infinitesimal Lorentz transform of a bispinor $ψ$ is written
\begin{equation}
	Λ ψ = \begin{bmatrix}
	Λ_{θ,η} & 0 \\
	0 & Λ_{θ,-η}
	\end{bmatrix} \begin{pmatrix} \phi_{\textsc{l}} \\ χ_{\textsc{r}} \end{pmatrix}
\end{equation}
in chiral representation.

%GEOMETRIC
\switchcolumn\paragraph{Geometric bispinors and quadrispinors} A bispinor is the even (resp. odd) part of a multivector $\Mc$, denoted $\Mc_{\eee}$ (resp. $\Mc_{\ooo}$), given by the following sum of a left and a right spinor 
\begin{eqnarray}
	\Mc_{\eee}	&≐& \Mc_{\0}+\Mc_{\2}\B_{3}ℐ_{\2}, \label{even}\\
	\Mc_{\ooo}	&≐& \Mc_{\3}\B_{3}ℐ_{\3} + \Mc_{\1}ℐ_{\1} ,  \label{odd} 
\end{eqnarray}
where $\Mc_{\bbmu}$ for $\bbmu = \0,\dots,\3$ are four different space spinors: $\Mc_{\0}$, $\Mc_{\3}$ are left chiral and $\Mc_{\1}$, $\Mc_{\2}$ are right chiral. Each term's chirality depends on the isoversor $ℐ_{\bbmu}$. The presence (or absence) of $\B_{3}$ produces the appropriate change of sign under charge conjugation, while the isoversor allows us to distinguish the left and right chiral components and also produce the appropriate behavior under boost (see below). For the above bispinors, the chirality operator is in the chiral representation
\begin{eqnarray} 
	γ^{5} \quad &\leftrightarrow& \quad  -ℐ_{\3}(\dots)ℐ_{\3}, \label{chiralop}
\end{eqnarray}
as we justify in the next paragraph. The eigenvalues of the chirality operator are $-1$ for the left spinors and $+1$ for right spinors. Accordingly, one has
\begin{center}
	\begin{tabular}{lr}
		\textsc{left} & \textsc{right} \\
		$\Mc_{\0}$, & $\Mc_{\2}\B_{3}ℐ_{\2}$,\\
		$\Mc_{\3}\B_{3}ℐ_{\3},\qquad$& $\Mc_{\1}ℐ_{\1}$
	\end{tabular} 
\end{center}
in the chiral representation. Let's emphasize that the quadrispinor $\Mc$ includes two pairs of particle/antiparticle, e.g., $(ν_\textsc{l},\overline{ν_\textsc{l}})$ and $(\overline{ν_\textsc{r}}, ν_\textsc{r})$ for neutrinos.
The infinitesimal Lorentz transform of a quadrispinors $\Mc$ writes (in chiral representation)
\begin{equation}
	\hat \Lc \Mc ≐ \Mc -  \sum_{k}\frac{θ_{k}}{2}\B_{k}\Mc + \sum_{k}\frac{η_{k}}{2}\B_{k} ℐ_{\3}\Mcℐ_{\3} \B_{3} 
	\label{infinitesimalLORENTZ}  
\end{equation}
in which the (opposite of the) chirality operator intrinsically produces the appropriate sign for each chiral component. For a given quadripsinor $\Mc$, the space spinors $\Mc_{\bbmu}$ for $\bbmu=\0,\1,\2,\3$ are called the spinorial components.

\end{paracol}

\begin{paracol}{2}[]
%MATRIX 
\paragraph{Chiral representation of Dirac matrices} The $γ$ matrices in chiral representation are \\ (Chiral basis)
\begin{equation}
	γ^{0} ≐ \begin{bmatrix} 0 & \1 \\ \1 & 0  \end{bmatrix}, \quad γ^{j} ≐ \begin{bmatrix} 0 &  σ^{j} \\ - σ^{j} & 0  \end{bmatrix}, \quad γ^{5}= \begin{bmatrix} -\1 & 0 \\ 0 & \1 \end{bmatrix} \label{gammas}
\end{equation}
for $j=1,2,3$, see \cite{Book_Maggiore2005} (pages 57--58). The fifth matrix is related to the four other matrices by
\begin{equation}
	γ^{5} ≐ ι γ^{0}γ^{1}γ^{2}γ^{3}. 
	\label{fifthgammamatrix}
\end{equation}
From the matrices $γ^{0}$ and $γ^{5}$, we can also define an operator required to switch from the chiral basis to the standard basis:
\begin{equation}
	γ^{0}γ^{5} = \begin{bmatrix} 0 & \1 \\ -\1 & 0  \end{bmatrix}. \label{gamma0gamma5}
\end{equation}
The matrices $γ^{0}$ and $γ^{5}$ have a square equal to $1$, and the matrices $γ^{j}$ have a square equal to $-1$. 

%GEOMETRIC
\switchcolumn 

\paragraph{Chiral representation of Dirac operators} Dirac's operators in geometric algebra, when acting on bispinors defined in Eqs.~(\ref{even}--\ref{odd}) of the right column, are \\ (Chiral basis)
\begin{eqnarray}
	γ^{0}	&\leftrightarrow&  ℐ_{\2} (\dots)\B_{3}, \\
	γ^{j}&\leftrightarrow&  \e_{j}(\dots)ℐ_{\3}  = ℐ_{\1}\B_{j}(\dots)\B_{3}^{2}ℐ_{\3}, \\
	γ^{5}	&\leftrightarrow&  -\e_{0}(\dots)ℐ_{\3}= -ℐ_{\3}(\dots)ℐ_{\3},   \\
	γ^{0}γ^{5}	&\leftrightarrow&  ℐ_{\1}(\dots)ℐ_{\3}\B_{3},
 \label{DoranChiral}
\end{eqnarray}
as one can check by direct computation. We remark that the combination of the last operator $γ^{0}γ^{5}$ with 
\begin{equation}
	σ_{j} \leftrightarrow \B_{j}(\dots)\B_{3}
\end{equation} 
leads to the $γ^{j}$ matrices for $j=1,2,3$. The chirality operator can  be computed using its definition in the left column Eq.~\eqref{fifthgammamatrix}:
\begin{eqnarray}
	γ^{5} &\leftrightarrow& (ℐ_{\2})\e_{1}\e_{2}\e_{3}(\dots)(ℐ_{\3})^{3}\B_{3}^{2}, \nonumber \\
	 && = -ℐ_{\3} (\dots) ℐ_{\3},
\end{eqnarray}
since $ℐ_{\1}≐ \e_{1}\e_{2}\e_{3}$. This justifies the definition of the chirality operator given in Eq.~\eqref{chiralop}.

\end{paracol}

\begin{paracol}{2}[]
% MATRIX
\paragraph{Standard representation of Dirac matrices} The $γ$ matrices in standard representation are \\ (Dirac basis)
\begin{equation}
	γ^{0}\,\!' 	= \begin{bmatrix} \1 & 0 \\ 0 & -\1 \end{bmatrix},\quad 
	γ^j\,\!' 			= \begin{bmatrix} 0 &  σ^j \\ - σ^j & 0 \end{bmatrix},\quad 
	γ^{5}\,\!' 	= \begin{bmatrix} 0 & \1 \\ \1 & 0 \end{bmatrix},
\end{equation}
for $j=1,2,3$, see \cite{Book_Maggiore2005} (page 59). With respect to the definition in the previous paragraph, we have $γ^{0}\,\!' = - γ^{5}$, $γ^j\,\!'= γ^j$ and $γ^{5}\,\!' = γ^{0}$. However, the following product is identical in the chiral and standard representations:
\begin{equation}
	γ^{0}\,\!'γ^{5}\,\!' = \begin{bmatrix} 0 & \1 \\ -\1 & 0  \end{bmatrix}.
\end{equation}
In standard representations as well, the matrices $γ^{0}\,\!'$ and $γ^{5}\,\!'$ have a square equal to $1$ while $ γ^{j}$ has a square equal to $-1$. They act on the bispinor obtained from the chiral bispinor using the operator $\mathsf{U}_{+}$ defined as
\begin{equation}
	\mathsf{U}_{±}≐ \frac{1}{\sqrt{2}}(\1 ± γ^{0}γ^{5}) =\frac{1}{\sqrt{2}} \begin{bmatrix}\1 & ±\1  \\ ∓\1 & \1  \end{bmatrix} .
\end{equation}
These operators are unitary, since $\mathsf{U}_{+}\mathsf{U}_{+}^{†}= \mathsf{U}_{+}\mathsf{U}_{-}=\1$, and can be defined with the gamma matrices of the standard or chiral representations. The action of $\mathsf{U}_{+}$ on a chiral Dirac bispinor $ψ$ gives the standard Dirac bispinor $ψ'$:
\begin{equation}
	ψ'≐ \mathsf{U}_{+} ψ =\frac{1}{\sqrt{2}} \begin{bmatrix}\1 & \1  \\ -\1 & \1  \end{bmatrix} \begin{pmatrix} \phi_{\textsc{l}} \\ χ_{\textsc{r}} \end{pmatrix} = \frac{1}{\sqrt{2}}\begin{pmatrix} \phi_{\textsc{l}} + χ_{\textsc{r}}\\ χ_{\textsc{r}}-\phi_{\textsc{l}} \end{pmatrix} .
\end{equation}
One recovers the chiral bispinor with $ψ=\mathsf{U}_{-}ψ'$.

%GEOMETRIC
\switchcolumn\paragraph{Standard representation of Dirac operators}  Dirac's operators in geometric algebra are \\ (Dirac basis)
\begin{eqnarray}
	γ^{0}\,\!' &\leftrightarrow&  \e_{0}(\dots)ℐ_{\3}= ℐ_{\3}(\dots)ℐ_{\3}, \\
	γ^{j} =γ^{j}\,\!'  &\leftrightarrow&  \e_{j}(\dots)ℐ_{\3} = ℐ_{\1}\B_{j}(\dots)ℐ_{\3}\B_{3}^{2}, \\
	γ^{5}\,\!'  &\leftrightarrow&  ℐ_{\2} (\dots)\B_{3}, \\
	γ^{0}\,\!'γ^{5}\,\!'  &\leftrightarrow&  ℐ_{\1}(\dots)ℐ_{\3}\B_{3}.
\label{DoranjStandard}
\end{eqnarray}
They act on quadrispinors in standard representation $\Mc' ≐ \mathsf{U}_{+}( \Mc) $ that are obtained from chiral quadrispinor $\Mc$ using 
\begin{eqnarray}
	\mathsf{U}_{±}(\Mc) 	&≐& \frac{\Mc±ℐ_{\1}\Mcℐ_{\3}\B_{3}}{\sqrt{2}} , \label{SwitchRepOp}
\end{eqnarray}
such that, as expected:
\begin{eqnarray}
	\Mc'_{\eee}&≐& \mathsf{U}_{+}(\Mc_{\eee}) \\ 
	&=& \frac{\Mc_{\2} + \Mc_{\0} }{\sqrt{2}} +\frac{\Mc_{\2} - \Mc_{0} }{\sqrt{2}}ℐ_{\2}\B_{3},  \\
	\Mc'_{\ooo}&≐& \mathsf{U}_{+}(\Mc_{\ooo})\\
	&=& \frac{\Mc_{\1} + \Mc_{\3} }{\sqrt{2}}ℐ_{\3}\B_{3} +\frac{\Mc_{\1} - \Mc_{3} }{\sqrt{2}}ℐ_{\1}. \nonumber \\
\end{eqnarray}
One recovers the chiral quadrispinor with $\Mc=\mathsf{U}_{-}(\Mc')$. Notice that using Section \ref{RepChoice} for representation change leads to $γ^{μ}\,\!' = \mathsf{U}_{+} \circ γ^{μ}\circ \mathsf{U}_{-}$. Like in the matrix framework, the operator associated with $γ^{5}\,\!'$ exchanges the two components of the geometric bispinors. 
However, in GA it also indicates the chirality of the spinor components of the quadrispinor since
\begin{equation}
	\Mc'_{\eee} = \Mc_{\0} \frac{1-ℐ_{\2}\B_{3}}{\sqrt{2}} + \Mc_{\2} \frac{1+ℐ_{\2}\B_{3}}{\sqrt{2}} \label{ChiralDecompositionEven}
\end{equation}
has a first (resp. second) term that is an eigenstate of $ℐ_{\2}(\dots)\B_{3}$ with eigenvalue $-1$ (resp. $+1$).
\end{paracol}

\begin{paracol}{2}[]
% MATRIX
\paragraph{Charge conjugation of Dirac bispinors} 
The \\ charge-conjugate bispinor is 
\begin{equation}
	ψ^\mathrm{c} ≐ -ι γ^{2}  \begin{pmatrix} \phi_{\textsc{l}}^{*} \\ χ_{\textsc{r}}^{*} \end{pmatrix} =   \begin{pmatrix} ι σ^{2} χ_{\textsc{r}}^{*} \\  -ι σ^{2} \phi_{\textsc{l}}^{*} \end{pmatrix}, \label{diracspinor-ChargeConjugate}
\end{equation} 
see Ref.~\cite{Book_Maggiore2005} (page 33). Since $γ^{2}$ is the same in chiral and standard representations, charge conjugation is defined in the same way for the two representations.

%GEOMETRIC
\switchcolumn\paragraph{Charge conjugation of quadrispinors} 
The definition of charge conjugation of Eqs.~(\ref{C-right}--\ref{C-left}) can be extended to an arbitrary quadrispinor $\Mc$ by
\begin{equation}
	\Mc^\mathrm{c} ≐ -\Mc ℐ_{\2}\B_{1}. \label{chargeconjugation}
\end{equation}
This is clearly an involution $\Mc=(\Mc^\mathrm{c})^\mathrm{c}$ that preserves the grade parity, e.g., the even grade part is transformed into an even multivector. 
Let's check directly that the extended charge conjugation of Eq.~\eqref{chargeconjugation} acts as expected: For the even part, we have
\begin{eqnarray} % RIGHT CHARGE CONJ
	\Mc^\mathrm{c}_{\eee}&≐& -(\Mc_{\0}+\Mc_{\2}\B_{3}ℐ_{\2}) ℐ_{\2}  \B_{1}, \label{chargeconj1} \\
							&≐& (\Mc_{\2}\B_{2})+(-\Mc_{\0}\B_{2})ℐ_{\2}\B_{3}.
\end{eqnarray}
For the odd part, we find
\begin{eqnarray} % RIGHT CHARGE CONJ
	\Mc^\mathrm{c}_{\ooo}&≐& -(\Mc_{\3}\B_{3}ℐ_{\3} + \Mc_{\1}ℐ_{\1}) ℐ_{\2}  \B_{1}, \\
							&≐& (\Mc_{\1}\B_{2})\B_{3}ℐ_{\3}+(-\Mc_{\3}\B_{2})ℐ_{\1}. \label{chargeconj4}
\end{eqnarray} 
We have used parentheses to emphasize the terms that can be interpreted as left and right components in the above equations. 

We get the definition of charge conjugation in the standard representation by computing 
\begin{equation}
	\Mc'\,\!^\mathsf{c} ≐ \mathsf{U}_{+}\left([\mathsf{U}_{-}(\Mc')]^\mathsf{c}\right) = \mathsf{U}_{+}\left(\Mc^\mathsf{c}\right)= \Mc^\mathsf{c}\,\!'.
\end{equation} 
By inspection of Eqs.~\eqref{SwitchRepOp} and \eqref{chargeconjugation}, the change of representation and the charge conjugation indeed commute, since $ℐ_{\3}\B_{3}$ commutes with $ℐ_{\2}\B_{1}$. Charge conjugation has the same definition in the chiral and standard representations.

\end{paracol}

\subsubsection{Wave equations for materially inhomogeneous quadrispinors}

\begin{paracol}{2}[]
	
\paragraph{Spacetime derivative and Dirac equation} In matrix algebra, the spacetime derivative uses the Dirac matrices in chiral or standard representation
\begin{eqnarray}
	\slashed{\partial} &≐& ∑_{μ}γ^{μ}∂_{μ} ,\\
	\slashed{\partial}'&≐& ∑_{μ}γ^{μ}\,\!'∂_{μ} ,
\end{eqnarray}
when denoting $\partial_{\mu}$ the partial derivative with respect to the spacetime coordinate $x^{\mu}$. Wave equation writes 
\begin{align}
	\imath ℏ \slashed{\partial} ψ &= mc ψ & \text{(chiral rep.),}\\
	\imath ℏ \slashed{\partial}' ψ' &= mc ψ' & \text{(standard rep.),}
\end{align}
where $ℏ$ is the Planck constant divided by $2π$ and $c$ the light speed in vacuum. With the electromagnetic field associated to vector potential $\bm a = \gamma^{μ} a_{μ}$ the wave equation writes in standard representation
\begin{equation}
	ℏ(\imath \gamma^\mu\,\!'\partial_\mu-mc)ψ' = e\gamma^\mu\,\!' a_\mu ψ'.
\end{equation}

In the following, we often use natural units for which $ℏ=1$ and $c=1$.

\switchcolumn
\paragraph{Spacetime derivative and wave equation} In geometric algebra, the spacetime derivative in chiral and standard representations are respectively
\begin{eqnarray}
	\slashed{\partial}(\dots) & \leftrightarrow & ℐ_{\2}∂_{0}(\dots)\B_{\3}+∑_{k}\e_{k}∂_{k}(\dots)ℐ_{\3},\nonumber  \\ \label{chiralderivative} \\
	\slashed{\partial}'(\dots) & \leftrightarrow & ∇(\dots)ℐ_{\3} = ∑_{μ} \e_{μ}∂_{μ}(\dots)ℐ_{\3}, \label{standardderivative}
\end{eqnarray}
when denoting $\partial_{\mu}$ the partial derivative with respect to the spacetime coordinate $x^{\mu}$, and using the conventional gradient operator of geometric algebra $∇≐ \sum_{μ} \e_{μ} ∂_{μ}$. We remark that the operators defined in Eqs.~(\ref{chiralderivative}--\ref{standardderivative}) preserve the grade parity. The wave equation is written in chiral representation
\begin{equation}
	-ℏℐ_{\2}∂_{0}\Mc+ℏ∑_{k}\e_{k}∂_{k}\Mcℐ_{\3}\B_{3} = mc \Mc,
\end{equation}
and in standard representation
\begin{equation}
	ℏ∇\Mc' ℐ_{\3}\B_{3} = m c \Mc'. \label{waveequationMassHomogenous}
\end{equation}
As expected, the even and odd parts evolve according to two decoupled wave equations corresponding to a pair of Dirac equations. We show in appendix \ref{WeylPauliWE} that the Weyl--Pauli wave Eqs.~(\ref{Pauli1}--\ref{Pauli2}) are recovered from the even part of the wave equation in chiral representation (the odd part would as well).
With the electromagnetic field associated with the vector potential $\bm a$, the wave equation was written in standard representation \cite{Doran2005}
\begin{equation}
	ℏ∇\Mc' ℐ_{\3}\B_{3} - ℏe\bm a\Mc' ℐ_{\3} = m c \Mc'.
	\label{wave-eq-abelian-gauge}
\end{equation}
However, we develop in Section~\ref{InternalDOF} a different approach for electroweak interaction that is incompatible with Eq.~\eqref{wave-eq-abelian-gauge}.

\xdef\mycounter{\arabic{equation}}

\end{paracol}

\setcounter{equation}{\mycounter}

\subsubsection{Advantages of the standard representation}

\paragraph{Gradient} Like in matrix algebra, many representations (different from the chiral and standard) are possible; see Section \ref{RepChoice}. The standard representation is convenient because the derivative operator defined with the Dirac matrices $\sum_{μ} γ^{μ}\,\!'∂_{μ}$ is the gradient operator of GA combined with the right multiplication by $ℐ_{\3}$. 

\paragraph{Chirality} We also noted that the eigenvalues for operator $γ^{5}\,\!'$ still determines the chirality of each spinorial component. This is surprising from the perspective of matrix algebra since the Dirac bispinor has upper and lower components mixing Weyl's spinors of different chirality.

\paragraph{Lorentz} Another advantage of the standard representation is that a general Lorentz transformation is generated by a spacetime rotor (i.e., a normalized even-grade multivector). It acts on the left on spinors (order 1) and by conjugacy action on other elements of the algebra (e.g., on vectors, on order 2 spinors). The infinitesimal Lorentz transformation in the standard representation is defined by switching to the chiral representation, applying the transformation of Eq.~\eqref{infinitesimalLORENTZ} and then switching back to the standard representation:
\begin{equation}
	\hat {\Lc}' (\dots) =  \mathsf{U}_{+} \hat {\Lc} (\mathsf{U}_{-} \dots).
\end{equation}
Performing this calculation leads to the expression of a finite Lorentz transformation in standard representation:
\begin{eqnarray}
	\hat \Lc' \Mc' &=& \mathsf{U}_{+}\circ\mathsf{U}_{-}(\Mc') -  \mathsf{U}_{+}\left[\sum_{k}\frac{θ_{k}}{2}\B_{k} \mathsf{U}_{-}(\Mc')\right] + \mathsf{U}_{+}\left[\sum_{k}\frac{η_{k}}{2}\B_{k} ℐ_{\3}\mathsf{U}_{-}(\Mc')ℐ_{\3} \B_{3} \right], \\
	\hat \Lc' \Mc'&=& \Mc' -  \sum_{k}\frac{θ_{k}}{2}\B_{k} \Mc' + \sum_{k}\frac{η_{k}}{2}ℐ_{\2}\B_{k} \Mc'_{\3},  \\
	\hat \Lc' \Mc'&=& \exp \left(-\sum_{k} θ_{k}\B_{k}/2 + \sum_{k}η_{k}ℐ_{\2}\B_{k}/2\right) \Mc', \label{LORENTZstd}
\end{eqnarray}
in which we have used $θ_{k} \ll 1$ and $η_{k} \ll 1$ for $k=1,2,3$, and
\begin{eqnarray}
	\B_{k}ℐ_{\3}\mathsf{U}_{-}(\Mc')ℐ_{\3} \B_{3} &=& \B_{k}ℐ_{\3}\frac{\Mc'-ℐ_{\1}\Mc'ℐ_{\3}\B_{3}}{\sqrt{2}}ℐ_{\3} \B_{3} \\ 
	&=&  \B_{k}\frac{ℐ_{\2}ℐ_{\1} \Mc'_{\3} ℐ_{\3}\B_{3}+ℐ_{\2}\Mc'}{\sqrt{2}}, \\
	&=& \B_{k}ℐ_{\2} \mathsf{U}_{+} (\Mc'), \\
	&=& \mathsf{U}_{-} (ℐ_{\2}\B_{k}\Mc').
\end{eqnarray}
As announced, we recognize that $\hat \Lc'$ is a spacetime rotor in Eq.~\eqref{LORENTZstd}.  Pure boosts in both representations are detailed further in Appendix~\ref{pureboost}.

\subsubsection{Projectors onto spinorial components in chiral and standard representations} 

Within the standard representation and the even subalgebra only, projectors have been introduced to collect left and right spinorial components \cite{Book_Doran2003, Hestenes1982vol12}. In this Section, we introduce projectors to extract from a quadrispinor the space spinor components, whatever the representation, and for the whole geometric algebra. On mass-homogeneous quadrispinors that we have considered up to now, spinorial components are associated with particles of matter or antimatter with a given left or right chirality.

\paragraph{Chiral representation} We define the projectors
\begin{equation}
	\mathsf{P}^{\aa}_{±}(\dots) ≐ \frac{(\dots)± ℐ_{\aa}(\dots)ℐ_{\aa}}{2} \quad \text{ for } \aa= \1,\2,\3, \label{aaProj}
\end{equation}
to extract the spinorial components since
\begin{eqnarray}
	\mathsf{P}^{\1}_{+}(\Mc) &=& \Mc_{\0}+ \Mc_{\1}ℐ_{\1}, \\
	\mathsf{P}^{\1}_{-}(\Mc) &=& \Mc_{\2}\B_{3}ℐ_{\2} + \Mc_{\3}\B_{3}ℐ_{\3} ,
\end{eqnarray}
where even and odd grade projections can be used in addition to extract just a single spinorial component:
\begin{eqnarray}
	\Mc_{\0} &=& \mathsf{P}^{\1}_{+}(\Mc_{\eee}), \\
	\Mc_{\1}ℐ_{\1} &=& \mathsf{P}^{\1}_{+}(\Mc_{\ooo}), \\
	\Mc_{\2}ℐ_{\2}\B_{3} &=& \mathsf{P}^{\1}_{-}(\Mc_{\eee}), \\
	\Mc_{\0}ℐ_{\3}\B_{3} &=& \mathsf{P}^{\1}_{-}(\Mc_{\ooo}). 
\end{eqnarray}
We remark that the even and odd grade projections write $(\dots)_{\eee} = \mathsf{P}^{\2}_{-}(\dots)$ and $(\dots)_{\ooo} = \mathsf{P}^{\2}_{+}(\dots) $ respectively.

\paragraph{Standard representation} To extract spinorial components in this representation, we define the projectors
\begin{equation}
	\P^{i}_{±}(\dots) ≐ \frac{(\dots)±ℐ_{\2}(\dots)\B_{i}}{2} \quad \underset{i=3}{\longrightarrow} \quad \P^{3}_{±}(\dots) ≐ \frac{(\dots)±ℐ_{\2}(\dots)\B_{3}}{2}, \label{StdChiralProj}
\end{equation}
in which we single out the third projector given the choice of complex number we did in Eq.~\eqref{scalprodSpinor}. For simplicity, we denote its application on $1$ by
\begin{equation}
	i_{±}≐\P^{i}_{±}(1) = \frac{1±ℐ_{\2}\B_{i}}{2}  \quad \underset{i=3}{\longrightarrow} \quad 3_{±} = \frac{1±ℐ_{\2}\B_{3}}{2},
\end{equation}
that is also a projector. We will use it most often acting from the right toward the multivector on its left, i.e., $(\dots)i_{±}$. Using this notation, we can use the same spinorial components $\Mc_{\bbmu}$ to write the quadrispinor $\Mc$ or $\Mc'$ in chiral or standard representation respectively:
\begin{eqnarray}
	\Mc &=& \Mc_{\0}+\Mc_{\2}\B_{3}ℐ_{\2} + \Mc_{\3}\B_{3}ℐ_{\3} + \Mc_{\1}ℐ_{\1}    \label{ChiralQuadrispinor}, \\
	\Mc'&=&  \left[ \Mc_{\0}3_{-} + \Mc_{\2}3_{+} -\Mc_{\3}3_{-}ℐ_{\1} + \Mc_{\1}3_{+}ℐ_{\1} \right] \sqrt{2}. \label{StandardQuadrispinor}
\end{eqnarray}
in which the second line (for the even part) is just a rewriting of Eq.~\eqref{ChiralDecompositionEven} using $3_{±}$. Quadrispinors in Eqs.~(\ref{ChiralQuadrispinor}--\ref{StandardQuadrispinor}) are mass-homogeneous: they include four particles with identical mass.

\begin{table}[t] \centering
{
\renewcommand{\arraystretch}{1.3}
	\begin{tabular}{|c|C|C|C|C|} \hline
		Chiral basis	&1&ℐ_{\2}\B_{3}&ℐ_{\3}\B_{3}& ℐ_{\1} \\ 
		Standard basis	&3_{-}&3_{+}&-3_{-}ℐ_{\1}&3_{+}ℐ_{\1} \\
		Chirality (absolute)	& \textsc{l}&\textsc{r}&\textsc{l}& \textsc{r}\\  \hline
		Spinorial component &\Mc_{\0}&\Mc_{\2}&\Mc_{\3}& \Mc_{\1} \\ 
		Matter First (MF)  	&f_\textsc{l}&\overline{f_\textsc{l}}& \overline{f_\textsc{r}}	&f_\textsc{r} \\
		Matter last	(ML)	&\overline{f_\textsc{r}}&f_\textsc{r}&f_\textsc{l}&\overline{f_\textsc{l}} \\ \hline
	\end{tabular}
} \caption[Basis decomposition of mass-homogeneous quadrispinor]{Basis decomposition of mass-homogeneous quadrispinor $\Mc$: A fermionic particle $f_{χ}$ (or antiparticle $\overline{f_{χ}}$) is described by a spinorial component $\Mc_{\bbmu}$ (i.e., a space spinor) that is associated with the basis element of spacetime algebra in the same column. The quadrispinors $\Mc$ and $\Mc'$ in chiral and standard representations, respectively, are the sum of the products of each spinorial component with its associated basis element (in the corresponding representation). The last two lines indicate the materiality of each of the quadrispinor components according to the type of the quadrispinor: for matter first (respectively last) quadrispinors, the first (respectively last) two spinorial components $\bbmu = \0,\1$ (respectively $\bbmu=\2,\3$) are associated with matter particles. \label{MassHomogenousTable}}
\end{table}

Given that the operators in the wave equations do not change the grade parity, we find the four projected wave equations in standard representation
\begin{eqnarray}
	∇(\Mc'_{\eee}3_{+}) ℐ_{\3}\B_{3} &=& m \Mc'_{\eee}3_{-}, \label{Weyls1}\\
	∇(\Mc'_{\eee}3_{-}) ℐ_{\3}\B_{3} &=& m \Mc'_{\eee}3_{+}, \\
	∇(\Mc'_{\ooo}3_{+}) ℐ_{\3}\B_{3} &=& m \Mc'_{\ooo}3_{-}, \\
	∇(\Mc'_{\ooo}3_{-}) ℐ_{\3}\B_{3} &=& m \Mc'_{\ooo}3_{+}. \label{Weyls4}
\end{eqnarray}
These equations are equivalent to Weyl's equations since $\Mc'_{\ppp}3_{±}$ are the GA equivalent of a Weyl's spinor in which the chirality is explicit.

\subsubsection{Matter first and matter last quadrispinors}

Due to the matter/antimatter symmetry under charge conjugation, the order of matter and antimatter spinorial components in a quadrispinor does not really matter as long as each component's chirality is correct. This allows for two different kinds of quadrispinors that we call \emph{matter first} (MF) and \emph{matter last} (ML) quadrispinors. MF quadrispinors have matter particles on spinorial components $\0,\1$ and antimatter particles on $\2,\3$. It is the opposite for ML quadrispinors: matter particles are on components $\2,\3$ and antimatter on $\0,\1$; see Table~\ref{MassHomogenousTable}.

However, to describe the electroweak and strong interactions, the isoversor associated with each spinorial component determines the charge and behavior of the corresponding particle under the chosen interaction. Operators used to describe the electroweak and strong interactions are provided in Section \ref{InternalDOF} based on the generators of the special unitary groups defined in Section \ref{GAST}. For the materiality inhomogeneous quadrispinors we have just introduced, the operators in the fundamental and conjugate representations must be composed with projection operators that extract the spinorial component with the right materiality. This approach is operational for the electroweak interaction, but becomes more sophisticated for the strong interaction. For this reason, we introduce in the next Section materiality-homogeneous quadrispinors on which interaction operators can act directly. 

\begin{table}
	\centering

\vspace{1em}
{\renewcommand{\arraystretch}{1.3}
\begin{tabular}{|c||C|C|C|C|} \hline
	Chiral basis 	&	1	&	ℐ_{\2}\B_{3}	&	ℐ_{\3}\B_{3}	&		ℐ_{\1} \\ 
	Standard basis &	3_{-}	&	3_{+}	&	-3_{-}ℐ_{\1}	&		3_{+}ℐ_{\1} \\ \hline 
	Fermions in $\Pc_{(c)}$ 				&	f_\textsc{l}		&	p_\textsc{r}			&	p_\textsc{l}		&	f_\textsc{r}		\\ 
	Anti-fermions in $\bar \Pc_{(c)}$ &	\overline{p_\textsc{r}}	&	\overline{f_\textsc{l}}	&	\overline{f_\textsc{r}}&	\overline{p_\textsc{l}}	\\
	\hline 
	Space spinors in $\Pc_{(c)}$ & \Mc_{\0}& \Nc_{\2}& \Nc_{\3}& \Mc_{\1}\\ 
	Space spinors in $\bar \Pc_{(c)}$ & \Nc_{\0}& \Mc_{\2}& \Mc_{\3}& \Nc_{\1}\\ \hline
\end{tabular}}
	\caption{Description of the relation between two mass homogenous quadrispinors $\Mc$ (MF type) and $\Nc$ (ML type) with the materiality homogenous quadrispinors $\Pc$ and $\bar \Pc$. For instance, the space spinor $\Nc_{\2}$ is associated to the right fermion $p_\textsc{r}$ and appear in front of $ℐ_{\2}\B_{3}$ inside $\Pc$ in chiral representation (resp. of $3_{+}$ in standard representation).\label{PandPBAR}}
\end{table}

\subsection{Quadrispinor of homogeneous materiality and associated wave equations}

In the previous Section, we introduced quadrispinors describing particles and antiparticles with different chirality, but identical mass. The latter mass homogeneity (of particles inside a quadrispinor) is particularly convenient to provide the dynamics with a unique wave equation whose free solution is easily accessible \cite{Book_Doran2003}. On the other hand, the fact that there exist two different kinds of quadrispinors, with opposite materiality of each spinorial component, drives us to consider the possibility of two different kinds of (third) isospin. In other words, the MF and ML quadrispinors translate in GA the concept of upper and lower components of an electroweak doublet.

Below, we define quadrispinors of homogeneous materiality whose spinorial components describe particles with different masses. Materiality-homogeneous quadrispinors are di-doublet quadrispinors or polychrome quadrispinors, denoted respectively $\Pc_{(c)}$ or $\Pc_{(\bbmu)}$. Di-doublet quadrispinors include a left and a right isospin doublet, all being leptons (assuming a right neutrino exists), or all being quarks (of the same color). Polychrome quadrispinors include a red, a green, and a blue quark and one lepton, all with the same weak isospin and chirality. Interaction operators acting on them mix particles appropriately: for instance, for the weak interaction, only left particles of opposite isospin are mixed. Similarly, for the strong interaction, only the quarks of different colors are mixed. Switching to quadrispinors of homogeneous materiality, mass operators emerge naturally to attribute the correct mass to each spinorial component.

\subsubsection{Di-doublet quadrispinors}

\label{MatterHomogenous}
\paragraph{From materiality-inhomogeneous to materiality-homogeneous} 
Let's consider a matter first quadrispinor $\Mc$ for fermions $f_{χ}$ and $\overline{f_{χ}}$ of mass $m_{f}$ and a matter last quadrispinor $\Nc$ for fermions $p_{χ}$ and $\overline{p_{χ}}$ of mass $m_{p}$, with $χ=\textsc{l,r}$. We denote, as usual, those quadrispinors $\Mc'$ and $\Nc'$ in the standard representation. We assume that the quadrispinors $\Mc$ and $\Nc$ (prime or not) include particles of the same color $c$ (red, green, blue for quarks or colorless for leptons). Based on those quadrispinors, the particle quadrispinor and the antiparticle quadrispinor are written as
\begin{eqnarray}
	\Pc_{(c)} &=& \mathsf{P}^{\1}_{+}(\Mc)  + \mathsf{P}^{\1}_{-}(\Nc),  \qquad \text{(particle)} \label{particle1}\\
	\bar \Pc_{(c)} &=& \mathsf{P}^{\1}_{-}(\Mc) +\mathsf{P}^{\1}_{+}( \Nc), \qquad \text{(anti-particle)} \label{antiparticle1}
\end{eqnarray}
in chiral representation and
\begin{eqnarray}
	\Pc'_{(c)} &=& \Mc' 3_{-} + \Nc' 3_{+},  \qquad \text{(particle)} \label{particle2}\\
	\bar \Pc'_{(c)} &=& \Mc' 3_{+} + \Nc' 3_{-}, \qquad \text{(anti-particle)} \label{antiparticle2}
\end{eqnarray}
in standard representation. In other words, $\mathsf{P}^{\1}_{+}$ [resp. $(\dots)3_{-}$] is the projector on matter spinorial components for matter first quadrispinors; $\mathsf{P}^{\1}_{-}$ [resp. $(\dots)3_{+}$] is the projector on matter spinorial components for matter last quadrispinors. Opposite signs are used to project onto antimatter spinorial components. The particle content of each quadrispinor is provided in Table~\ref{PandPBAR}.
We can recover the (mass-homogeneous) quadrispinors using
\begin{eqnarray}
	\Mc &=& \mathsf{P}^{\1}_{+}(\Pc_{(c)})  + \mathsf{P}^{\1}_{-}(\bar \Pc_{(c)}),  \qquad \text{(matter first)} \label{particle3}\\
	\Nc &=& \mathsf{P}^{\1}_{-}(\Pc_{(c)}) +\mathsf{P}^{\1}_{+}( \bar \Pc_{(c)}), \qquad \text{(matter last)} \label{antiparticle3}
\end{eqnarray}
in chiral representation and
\begin{eqnarray}
	\Mc' &=& \Pc'_{(c)} 3_{-} + \bar \Pc'_{(c)} 3_{+},\qquad \text{(matter first)} \label{particle4} \\
	\Nc' &=& \Pc'_{(c)} 3_{+} + \bar \Pc'_{(c)} 3_{-},\qquad \text{(matter last)} \label{particle4}
\end{eqnarray}
in standard representation. We remark that materiality-homogeneous quadrispinors have independent spinorial components in general: the overbar is a notation indicating that the quadrispinor is composed of antimatter particles, not an involution that produces the antimatter partner from the quadrispinor for matter particles.

As already mentioned, bi-doublet quadrispinors are labeled with subscript $c=r,b,g,w$ as color index: $w$ for white leptons and $r,b,g$ for red, blue and green quarks respectively. Table \ref{ParticleNotation} summarizes these notations. For instance, a leptonic quadrispinor associated with the fourth row ($c=w$) in the sub-table entitled ``Fermions'' is written
\begin{equation}
	\Pc_{(w)}=\underbrace{\Pc_{w\0}}_{ν_{\textsc{l}}} +\overbrace{\Pc_{w\2}ℐ_{\2}\B_{3}}^{e_\textsc{r}}+\underbrace{\Pc_{w\3}ℐ_{\3}\B_{3}}_{e_\textsc{l}}+\overbrace{\Pc_{w\1}ℐ_{\1}}^{ν_\textsc{r}},
\end{equation}
in which particles for each spinorial component are in reading order $ν_\textsc{l}$, $e_\textsc{r}$, $e_\textsc{l}$ and $ν_\textsc{r}$, with the underbrace (resp. overbrace) emphasizing the left (resp. right) weak doublet. Similarly, a quadrispinor for red quarks is, for instance
\begin{equation}
	\Pc_{(r)}=\underbrace{\Pc_{r\0}}_{u^{r}_{\textsc{l}}} +\overbrace{\Pc_{r\2}ℐ_{\2}\B_{3}}^{d^{r}_\textsc{r}}+\underbrace{\Pc_{r\3}ℐ_{\3}\B_{3}}_{d^{r}_\textsc{l}}+\overbrace{\Pc_{r\1}ℐ_{\1}}^{u^{r}_\textsc{r}},
\end{equation}
in which particles for each spinorial component are in reading order $u_\textsc{l}^{r}$, $d_\textsc{r}^{r}$, $d_\textsc{l}^{r}$ and $u_\textsc{r}^{r}$.

\begin{table} \centering
	\renewcommand{\arraystretch}{1.3} 
	\begin{tabular}{|C|C|C|C|C|} 
	\hline
		\text{Rep.}		& \multicolumn{4}{c|}{Eigenstates} \\ \hline
		\text{Chiral}	& 1	& ℐ_{\2}\B_{3} & ℐ_{\3}\B_{3} 	& ℐ_{\1} \\ \hline
		\text{Std.}	&3_{-} & 3_{+} &  -3_{-}ℐ_{\1} & 3_{+}ℐ_{\1}  \\ \hline
		& \multicolumn{4}{c|}{Fermions} \\ \hline
		\Pc_{c\bbmu} & \bbmu = \0 & \bbmu =\2 & \bbmu= \3 & \bbmu=\1 \\ \hline 
		c=r &	u_{\textsc{l}}^{r}\rc	&	d_{\textsc{r}}^{r}\rc	& 	d_{\textsc{l}}^{r}\rc 	& u_{\textsc{r}}^{r}\rc		\\ \hline
		c=b &	u_{\textsc{l}}^{b}\bc	&	d_{\textsc{r}}^{b}\bc	&	d_{\textsc{l}}^{b}\bc	& u_{\textsc{r}}^{b}\bc			\\ \hline
		c=g &	u_{\textsc{l}}^{g}\gc		&	d_{\textsc{r}}^{g}\gc		&	d_{\textsc{l}}^{g}\gc		& 	u_{\textsc{r}}^{g}\gc		\\ \hline
		c=w &	ν_{\textsc{l}}		&	e_{\textsc{r}}		&	e_{\textsc{l}}		& ν_{\textsc{r}}			\\ \hline 
						& \multicolumn{4}{c|}{Weak charges} \\ \hline
		\mathsf{T}^\textsc{w}_{\3}	&	1/2	&	0		&	-1/2	&	0	\\
		\mathsf{Y}^\textsc{w}		&	1/3	&	-2/3	&	1/3	&	4/3	\\ \hline
		\mathsf{T}^\textsc{w}_{\3}	&	1/2	&	0		&	-1/2	&	0	\\
		\mathsf{Y}^\textsc{w}		&	1/3	&	-2/3	&	1/3	&	4/3	\\ \hline{}
		\mathsf{T}^\textsc{w}_{\3}	&	1/2	&	0		&	-1/2	&	0	\\
		\mathsf{Y}^\textsc{w}		&	1/3	&	-2/3	&	1/3	&	4/3	\\ \hline{}
		\mathsf{T}^\textsc{w}_{\3}	&	1/2	&	0		&	-1/2	&	0	\\
		\mathsf{Y}^\textsc{w}		&	-1		&	-2		&	-1		&	0	\\ \hline 
						& \multicolumn{4}{c|}{Strong charges} \\ \hline
		\mathsf{T}^\textsc{c}_{\3}	&	1/2	&	1/2	&	1/2	&	1/2	\\
		\mathsf{Y}^\textsc{c}		&	1/3	&	1/3	&	1/3	&	1/3	\\ \hline
		\mathsf{T}^\textsc{c}_{\3}	&	0		&	0		&	0		&	0		\\
		\mathsf{Y}^\textsc{c}		&	-2/3	&	-2/3	&	-2/3	&	-2/3	\\ \hline
		\mathsf{T}^\textsc{c}_{\3}	&	-1/2	&	-1/2	&	-1/2	&	-1/2	\\
		\mathsf{Y}^\textsc{c}		&	1/3	&	1/3	&	1/3	&	1/3	\\ \hline
		\mathsf{T}^\textsc{c}_{\3}	&	0		&	0		&	0		&	0	\\
		\mathsf{Y}^\textsc{c}		&	0		&	0		&	0		&	0	\\ \hline
	\end{tabular}
	\begin{tabular}{|C|C|C|C|C|C|C|} 
	\cline{1-5}
		\text{Rep.}		& \multicolumn{4}{c|}{Eigenstates}  \\ \cline{1-5}
		\text{Chiral}	& 1	& ℐ_{\2}\B_{3} & ℐ_{\3}\B_{3} 	& ℐ_{\1} \\ \cline{1-5}
		\text{Std.}	&3_{-} & 3_{+} &  -3_{-}ℐ_{\1} & 3_{+}ℐ_{\1}  \\ \cline{1-5}
		& \multicolumn{4}{c|}{Anti-Fermions} \\ \cline{1-5}
		\bar \Pc_{c\bbmu} & \bbmu = \0 & \bbmu =\2 & \bbmu= \3 & \bbmu=\1 \\ \hline 
		c=b &	\bar u_{\textsc{l}}^{b}\bc		&	\bar d_{\textsc{r}}^{r}\bc	& 	\bar d_{\textsc{l}}^{r}\bc & \bar u_{\textsc{r}}^{r}\bc	& 1	& 3_{-}\\ \hline
		c=r &	\bar u_{\textsc{l}}^{r}\rc		&	\bar d_{\textsc{r}}^{r}\rc	&	\bar d_{\textsc{l}}^{r}\rc	& \bar u_{\textsc{r}}^{r}\rc	& ℐ_{\2}\B_{3}&3_{+}\\ \hline
		c=g &	\bar {ν}_{\textsc{l}}			&	\bar e_{\textsc{r}}			&	\bar e_{\textsc{l}}			& \bar {ν}_{\textsc{r}}			& ℐ_{\3}\B_{3} & -3_{-}ℐ_{\1}\\ \hline
		c=w &	\bar u_{\textsc{l}}^{g}\gc		&	\bar d_{\textsc{r}}^{g}\gc	&	\bar d_{\textsc{l}}^{g}\gc	& \bar u_{\textsc{r}}^{g}\gc	& ℐ_{\1} & 3_{+}ℐ_{\1}	\\ \hline
						& \multicolumn{4}{c|}{Weak charges}  \\ \cline{1-5}
		\mathsf{T}^\textsc{w}_{\3}	&	-1/2	&	0		&	1/2	&	0	\\
		\mathsf{Y}^\textsc{w}		&	-1/3	&	2/3	&	-1/3	&	-4/3	\\ \cline{1-5}
		\mathsf{T}^\textsc{w}_{\3}	&	-1/2	&	0		&	1/2	&	0	\\
		\mathsf{Y}^\textsc{w}		&	-1/3	&	2/3	&	-1/3	&	-4/3	\\ \cline{1-5}
		\mathsf{T}^\textsc{w}_{\3}	&	-1/2	&	0		&	1/2	&	0	\\
		\mathsf{Y}^\textsc{w}		&	1		&	2		&	1		&	0	\\ \cline{1-5}
		\mathsf{T}^\textsc{w}_{\3}	&	-1/2	&	0		&	1/2	&	0	\\
		\mathsf{Y}^\textsc{w}		&	-1/3	&	2/3	&	-1/3	&	-4/3	\\ \cline{1-5}
						& \multicolumn{4}{c|}{Strong charges} \\ \cline{1-5}
		\mathsf{T}^\textsc{c}_{\3}	&	0		&	0		&	0		&	0		\\
		\mathsf{Y}^\textsc{c}		&	2/3	&	2/3	&	2/3	&	2/3	\\ \cline{1-5}
		\mathsf{T}^\textsc{c}_{\3}	&	-1/2	&	-1/2	&	-1/2	&	-1/2	\\
		\mathsf{Y}^\textsc{c}		&	-1/3	&	-1/3	&	-1/3	&	-1/3	\\ \cline{1-5}
		\mathsf{T}^\textsc{c}_{\3}	&	0		&	0		&	0		&	0	\\
		\mathsf{Y}^\textsc{c}		&	0		&	0		&	0		&	0	\\ \cline{1-5}
		\mathsf{T}^\textsc{c}_{\3}	&	1/2	&	1/2	&	1/2	&	1/2	\\
		\mathsf{Y}^\textsc{c}		&	-1/3	&	-1/3	&	-1/3	&	-1/3	\\ \cline{1-5}

	\end{tabular}
	\caption{Sub-tables are separated by double bars. For the last three subtables, cells at the same location are for the same particle state. Left (right) sub-tables are for eigenstates associated with operators belonging to the fundamental (conjugate) representation. (Eigenstates sub-table) Eigenstates for SU(3) isoversor in chiral and standard representations are given in the second and third lines, respectively. (Fermions subtable) The space spinors $\Pc_{(c\bbmu)}$ appear in product with the previously mentioned eigenstates, e.g., $\Pc_{(b\0)}$ is the space spinor for the blue left up quark $u_\textsc{l}^{b}$. This space spinor is in product with eigenstate $ℐ_{\2}\B_{3}$ in the polychrome quadrispinor $\Pc_{(\0)}$, while it is in product with eigenstate $1$ in the bi-doublet quadrispinor $\Pc_{(b)}$. In other words, columns are for polychrome quadrispinors with given isospin and chirality. Rows are for di-doublet quadrispinors of various isospins and chirality. (Weak charges sub-table) Electroweak charges associated with each particle and hence with each $\Pc_{(c\bbmu)}$ component. (Strong charge sub-table) Color charges associated to each particle and hence to each $\Pc_{(c\bbmu)}$ component.  \label{ParticleNotation}}
\end{table}

\paragraph{Free wave equation for materiality-homogeneous quadrispinors}
For simplicity, we consider only the standard representation. The wave equations for two quadrispinors $\Mc'$ of MF type and $\Nc'$ of ML type are written as 
\begin{eqnarray}
	∇\Mc' ℐ_{\3}\B_{3} &=& m_{f} \Mc', \label{WE1}\\
	∇\Nc' ℐ_{\3}\B_{3} &=& m_{p} \Nc'. \label{WE2}
\end{eqnarray}
The wave equations for the quadrispinors $\Pc'$ and $ \bar \Pc'$ are obtained by chiral projections using Eq.~\eqref{StdChiralProj} on Eqs.~(\ref{WE1}--\ref{WE2})
\begin{eqnarray}
	∇(\Mc'3_{κ}) ℐ_{\3}\B_{3} &=& m_{f} \Mc'3_{-κ}, \\
	∇(\Nc'3_{-κ}) ℐ_{\3}\B_{3} &=& m_{p} \Nc'3_{κ}.
\end{eqnarray}
The sum of these equations yields
\begin{equation}
	∇(\Mc'3_{κ}+\Nc'3_{-κ}) ℐ_{\3}\B_{3} = m_{f} \Mc'3_{-κ}+m_{p}\Nc'3_{κ} = (\Mc'3_{-κ}+\Nc'3_{κ})(3_{-κ} m_{f}+ 3_{κ} m_{p}) 
	. \label{WEkappa}
\end{equation}
When taking $κ=∓1$, we get the following pair of wave equations
\begin{eqnarray}
	∇\Pc'_{(c)} ℐ_{\3}\B_{3} &=&  \bar \Pc'_{(c)} (3_{+} m_{f}+ 3_{-} m_{p}), %= \bar \Pc' M_{-1}^{fp} , 
	\label{WEP}\\
	∇\bar \Pc'_{(c)} ℐ_{\3}\B_{3} &=& \Pc'_{(c)}(3_{-} m_{f}+ 3_{+} m_{p}). %= \Pc M_{+1}^{fp}.	
	\label{WEPbar}
\end{eqnarray}
To attribute the right mass to each particle in the wave equation, we introduce a mass bivector (acting on the right of quadrispinors), namely
\begin{equation}
	M_{κ}^{f p} ≐ \frac{1+κℐ_{\2}\B_{3}}{2} m_{f}+  \frac{1-κℐ_{\2}\B_{3}}{2} m_{p} = \frac{1}{2} (m_{f} + m_{p}) + \frac{κ}{2}ℐ_{\2}\B_{3}(m_{f} - m_{p}).
\end{equation}
Then, the wave equation is written for materiality-homogeneous quadrispinors
\begin{eqnarray}
	∇\Pc'_{(c)} ℐ_{\3}\B_{3} &=&  \bar \Pc'_{(c)} M_{+1}^{fp}, \label{WEPmassop}\\
	∇\bar \Pc'_{(c)} ℐ_{\3}\B_{3} &=& \Pc'_{(c)} M_{-1}^{f p}.	\label{WEPbarmassop}
\end{eqnarray}
Using materiality-homogeneous quadrispinors such as given above is convenient to consider the gauge symmetry associated with the electroweak interaction. Below, we do the same for the strong interaction.

\subsubsection{Polychrome quadrispinors }

\paragraph{Quadrispinors with homogeneous isospin and chirality} In the previous Section, (matter) particles were attributed to different quadrispinors based on their color (or colorless) character, and this was indicated with subscript $(c)$. We define below quadrispinors (still with homogeneous materiality) for four (matter) particles with the same isospin and chirality. We use the following correspondence between subscript $\bbmu=\0,\1,\2,\3$ and third isospin $τ=\uparrow,\downarrow$ and chirality $χ=\textsc{l,r}$:
\begin{eqnarray}
	\0 &=& \uparrow \textsc{l}, 	\qquad \1 = \quad \uparrow \textsc{r}, \\
	\3 &=& \downarrow \textsc{l},	\qquad \2 = \quad \downarrow \textsc{r}.
\end{eqnarray}

A polychrome quadrispinor (for three down left quarks and a left electron) associated with the $\bbmu$th column in the fermions sub-table of Table \ref{ParticleNotation} is
\begin{align}
	\Pc_{(\0)}&=\overbrace{\Pc_{r\0}+\Pc_{b\3}ℐ_{\2}\B_{3}+\Pc_{g\0}ℐ_{\3}\B_{3}}^{\text{quarks}}+ \overbrace{\Pc_{w\0}ℐ_{\1}}^{\text{leptons}} && \text{(Up left particles)}, \\ 
	\Pc_{(\1)}&=\Pc_{r\1}+\Pc_{b\1}ℐ_{\2}\B_{3}+\Pc_{g\1}ℐ_{\3}\B_{3}+\Pc_{w\1}ℐ_{\1} && \text{(Up right particles)}, \\ 
	\Pc_{(\2)}&=\Pc_{r\2}+\Pc_{b\2}ℐ_{\2}\B_{3}+\Pc_{g\2}ℐ_{\3}\B_{3}+\Pc_{w\2}ℐ_{\1} && \text{(Down right particles)}, \\ 
	\Pc_{(\3)}&=\Pc_{r\3}+\Pc_{b\3}ℐ_{\2}\B_{3}+\Pc_{g\3}ℐ_{\3}\B_{3}+\Pc_{w\3}ℐ_{\1} && \text{(Down left particles)}. 
\end{align}
The basis isoversors associated with each spinorial component appear on the right of the anti-fermion subtable. We use parentheses for the subscript associated with a row (column) to indicate that $\Pc_{(\bbmu)}$ (respectively $\Pc_{(c)}$) are quadrispinors with components of isospin and chirality associated to $\bbmu$ (respectively quadrispinors of color $c$), and not just space spinor components.

\paragraph{Free wave equation for polychrome quadrispinors} In standard representation, the wave equation for quadrispinor $\Pc_{(\bbmu)}$ of homogenous isospin and chirality given by index $\bbmu=\0,\1,\2,\3$ (see below) is
\begin{eqnarray}
	∇ \Pc_{(\bbmu)}' ℐ_{\3}\B_{3} &=& \bar{\mathsf{m}}'_{(\bbmu)} \bar {\Pc}_{(\bbmu)} \label{QCD-WE1}, \\
	∇ \bar{\Pc}_{(\bbmu)}' ℐ_{\3}\B_{3} &=& \mathsf{m}'_{(\bbmu)} \Pc_{(\bbmu)},\label{QCD-WE2}
\end{eqnarray}
where the mass operators are defined by
\begin{eqnarray}
	\bar{\mathsf{m}}'_{(\bbmu)} &=& m_{\ell} \mathsf{P}'_{\3} +m_{q}\mathsf{P}_{\3}^{\perp}\,\!', \label{massop}\\
	\mathsf{m}'_{(\bbmu)} &=& m_{\ell} \mathsf{P}'_{\1} +m_{q}\mathsf{P}_{\1}^{\perp}\,\!',  \label{barmassop}
\end{eqnarray}
with $\ell=ν,e$ and $q=u,d$ being lepton and quark labels providing the mass according to index $\bbmu$ (mass of up isospin particles for $\bbmu=\0,\1$ and down isospin for $\bbmu =\2,\3$). The projectors for matter and antimatter quadrispinors are different because leptons, neutral for the strong interaction, are on different spinorial components in fundamental and conjugate representations of isoversor $SU(3)$. Obviously, the wave equations Eqs.~(\ref{QCD-WE1}-\ref{QCD-WE2}) satisfied by polychrome quadrispinors lead to Weyl--Pauli wave equations when projected onto the different spinorial components. 

Finally, let's remark that wave Eqs.~(\ref{QCD-WE1}--\ref{QCD-WE2}) can describe free quarks of different generations by using a linear combination of $\mathsf{P}'_{\bbmu}$ with linear coefficients associated to the mass of quarks (and the lepton) on each spinorial component.

\subsection{Partial conclusion}

In this Section, in addition to recovering the wave equation for bispinors in GA~\cite{Book_Hestenes2015}, we have clarified the notion of representation of the $γ$ operators. This was possible thanks to a precise analogy between matrix and geometric frameworks for chirality, charge conjugation and other involutions. The obtained wave equation holds for quadrispinors of homogeneous mass: all particles it describes must have the same mass. The precise notion of chirality, intrinsically defined according to the isoversor associated with each space spinor component motivates the introduction of the di-doublet quadrispinors describing in one object both left and right isospin doublets, hence assuming the existence of a right neutrino. This was only possible by separating matter and antimatter particles into different quadrispinors and also by introducing mass operators to attribute the right mass to each spinorial component of the quadrispinor. Since the sixteen particles of a generation were separated by color into four different quadrispinors, we also defined another splitting of the same fourteen particles into four polychrome quadrispinors in which all spinorial components have the same chirality and third isospin. Given those two types of quadrispinors, we define in the next Section the operator providing the different particle charges and elementary aspects of gauge transformation.

\newpage
\section{Electroweak and strong interactions of massless fermions}
\label{InternalDOF}

We show below that using the two different types of quadrispinors (di-doublet or polychrome), the same isoversor SU(3) group generates the gauge symmetry of the weak and strong interactions. It is clear that the operators of the fundamental representation of isoversor SU(3) generate a color mixing of the down quarks in the polychrome quadrispinor $\Pc_{(\3)}$ while having no action on the left electron, nor on the weak isospin and chirality of the quarks. Similarly, the operators $λ_{\1,\2,\3,}$ are associated to weak isospin, with $λ_{\1,\2}$ mixing left chiral states and $λ_{3}$ being the third isospin operator. Those operators acting on the di-doublet quadrispinor $\Pc_{(w)}$ do not modify the color charge of the particles (actually here colorless leptons).

The Higgs mechanism responsible for the symmetry breaking is not considered. Neither is the Yukawa coupling to the Higgs field generating the fermion masses. In this massless case, as is well known, the wave equations display the gauge symmetries associated with electroweak and strong interactions. Interestingly, gauge transformation in our geometric algebra framework corresponds to a local change of representation of Dirac operators. 

\subsection{Electroweak interaction for di-doublet quadrispinors}
\label{EWwithout}
Exchange of electroweak bosons is possible between Fermions appearing as the spinorial components of a di-doublet quadrispinor. Those Fermions have the same color, but different chirality and isospin. Given the color homogeneity (colorless or not), the Baryon minus Lepton number operator is proportional to the identity operator (in both standard and chiral representations) when acting on a di-doublet quadrispinor
\begin{eqnarray}
	(\mathsf{Ba-Le})(\Pc_{(c)}) &=& \left \{ \begin{array}{l} (1/3)\Pc_{(c)} \quad \text{ for } c= r,b,g, \\
	(-1)\,\Pc_{(c)} \quad \text{ for } c= w ,
	\end{array}\right.
\end{eqnarray}
opposite signs are obtained for quadrispinor $\bar \Pc_{(c)}$ describing antimatter particles.
Then, the three weak vector bosons $\bm w_{\ii}$ are associated to the weak isospin operators $\mathsf{T}^\textsc{w}_{\ii}$ and the vector boson $\bm b$ to the hypercharge operator $\mathsf{Y}^\textsc{w}$ defined respectively as
\begin{eqnarray}
	\mathsf{T}^\textsc{w}_{\ii} &=& λ_{\ii}/2 \\ 
	\mathsf{Y}^\textsc{w} &=& \mathsf{Ba-Le}+\bar{λ}_{\3},
\end{eqnarray}
with the lambda operators defined in Table \ref{IsoversorGroup}.
As emphasized by the notation, the vector bosons $\bm w_{\ii}$ and $\bm b$ are spacetime vectors of $\mathcal{G}(1,3)$. Their corresponding gauge operators $\mathsf{T}^\textsc{w}_{\ii}$ and $\mathsf{Y}^\textsc{w}$ act on di-doublet quadrispinors for matter particles in chiral representation. We add primes to the $λ$ operators when working with the standard representation and the superscript $\textsc{w}$ in this case. Conjugate operators are used when applied to quadrispinors for antimatter particles, i.e., $ \bar{\mathsf{T}}^\textsc{w}_{\ii} = \frac{\bar{λ}_{\ii}}{2} $ with $\ii=\1,\2,\3$ and $\bar{\mathsf{Y}}^\textsc{w}  = \mathsf{Ba-Le}+λ_{\3}$ when acting on $\bar \Pc_{(c)}$. 
We emphasize that the charge operator defined in Eq.~\eqref{chargeSU3isoversor} is different from the electric charge operator defined using the weak isospin and hypercharge, i.e.,  $ \mathsf{Q}_\mathrm{e} ≐ \mathsf{T}^\textsc{w}_{\3} + \mathsf{Y}^{\textsc{w}}/2$.

\begin{figure} \centering
	\includegraphics[height=8cm]{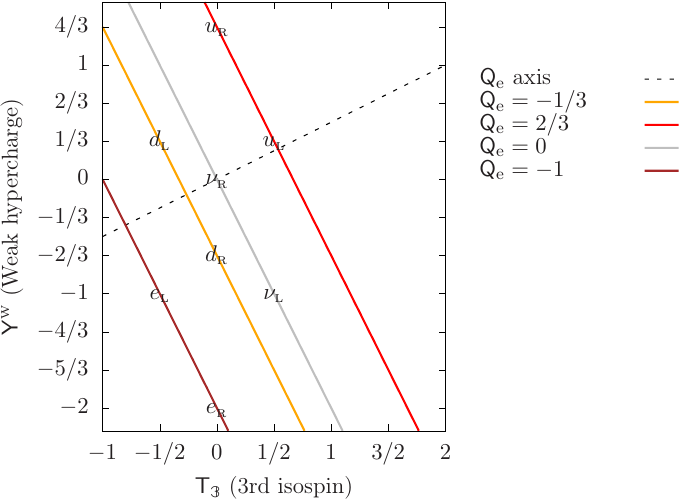} \hspace{0.5cm}
	\includegraphics[height=8cm]{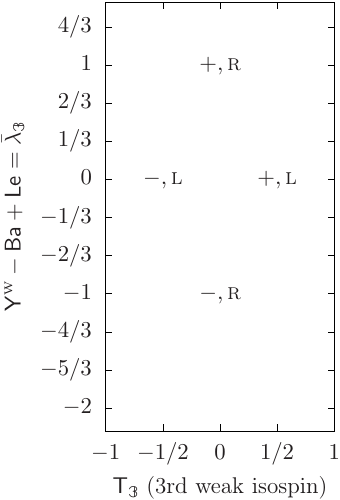}
	\caption{(Left) Eight eigenstates of the electroweak interaction in the plane of third weak isospin $\mathsf{T}^\textsc{w}_{\3}$ and weak hypercharge $\mathsf{Y}^\textsc{w}$. The dashed line is the direction of the electric charge axis $\mathsf{Q}_\mathrm{e}$. Particles on the same solid line (orthogonal to the dashed line) have equal electric charges. (Right) Isospin up ($+$) or down ($-$) and chirality ($\textsc{l,r}$) diagram associated to the particles on the left diagram, considering only their weak properties (independently of their leptonic or hadronic nature) \label{WEAKweightdiag}}
\end{figure}

We remark that the weak hypercharge and third isospin are the same for the lepton and quarks up to the contribution due to the baryon minus lepton number. As a consequence, the particles on the left of Figure \ref{WEAKweightdiag}, associated with eight different points, map onto only four points on the right of Figure \ref{WEAKweightdiag}: the left states at $(±1/2,0)$ and the right states $(0,±1)$, in which the plus (minus) is for up (resp. down) isospin states. By construction, the baryon minus lepton number operator differentiates the two types (lepton or quark) of isospin eigenstates.

Including electroweak interaction inside the wave equations (\ref{WEPmassop}--\ref{WEPbarmassop}) for massless Fermions yields
\begin{eqnarray}
	∇ \Pc_{(c)}'ℐ_{\3}\B_{3}+\bm b \mathsf{Y}'(\Pc_{(c)}')ℐ_{\3}+∑_{\ii}\bm w_{\ii} \mathsf{T}'_{\ii}(\Pc_{(c)}')ℐ_{\3} &=& 0%
	, \label{Weak-WE-Gauge1} \\
∇\bar\Pc_{(c)}'ℐ_{\3}\B_{3}+\bm b \bar{\mathsf{Y}}'(\bar \Pc_{(c)}')ℐ_{\3}+∑_{\ii}\bm w_{\ii} \bar{ \mathsf{T}}'_{\ii}(\bar{\Pc}_{(c)}')ℐ_{\3} &=&  0 
,\label{Weak-WE-Gauge2}
\end{eqnarray}
for quarks of color $c=r,b,g$ and colorless leptons when $c=w$.
Let's mention that the Lorentz covariance of Eqs.~(\ref{Weak-WE-Gauge1}--\ref{Weak-WE-Gauge2}) is granted by the fact that the Lorentz transformation of Eq.~\eqref{infinitesimalLORENTZ} (or its exponentiation) commutes with the four generators of the electroweak gauge group.

\subsection{Strong interaction for polychrome quadrispinors}
\label{QCDwithout}

Exchange of gluons is possible between Fermions appearing as the spinorial components of polychrome quadrispinors. Each component has a different color, but all particles within the quadrispinor have the same chirality and isospin. Therefore, using the whole isoversor SU(3) group is possible for massless fermions. We remark that the leptons are associated with the $ℐ_{\1}$ component in the fundamental representation and with the $ℐ_{\3}$ component in the conjugate representation. Those isoversors are in the kernel of all the generators of their corresponding representations. This explains why, as required, they do not interact with gluons in our framework. Then, the Baryon minus Lepton number operator is written 
\begin{eqnarray}
	(\mathsf{Ba-Le})(\Pc_{(\bbmu)}) &=& \frac{1}{3} K_{\1}(\Pc_{(\bbmu)}), \\
	(\overline{\mathsf{Ba}}-\overline{\mathsf{Le}})(\bar \Pc_{(\bbmu)}) &=& -\frac{1}{3} K_{\3}(\bar \Pc_{(\bbmu)}), 
\end{eqnarray}
when acting on polychrome quadrispinors, see Eqs.~(\ref{K11}--\ref{K33})

Including strong interaction inside the wave equations (\ref{QCD-WE1}--\ref{QCD-WE2}) for massless Fermions yields
\begin{eqnarray}
	∇ \Pc_{(\bbmu)}' ℐ_{\3}\B_{3}+∑_{\bb} \bm g_{\aa} \mathsf{T}'_{\aa}(\Pc_{(\bbmu)}')ℐ_{\3} &=& 0 %\bar{\mathsf{m}}'_{(\bbmu)} \bar {\Pc}_{(\bbmu)} 
	\label{QCD-WE-Gauge1}, \\
	∇ \bar{\Pc}_{(\bbmu)}' ℐ_{\3}\B_{3}+∑_{\bb} \bm g_{\aa} \bar{\mathsf{T}}'_{\aa}(\bar\Pc_{(\bbmu)}')ℐ_{\3} &=& 0% \mathsf{m}'_{(\bbmu)} \Pc_{(\bbmu)}
	.\label{QCD-WE-Gauge2}
\end{eqnarray}
Eqs.~(\ref{Weak-WE-Gauge1}--\ref{Weak-WE-Gauge2}) and (\ref{QCD-WE-Gauge1}--\ref{QCD-WE-Gauge2}) for massless fermions exhibit a global U(1) and a local gauge symmetry based on isoversor SU(3), as we show in the next Section. Finally, let's mention that the Lorentz covariance of Eqs.~(\ref{QCD-WE-Gauge1}--\ref{QCD-WE-Gauge2}) is granted by the fact that the Lorentz operators of Eqs.~(\ref{LorentzLeft}--\ref{LorentzRight}) commute with the generators of isoversor SU(3). Given that the polychrome quadrispinors have homogeneous chirality, one must not use the Lorentz transformation of Eq.~\eqref{LORENTZstd}, but the appropriate one among Eqs.~(\ref{LorentzLeft}--\ref{LorentzRight}) according to the quadrispinor chirality (extrinsic in this case).

\subsection{Global and local gauge invariance}

\paragraph{Invariance under constant phase change} 
The wave Eqs.~(\ref{WEPmassop}--\ref{WEPbarmassop}) are invariant under the overall phase transformation
\begin{eqnarray}
	\Pc'_{(z)}&\rightarrow& e^{θℐ_{\2}} \Pc'_{(z)}, \label{global-phase1} \\
	\bar \Pc'_{(z)}&\rightarrow& e^{-θℐ_{\2}} \bar \Pc'_{(z)}, \label{global-phase2}
\end{eqnarray}
with $z=w,r,g,b,\0,\1,\2,\3$, as one can check directly using 
\begin{eqnarray}
	e^{±θℐ_{\2}}∇ &=& ∇ e^{∓θℐ_{\2}} ,
\end{eqnarray}
with $θ$ a constant. Let's remark that we arbitrarily chose a left transformation, although a right transformation is also possible. As usual, using a vector potential as in Eq.~\eqref{wave-eq-abelian-gauge} allows transforming this global symmetry into a local one. The wave Eqs.~(\ref{QCD-WE1}--\ref{QCD-WE2}) are also invariant under the overall phase transformation of Eqs.~(\ref{global-phase1}--\ref{global-phase2}) upon the following substitution of mass operators
\begin{eqnarray}
	\mathsf{m}'_{(\bbmu)}(\dots) &\rightarrow&e^{-θℐ_{\2}}\mathsf{m}'_{(\bbmu)}(e^{θℐ_{\2}} \dots ), \\
	\bar{\mathsf{m}}'_{(\bbmu)}(\dots) &\rightarrow& e^{θℐ_{\2}} \bar {\mathsf{m}}'_{(\bbmu)}(e^{-θℐ_{\2}} \dots ).
\end{eqnarray}

\paragraph{Local gauge transformation}

We introduce the infinitesimal transformations 
\begin{equation}
	\mathsf{O}_{±} = \mathsf{Id} ± ∑_{\bb} θ_{\bb} λ'_{\bb}, \label{UnitaryOp}
\end{equation}
for $θ_{\bb} \ll 1$.
The vector potentiel $\bm a_{\bb}$ is associated to the operator $λ'_{\bb}$ of the gauge group to produce a gauge potential operator
\begin{equation}
	\mathsf{A}_{\bb} = \bm a _{\bb} λ'_{\bb} = \sum_{μ=0}^{3}  a _{\bb}^{μ} \e_{μ} λ'_{\bb} , \qquad \text{with } a _{\bb}^{μ} ≐ \bm a_{\bb} ⋅ \e^{μ}.
\end{equation}
This vector potential $\bm a_{\bb}$ generically denotes one of the (boson) spacetime vectors $\bm b, \bm w_{\ii}$ or $\bm g_{\aa}$ defined in the previous Section.
We define the gauge transformation by the corresponding changes on the quadrispinor, the derivative operator (standard representation), the vector potential, and its associated operator
\begin{eqnarray}
	\Mc' & \longrightarrow & \check \Mc = \mathsf{O}_{+} (\Mc'), \label{gauge-trans-1}\\
	∇ (\quad)ℐ_{\3} & \longrightarrow & \mathsf{O}_{+} [\dot {∇} \mathsf{O}_{-} (\dot \quad)ℐ_{\3} ],\\
	\bm a_{\bb} &\longrightarrow& \check{\bm a}_{\bb} = \bm a_{\bb} + g^{-1} ∇ θ_{\bb}, \\
	\mathsf{A}_{\bb} &\longrightarrow&  \check{\mathsf{A}}_{\bb}  =  \mathsf{O}_{+} \circ \left(\check{\bm a}_{\bb} λ'_{\bb} \circ \mathsf{O}_{-}(\dots)ℐ_{\3} \right), \label{gauge-trans-4}
\end{eqnarray}
respectively. The two overdots indicate that the gradient shall apply to the term inside the parenthesis only.
We remark that the above transformation rules lead to
\begin{equation}
	∇ \Mc' ℐ_{\3} \longrightarrow \mathsf{O}_{+} \left\{\dot {∇} \mathsf{O}_{-} \dot{ \overbracket{\mathsf{O}_{+} (\Mc')}}ℐ_{\3} \right\} = \mathsf{O}_{+} \left[∇ \Mc'ℐ_{\3} + ∑_{\bb}(∇ θ_{\bb}) λ'_{\bb}(\Mc')ℐ_{\3} \right] , \label{gradTERM}
\end{equation}
since
\begin{eqnarray}
	\mathsf{O}_{+} \left\{\dot {∇} \mathsf{O}_{-} \dot{ \overbracket{\mathsf{O}_{+} (\Mc')}} ℐ_{\3} \right\} &=& \mathsf{O}_{+} \left\{ \dot {∇} \mathsf{O}_{-} \left[ \dot{\Mc'} + ∑_{\bb}\dot{θ_{\bb}} λ'_{\bb}(\Mc') + ∑_{\bb}θ_{\bb}  λ'_{\bb}(\dot \Mc') \right] ℐ_{\3}\right\},\\
	&=& \mathsf{O}_{+} \left\{\dot {∇} \mathsf{O}_{-} \left[ \mathsf{O}_{+} (\dot{\Mc'}) + ∑_{\bb}\dot{θ_{\bb}} λ'_{\bb}(\Mc') \right]ℐ_{\3}\right\}, \\
	&=& \mathsf{O}_{+} \left[ ∇ \Mc' ℐ_{\3} + ∑_{\bb}(∇ θ_{\bb}) λ'_{\bb}(\Mc')ℐ_{\3} \right] ,
\end{eqnarray}
where the last equality holds at first order in $θ_{\bb}$. Indeed, one has $ \mathsf{O}_{-} \circ  \mathsf{O}_{+} =  \mathsf{Id}$ at first order and $ \mathsf{O}_{-}\simeq \mathsf{Id} $ when applied on a first order term. Similarly, we find
\begin{eqnarray}
	\mathsf{A}_{\bb} (\Mc') \longrightarrow  \check{\mathsf{A}}_{\bb}(\check{\Mc'}) &=&  \mathsf{O}_{+} \circ \left[ (\bm a_{\bb} + g^{-1}∇ θ_{\bb}) λ'_{\bb}(\Mc') ℐ_{\3} \right], \\
	&=& \mathsf{O}_{+} \left[\mathsf{A}_{\bb} (\Mc')ℐ_{\3} + g^{-1}(∇ θ_{\bb}) λ'_{\bb}(\Mc')ℐ_{\3} \right] , \label{gaugefieldTERM}
\end{eqnarray}
such that
\begin{eqnarray}
	∇ \Mc' ℐ_{\3} - g ∑_{\bb} \mathsf{A}_{\bb}(\Mc')ℐ_{\3} 
	&\longrightarrow &  \mathsf{O}_{+} [∇ \Mc' ℐ_{\3} - g ∑_{\bb}\mathsf{A}_{\bb}(\Mc') ℐ_{\3}],
\end{eqnarray}
from Eqs.~\eqref{gradTERM} and \eqref{gaugefieldTERM}. From the above calculation, we can conclude that the wave equations are invariant under a (non-abelian) gauge transformation when the fermion fields are massless. More generally, we have seen that the wave equation with materiality inhomogenous quadrispinors involves a real mass term. Such a simple term commutes with any generators of a gauge group and therefore the wave equation displays gauge invariance even for a positive mass. In this case, those operators mix the generator of the fundamental and conjugate representations, using projectors to apply them on spinorial components with appropriate materiality, see Eqs.~(\ref{particle1}--\ref{antiparticle2}). Then, the gauge transformed wave equation write
\begin{equation}
	\mathsf{O}_{+} \left(∇ \Mc' ℐ_{\3} \B_{3} - g ∑_{\bb} \mathsf{A}_{\bb}(\Mc') ℐ_{\3} \right) = m \mathsf{O}_{+}(\Mc').
\end{equation}
Here, as aforementioned, the gauge transformation can be seen as a particular and local change of representation (of the gamma operators) that requires a gauge vector potential and a symmetry group. The fact that the electroweak interaction is limited to the continuous group U(1)$ \otimes$SU(2) comes solely from the fact that it is meaningless to mix, through a gauge transformation, fields of different chirality. The rearrangement of fermions into polychrome quadrispinors then allows us to use the full isoversor SU(3) group. 
Let's remark that processes that simultaneously involve strong and electroweak bosons are possible in principle: one must use the projection operators of Section \ref{ProjIsoversor} to define polychrome quadrispinors from the four di-doublet quadrispinors (or vice-versa if working with di-doublet quadrispinors).

\section{Conclusion}

Additional dimensions, relying on new internal or external degrees of freedom, are commonly explored in modern physics as attempts to explain not yet understood physical phenomena. In this work, we took the opposing view of additional dimensions: We focused on spacetime only. This spacetime approach was a fruitful direction to question the origin of the different particles from symmetry arguments. In this approach, exactly four particles (e.g., $ν_\textsc{l},\overline{\nu_\textsc{l}},\overline{\nu_\textsc{r}},ν_\textsc{r}$) of opposite chirality and materiality must have the same mass: this comes from the fact that the most general elements in the algebra are quadrispinors that have given symmetry between matter and antimatter in the rest frame (see appendix \ref{derivWaveEq}). 
On the same idea, the symmetry between matter--antimatter of the theory suggests the existence of two different types of quadrispinors (the so-called matter first and matter last) which lead naturally to weak isospin doublets when separating matter and antimatter in two different quadrispinors. 
Likewise, disposing of a three-dimensional unitary group generating a symmetry inside the geometric algebra of spacetime also reverses the traditional way of thinking. Usually, this symmetry is assumed to explain the observation of three colored quarks. On the contrary, given that the symmetry group exists in spacetime, colored particles must exist, as they are made possible by this symmetry inside the geometric algebra of spacetime. Interestingly, the fundamental (and conjugate) isoversor and bivector representations of the special unitary groups in the geometric algebra of spacetime display a null weight in the plane of the eigenvalues of Cartan generators. This produces effectively three-dimensional representations, although the operators apply to four real-component objects, i.e., quadrispinors or space spinors. The null weight corresponds to a lepton in the case of strong interaction, and to a right neutrino in the case of electroweak interaction, identifying in this framework a common mechanism preventing leptons from interacting strongly and right neutrinos weakly. With this viewpoint that we adopted here, leptons and quarks differ by the way they are arranged into quadrispinors, changing their properties with respect to gauge transformations. Alternatively, the striking entanglement between the special unitary groups of dimension two and three in geometric algebra, as exhibited in Figures \ref{SU2bivectorANDisoversor}--\ref{4*44b}, lets us imagine a better understanding of the lepton and quark similarities and differences. 
In any case, studying the plethora of particle fields and their transformations from the perspective of geometric algebra of spacetime significantly constrains the possibilities while leaving room for many variations and hopefully improvements.   

Considerable work must be done to determine if an approach based on geometric algebra can model elementary particles at the level of achievement of the Standard Model of particle physics and its extensions. Even so, the corresponding approach should not only facilitate the fundamental understanding of the theory (as it does, for instance, by removing extrinsic properties such as chirality), but should also improve the methods and techniques for predicting scattering amplitudes. %\cite{Lewis2001vol40} 
At a more basic level, we will investigate in future works the description of Boson fields directly within the geometric algebra of spacetime. For fermions, the understanding of generation mixing all inside the geometric algebra of spacetime is challenging: We did not manage to de-correlate the dynamical spin-momentum degrees of freedom from the generation degrees of freedom in our spacetime approach. However, several facts indicate that encompassing flavor in spacetime might be achieved: First, there is rotational freedom in the choice of the unit space bivector $\B_{3}$ replacing the complex number $\imath$. Second, and in the same line, we did not use the complete continuous bivector SU(3) group, as spin physics only requires the SU(2) subgroup. Third, mass operators naturally emerge in our framework: the misalignment of the weak and mass eigenbasis is quite natural if the bivector playing the role of a complex number in the isoversor SU(3) group is not $\B_{3}$. In this direction, the Koide formula involving the three masses of a particle replica in each generation may play a role in the understanding of flavor within spacetime algebra.
More generally, the mechanisms of fermion mass generation from Yukawa-type interactions with the Higgs field should be translated to the geometric algebra of spacetime, together with the symmetry breaking produced by the Higgs mechanism and leading to the massive electroweak bosons.
The fact that gauge transformations in our framework produce local change of representation for the Dirac operator (beyond chiral and standard representation) is also inspiring. Local curvature of spacetime through gravitational interaction will also modify the set of basis vectors locally, and hence versors, enabling the gravitational generation of matter. Finally, we explored in detail the geometric algebra of spacetime $\mathcal{G}(1,3)$ due to its relation to the algebra of Dirac matrices. The algebra with the opposite signature $\mathcal{G}(3,1)$ is equally legitimate. Particle fields belonging to this spacetime will display a different physics, as the underlying group of basis versors is the central product of two quaternion groups. Those particle fields, living in another algebra, may not interact with the fields of $\mathcal{G}(1,3)$, except probably through gravitational interaction. Given their potentially different nature, these fields are credible dark matter candidates.

%\newpage

\section*{Acknowledgment}

I thank Méril Reboud for insightful discussion on particle physics, within and beyond the Standard Model. I'd like to thank also Salvador Rosauro and Panagiotis Marinellis.

\appendix

\section{Fundamental representation of SU(3) in matrix algebra}
\label{GuellMann}

For the sake of completeness, I provide below the Gell-Mann generators of the special unitary group of dimension three in its fundamental representation:
\begin{equation}
\begin{aligned}
	T_{0}=\frac{\lambda_{8}}{2}=\frac{1}{2\sqrt{3}}&\begin{bmatrix}
		 1&0&0\\
		 0&1&0\\
		 0&0&-2\\
	 \end{bmatrix},   &
	 \textcolor{gray}{U_{0}=\frac{1}{2\sqrt{3}}}&\textcolor{gray}{\begin{bmatrix}
		 -2&0&0\\
		 0&1&0\\
		 0&0&1\\
	 \end{bmatrix}},&
	 \textcolor{gray}{V_{0}=\frac{1}{2\sqrt{3}}}&\textcolor{gray}{\begin{bmatrix}
		 1&0&0\\
		 0&-2&0\\
		 0&0&1\\
	 \end{bmatrix}},\\
	 T_{1}=\frac{\lambda_{1}}{2}=\frac{1}{2}&\begin{bmatrix}
		 0&1&0\\
		 1&0&0\\
		 0&0&0\\
	 \end{bmatrix},   &
	 U_{1}=\frac{\lambda_{6}}{2}=\frac{1}{2}&\begin{bmatrix}
		 0&0&0\\
		 0&0&1\\
		 0&1&0\\
	 \end{bmatrix},&
	 V_{1}=\frac{\lambda_{4}}{2}=\frac{1}{2}&\begin{bmatrix}
		 0&0&1\\
		 0&0&0\\
		 1&0&0\\
	 \end{bmatrix},\\
	 T_{2}=\frac{\lambda_{2}}{2}=\frac{1}{2}&\begin{bmatrix}
		 0&-i&0\\
		 i&0&0\\
		 0&0&0\\
	 \end{bmatrix},  &
	 U_{2}=\frac{\lambda_{7}}{2}=\frac{1}{2}&\begin{bmatrix}
		 0&0&0\\
		 0&0&-i\\
		 0&i&0\\
	 \end{bmatrix}, &	 
	 -V_{2}=\frac{\lambda_{5}}{2}=\frac{1}{2}&\begin{bmatrix}
		 0&0&-i\\
		 0&0&0\\
		 i&0&0\\
	 \end{bmatrix}, \\
	 T_{3}=\frac{\lambda_{3}}{2}=\frac{1}{2}&\begin{bmatrix}
		 1&0&0\\
		 0&-1&0\\
		 0&0&0\\
	 \end{bmatrix} ,& 
	 \textcolor{gray}{U_{3}=\frac{1}{2}}& \textcolor{gray}{\begin{bmatrix}
		 0&0&0\\
		 0&1&0\\
		 0&0&-1\\
	 \end{bmatrix}}, &
	 -\textcolor{gray}{V_{3}=\frac{1}{2}}&\textcolor{gray}{\begin{bmatrix}
		 -1&0&0\\
		 0&0&0\\
		 0&0&1\\
	 \end{bmatrix}}.
\end{aligned}
\end{equation}
The matrices in gray are linearly dependent on the matrices written in black. The complex conjugate representation is $\bar {λ}_{a} = -λ_{a}^{*}$. 
The generator of SO(3) are 
\begin{align}
	-2iU_{2} & = K_{1} = \begin{bmatrix}
		 0&0&0\\
		 0&0&-1\\
		 0&1&0\\
	 \end{bmatrix} &
	-2iV_{2} & = K_{2} = \begin{bmatrix}
		 0&0&1\\
		 0&0&0\\
		 -1&0&0&\\
	 \end{bmatrix} &
	-2iT_{2} & = K_{3} = \begin{bmatrix}
		 0&-1&0\\
		 1&0&0\\
		 0&0&0\\
	 \end{bmatrix}
\end{align}
We remark that, compared to the Gell-Mann matrices, the choice of the generators $T_{μ}$, $U_{μ}$, $V_{μ}$ has a sign convention that is more consistent with the three-dimensional rotation group since
$-i λ_{7} = K_{1}$, $i λ_{5} = K_{2}$ and $-i λ_{2} = K_{3}$. Then, a rotation of angle $θ$ with respect to unit vector $\bm k = (k_{1},k_{2},k_{3})$ is achieved with matrix
\begin{equation}
	\bm R = I + \sin (θ ) \bm k ⋅ \bm K +(1-\cos θ) (\bm k ⋅ \bm K)^{2}
\end{equation}
when denoting $\bm K= (K_{1},K_{2},K_{3})$. This is the matrix form of Rodrigues' rotation formula.

\section{Pure boost in standard and chiral representations}
\label{pureboost}

Boosts are special Lorentz transformations without rotation. They relate two reference frames, one translating at constant velocity with respect to the other one. One can consider, for instance, a particle of mass $m$ in its rest frame, and an observer's frame where the particle has a momentum bivector $\bm P$ (describing the plane orthogonal to the space vector of linear momentum $\vec p $). The rotor boosts $\hat \Bc'(\bm P)$ in standard representation is defined by
\begin{eqnarray}
	\hat \Bc'_{\bm P}&=&\cosh \eta/2 \left( 1 + \frac{ ℐ_{\2}\bm P}{|\bm P|} \tanh\eta/2 \right), 
\end{eqnarray}
allows us to transform the fields from one frame to another.
Using standard hyperbolic trigonometry and the definition of kinetic energy $\varepsilon = \gamma m c^{2}$ in special relativity in terms of the Lorentz factor:
\begin{equation}
	\gamma = \frac{\varepsilon}{mc^{2}}  =\cosh \eta = 2 \cosh^{2}(η/2)-1 = 1+ 2\sinh^{2}(η/2),\label{cosh}
\end{equation}
leads to
\begin{eqnarray}
	\cosh \eta/2 &=& \sqrt{\frac{\varepsilon+mc^{2}}{2mc^{2}}}, \\
	\sinh \eta/2 &=& \sqrt{\frac{\varepsilon-mc^{2}}{2mc^{2}}}, \label{sinhhalfeta} \\
	\tanh \eta/2 &=& \sqrt{\frac{\varepsilon-mc^{2}}{\varepsilon+mc^{2}}} = \frac{\sqrt{\varepsilon^{2}-m^{2}c^{4}}}{\varepsilon+mc^{2}} = \frac{|\bm P|c}{\varepsilon+mc^{2}},
\end{eqnarray}
where we use $\varepsilon^{2}=m^{2}c^{4}+|\bm P|^{2}c^{2}$. Then, the boost in standard representation is written in terms of the particle energy, mass, and momentum (bivector) only
\begin{equation}
	\hat \Bc'_{\bm P} (\dots)=\sqrt{\frac{\varepsilon+mc^{2}}{2mc^{2}}}\left( 1 +  \frac{ ℐ_{\2}\bm P c}{\varepsilon+mc^{2}} \right) = \frac{\varepsilon + mc^{2} +  ℐ_{\2}\bm Pc}{\sqrt{2mc^{2}(\varepsilon+mc^{2})}} (\dots). \label{boostrotors2}
\end{equation} 
In chiral representation, the boost writes
\begin{equation}
	\hat \Bc_{\bm P} (\dots)=\frac{(\varepsilon + mc^{2})(\dots) +  ℐ_{\2}\bm Pcℐ_{\3}(\dots)ℐ_{\3}\B_{3}}{\sqrt{2mc^{2}(\varepsilon+mc^{2})}}. 
\end{equation}
It is useful to note for later that
\begin{equation}
	\frac{\varepsilon}{mc^{2}}  =\cosh \eta =  \sqrt{1-\sinh^{2} η}  \qquad \Rightarrow \qquad \sinh^{2} η = \frac{ε^{2}-m^{2}c^{4}}{m^{2}c^{4}} = \frac{|\bm P |^{2}c^{2}}{m^{2}c^{4}},
\end{equation}
meaning that 
\begin{equation}
	\sinh η = \frac{|\bm P|}{mc} \label{sinh}
\end{equation}
assuming positive rapidity.

\section{Recovering Weyl--Pauli wave equations}
\label{WeylPauliWE}

We consider the even part of the wave equation in chiral representation only and then focus on $\Mc_{\eee} = \Mc_{0}+\Mc_{\2}ℐ_{\2}\B_{3}$, in which $\Mc_{0}$ can be seen as the first spinorial component and $\Mc_{\2}$ as the second. The $γ_{0}$ operator acting on $\Mc_{\eee}$ gives
\begin{eqnarray}
	γ^{0}(\Mc_{\eee}) &=& ℐ_{\2}(\Mc_{\0}+\Mc_{\2}ℐ_{\2}\B_{3})\B_{3}\\
	 &=& \Mc_{\2}+\Mc_{\0}ℐ_{\2}\B_{3}.
\end{eqnarray}
The $γ^{j}$ operator acting on $\Mc_{\eee}$ gives
\begin{eqnarray}
	γ^{j}(\Mc_{\eee}) &=& -ℐ_{\1}\B_{j} (\Mc_{\0}+\Mc_{\2}ℐ_{\2}\B_{3})ℐ_{\3} \\
	 &=& \underbrace{\B_{j}\Mc_{\2}\B_{3}}_{σ^{j} \text{ action} }-\underbrace{\B_{j}\Mc_{\0}\B_{3}}_{-σ^{j} \text{ action}} ℐ_{\2}\B_{3}. \nonumber \\
\end{eqnarray}
Hence the even part of the wave equation in chiral representation is
\begin{eqnarray}
	γ^{0}(∂_{0}\Mc_{\eee}) \B_{3} + γ^{j}(∂_{j}\Mc_{\eee})\B_{3} &=& m \Mc_{\eee}, \\
	∂_{0}\Mc_{\2}\B_{3}+∂_{0}(\Mc_{\0}ℐ_{\2}\B_{3})\B_{3} + (\B_{j}∂_{j}\Mc_{\2}\B_{3})\B_{3}-(\B_{j}∂_{j}\Mc_{\0}\B_{3})\B_{3} ℐ_{\2}\B_{3}.&=& m(\Mc_{0}+\Mc_{\2}ℐ_{\2}\B_{3}), \nonumber \\
\end{eqnarray}
from which we can extract the space spinor parts in the product with $ℐ_{\2}\B_{3}$ and the one in the product with $1$, leading to the Weyl and Pauli wave equations:
\begin{eqnarray}
	∂_{0}\Mc_{\2}\B_{3} +(\B_{j}∂_{j}\Mc_{\2}\B_{3})\B_{\3} &=& m \Mc_{\0}, \\
	∂_{0}\Mc_{\0}\B_{3} - (\B_{j}∂_{j}\Mc_{\0}\B_{3})\B_{\3} &=& m \Mc_{\2},
\end{eqnarray}
as obtained by direct translation from the matrix framework at the end of Section~\ref{WeylSpinors}.

\section{Derivation of the wave equation}\label{derivWaveEq}

From the fact that the left and right spinors are identical up to a sign in the center-of-mass frame, one can boost this spinor identity to any other frame, as explained in Ref.~\cite{Steane2013vol}. The first quantification of this identity leads (as expected) to the Dirac wave equation. We follow this route below in geometric algebra and for quadrispinors in the standard representation.

In their center of mass frame, a pair of particle and antiparticle of mass $m$ have space spinors $\Mc_{\0}(0)$ and $\Mc_{\2}(0)$ respectively (or $\Mc_{\3}(0)$ and $\Mc_{\1}(0)$ equivalently). In this frame, the space spinors are equal ($χ=+1$) or opposite ($χ=-1$) \cite{Steane2013vol}
\begin{eqnarray}
	\Mc_{\0}(0) &=& χ\Mc_{\2}(0), \\
	\Mc_{\3}(0) &=& χ\Mc_{\1}(0). 
\end{eqnarray}
This can be written with a quadrispinor $\Mc'(0)$ in standard representation as
\begin{equation}
	\Mc'(0) = χ ℐ_{\3}\Mc'(0)ℐ_{\3} 
\end{equation}
since for 
\begin{equation}
	\Mc'(0)=\Mc_{\0}(0)3_{-}+\Mc_{\2}(0)3_{+}-\Mc_{\3}(0)3_{-}ℐ_{\1}+\Mc_{\1}(0)3_{+}ℐ_{\1},
\end{equation}
we have
\begin{eqnarray}
	ℐ_{\3}\Mc'(0)ℐ_{\3} &=& ℐ_{\3}[\Mc_{\0}(0)3_{-}+\Mc_{\2}(0)3_{+}-\Mc_{\3}(0)3_{-}ℐ_{\1}+\Mc_{\1}(0)3_{+}ℐ_{\1}]ℐ_{\3}, \\
	&=& \Mc_{\0}(0)3_{+}+\Mc_{\2}(0)3_{-}+\Mc_{\3}(0)3_{+}ℐ_{\1}-\Mc_{\1}(0)3_{-}ℐ_{\1}.
\end{eqnarray}
We defined by $\Mc(\bm P)$ the quadrispinor obtaing from a boost of momentum space bivector $\bm P$
\begin{eqnarray}
	\Mc'(\bm P) &≐& \hat \Bc'_{\bm P} \Mc'(0) = \frac{\varepsilon + mc^{2} +  ℐ_{\2}\bm Pc}{\sqrt{2mc^{2}(\varepsilon+mc^{2})}}\Mc'(0),\\
	ℐ_{\3}\Mc'(\bm P)ℐ_{\3} &≐& %\hat \Bc'_{-\bm P} ℐ_{\3}\Mc'(0)ℐ_{\3} =
	\frac{\varepsilon + mc^{2} +  ℐ_{\2}\bm Pc}{\sqrt{2mc^{2}(\varepsilon+mc^{2})}}ℐ_{\3}\Mc'(0)ℐ_{\3}
\end{eqnarray}
in which the second equation is an immediate consequence of the first equation. Then
we have on one side
\begin{eqnarray}
(ε+mc^{2}-ℐ_{\2}\bm P c) \Mc'(\bm P ) &=& (ε+mc^{2}-ℐ_{\2}\bm P c) \frac{\varepsilon + mc^{2} +  ℐ_{\2}\bm Pc}{\sqrt{2mc^{2}(\varepsilon+mc^{2})}}\Mc'(0) \\ &=& \frac{(\varepsilon + mc^{2} )^{2} -(  ℐ_{\2}\bm Pc)^{2}}{\sqrt{2mc^{2}(\varepsilon+mc^{2})}} \Mc'(0)
\end{eqnarray}
and on the other side
\begin{eqnarray}
	(ε+mc^{2}+ℐ_{\2}\bm P c) ℐ_{\3}\Mc'(\bm P )ℐ_{\3} &=& (ε+mc^{2}+ℐ_{\2}\bm P c) \frac{\varepsilon + mc^{2} -  ℐ_{\2}\bm Pc}{\sqrt{2mc^{2}(\varepsilon+mc^{2})}}ℐ_{\3}\Mc'(0)ℐ_{\3} \\ &=& \frac{(\varepsilon + mc^{2} )^{2} -(  ℐ_{\2}\bm Pc)^{2}}{\sqrt{2mc^{2}(\varepsilon+mc^{2})}} ℐ_{\3}\Mc'(0)ℐ_{\3}
\end{eqnarray}
We conclude that
\begin{eqnarray}
	(ε+mc^{2}-ℐ_{\2}\bm P c) \Mc'(\bm P) &=& χ (ε+mc^{2}+ℐ_{\2}\bm P c)ℐ_{\3} \Mc'(\bm P) ℐ_{\3}\label{WEprelude1} 
\end{eqnarray}
Similarly, on one side
\begin{eqnarray}
(ε-mc^{2}-ℐ_{\2}\bm P c) \Mc'(\bm P ) &=& (ε-mc^{2}-ℐ_{\2}\bm P c) \frac{\varepsilon + mc^{2} +  ℐ_{\2}\bm Pc}{\sqrt{2mc^{2}(\varepsilon+mc^{2})}}\Mc'(0) \\ &=& \frac{\varepsilon^{2} -(  mc^{2} +  ℐ_{\2}\bm Pc)^{2}}{\sqrt{2mc^{2}(\varepsilon+mc^{2})}} \Mc'(0)
\end{eqnarray}
and on the other side
\begin{eqnarray}
	(-ε+mc^{2}-ℐ_{\2}\bm P c) ℐ_{\3}\Mc'(\bm P )ℐ_{\3} &=& (-ε+mc^{2}-ℐ_{\2}\bm P c) \frac{\varepsilon + mc^{2} -  ℐ_{\2}\bm Pc}{\sqrt{2mc^{2}(\varepsilon+mc^{2})}}ℐ_{\3}\Mc'(0)ℐ_{\3} \\ &=& \frac{-\varepsilon^{2} +(mc^{2}-  ℐ_{\2}\bm Pc)^{2}}{\sqrt{2mc^{2}(\varepsilon+mc^{2})}} ℐ_{\3}\Mc'(0)ℐ_{\3}.
\end{eqnarray}
Since
\begin{equation}
	-\varepsilon^{2} +(mc^{2}-  ℐ_{\2}\bm Pc)^{2} = -2mc^{3}ℐ_{\2}\bm P = \varepsilon^{2} -(  mc^{2} +  ℐ_{\2}\bm Pc)^{2},
\end{equation}
we conclude that
\begin{eqnarray}
	(ε-mc^{2}-ℐ_{\2}\bm P c) \Mc'(\bm P) &=& χ (-ε+mc^{2}-ℐ_{\2}\bm P c)ℐ_{\3} \Mc'(\bm P) ℐ_{\3} . \label{WEprelude2}
\end{eqnarray}
The sum and difference of Eqs.~(\ref{WEprelude1}--\ref{WEprelude2}) leads to two versions of the same equation
\begin{eqnarray}
	(ε-ℐ_{\2}\bm P c) \Mc'(\bm P) &=& χ mc^{2}ℐ_{\3} \Mc'(\bm P) ℐ_{\3} \\
	mc^{2} \Mc'(\bm P) &=& χ (ε+ℐ_{\2}\bm P c)ℐ_{\3} \Mc'(\bm P) ℐ_{\3} 
\end{eqnarray}
that, using $\bm pc =ℐ_{\3}ε-ℐ_{\1}\bm P c $, we write as
\begin{equation}
		 χ\bm p \Mc'(\bm P) ℐ_{\3} = mc \Mc'(\bm P) \label{momentum-constraint}
\end{equation}
In quantum mechanics based on complex algebra, the first quantization step replaces the linear momentum according to $\bm p \leftrightarrow -\imath ℏ ∇ $. The equivalent in geometric algebra is $\bm p \leftrightarrow - ℏ∇(\dots)\B_{3} $ as the complex number acting on space spinor is the bivector $\B_{3}$ applied on the right. This yields for $χ=-1$
\begin{equation}
		 ℏ ∇ \Mc' ℐ_{\3}\B_{3} = mc \Mc' \label{FourierWE}
\end{equation}
corresponding exactly to wave Eq.~\eqref{waveequationMassHomogenous}. The sign $χ=+1$ is equally possible and just changes the way linear momentum and spinors are related. 

We remark that $χ$ could depend on the grade parity of the subparts of the multivector $\Mc$. For instance, the time reversal defined in Section \ref{SecInvolutions} transforms each component of a quadrispinor in standard representation as
\begin{eqnarray}
	(\Mc_{\0}3_{-})^\mathsf{t} &=& +ℐ_{\3}(\Mc_{\0}3_{-})ℐ_{\3} =\Mc_{\0}3_{+},  \\
	(\Mc_{\2}3_{+})^\mathsf{t} &=& +ℐ_{\3}(\Mc_{\2}3_{+})ℐ_{\3} = \Mc_{\2}3_{-},\\
	(-\Mc_{\3}3_{-}ℐ_{\1})^\mathsf{t} &=& -ℐ_{\3}(-\Mc_{\3}3_{-}ℐ_{\1})ℐ_{\3} = -\Mc_{\3}3_{+}ℐ_{\1}, \\
	(\Mc_{\1}3_{+}ℐ_{\1})^\mathsf{t} &=&- ℐ_{\3}(\Mc_{\1}3_{+}ℐ_{\1})ℐ_{\3} = \Mc_{\1}3_{-}ℐ_{\1}.
\end{eqnarray}
Then, a quadrispinor that is its own time reverse ($\Mc'=(\Mc')^\mathsf{t}$) has components verifying $\Mc_{\0} = \Mc_{\2}$ and $\Mc_{\1}=-\Mc_{\3}$, i.e., the sign $χ$ above is given by $χ(\ppp)$ with $\ppp=\eee,\ooo$ the grade parity of the basis element on which is placed the spinorial component. Therefore, in this case, the wave equation transforms into any frame (obtained by a Lorentz boost) the relation between the two space spinors in the even (or odd) subalgebra that imposes invariance under time reversal in the center of mass frame. \\

We remark that the wave equation for the quadrispinor describes two particle-antiparticle pairs. Particles in a pair have opposite linear momentum, but momenta of particles in different pairs shall not be related in general. This explains why the wave equation for free fermions is solved in the next appendix after even- and odd-grade projecting the wave equation.

\section{Plane wave solutions and spacetime vectors of spin-momentum}
\label{spin-momentum}

In this appendix, we use a plane wave Ansatz for the even and odd parts of a quadrispinor field to solve the wave equation in standard representation.
According to the phase $β=0$ or $π$, the plane wave is associated with a matter or an antimatter particle. The field with $β=0$ ($β=π$) is associated with the particle (antiparticle) state since the zeroth component of the spacetime momentum is positive (negative) in this case. A general solution describes a particle-antiparticle pair in a linear superposition with coefficients $ \cos β/2 $ and $ \sin β/2$ respectively \cite{Hestenes1992vol2, Lewis2001vol40}. Next, the solution in standard representation is used to determine the corresponding plane wave in chiral representation. We provide the method to directly obtain the spacetime vectors of spin and momentum that we recover in this representation. Then, the parameters in the plane wave solution are all connected to spin and linear momentum. Finally, we consider the inverse problem of finding the pair of space spinors associated with the spacetime vectors of spin and momentum.

\paragraph{Plane wave solution of the wave equation in standard representation} 
Let's consider a quadrispinor $\Mc'$ and the even ($\ppp=\eee$) and odd ($\ppp=\ooo$) grade projections of the wave equation in standard representation
\begin{eqnarray}
	ℏ∇\Mc'_{\ppp} ℐ_{\3}\B_{3} &=& mc \Mc'_{\ppp}. \label{WEappendix}
\end{eqnarray}
Let's write a plane wave Ansatz for the geometric bispinors (even and odd projection of $\Mc'$) as 
\begin{eqnarray}
	\Mc_{\eee}'(ρ,β,η,\hat{\bm K},\bm k) &=& \sqrt{ρ} e^{ℐ_{\2}β/2}e^{ℐ_{\2} \hat{\bm K} η/2 } e^{-\B_{3}(\bm x ⋅\bm k)} , \label{STDgeobispinorEVENup} \\
	\Mc_{\ooo}'(ρ,β,η,\hat{\bm K},\bm k) &=& \sqrt{ρ} e^{ℐ_{\2}β/2}e^{ℐ_{\2} \hat{\bm K} η/2 } e^{-\B_{3}(\bm x ⋅\bm k)} ℐ_{\3}\B_{3}, \label{STDgeobispinorODDup} 
 \end{eqnarray}
in which $ρ$ is a positive real number, $η$ a real number, $\hat{\bm K}$ a unit space bivector,  $\bm k$ and $\bm x$ spacetime vectors. We chose identical parameters in the Ansatz for the even and odd grade parts to run the calculation simultaneously, although these two parts are independent in general. Since $ℐ_{\3}\B_{3}(ℐ_{\3}\B_{3})^\mathsf{r}=1$, we have
\begin{eqnarray}
	\Mc'_{\ppp}\Mc_{\ppp}'\,\!^\mathsf{r}=ρe^{ℐ_{\2}β},
\end{eqnarray}
Then, the Ansatz in the wave equation gives
\begin{eqnarray}
	- ℏ\bm k\Mc'_{\ppp} ℐ_{\3}\B_{3}^{2} &=& mc \Mc'_{\ppp} \\
	\bm p\Mc'_{\ppp}\Mc_{\ppp}'\,\!^\mathsf{r}  &=& mc \Mc'_{\ppp}ℐ_{\3}\Mc_{\ppp}'\,\!^\mathsf{r}, \\
	\bm p ρe^{ℐ_{\2}β}  &=& mc \Mc'_{\ppp}ℐ_{\3}\Mc_{\ppp}'\,\!^\mathsf{r},  \label{momentum-planewave}
\end{eqnarray}
where we define the spacetime vector of linear momentum as usual by $\bm p = ℏ\bm k$. This reduces naively to
\begin{equation}
	\bm p(β,\ppp)   =  \Mc'_{\ppp}ℐ_{\3}\Mc_{\ppp}'\,\!^\mathsf{r}   mcρ^{-1}e^{-ℐ_{\2}β }, \label{generalmomentum}
 \end{equation} 
that is, in fact, the sum of a spacetime vector and trivector. 
Eq.~\eqref{momentum-planewave} holds for a spacetime vector of linear momentum for $β=0,π$ whatever the grade parity $\ppp=\eee,\ooo$. Among these four cases, there are two pairs of solutions with opposite energies (first component of the momentum spacetime vector): 
\begin{align}
	\bm p(0,\ppp) &= + mcρ^{-1} \Mc_{\ppp}'\e_{0}\Mc_{\ppp}'\,\!^\mathsf{r}, && (\text{Matter solution}) \\
	\bm p(π,\ppp) &= - mcρ^{-1} \Mc_{\ppp}'\e_{0}\Mc_{\ppp}'\,\!^\mathsf{r},  && (\text{Antimatter solution})
\end{align}
$\Mc_{\eee}'$ and $\Mc_{\ooo}'$ in Eqs.~(\ref{STDgeobispinorEVENup}--\ref{STDgeobispinorODDup}) differ by a right multiplication by $ℐ_{\3}\B_{3}$ only. However, we have  $(ℐ_{\3}\B_{3})\e_{0}(ℐ_{\-} \B_{3})^\mathsf{r}=\e_{0}$ and then the term $\Mc_{\ppp}'\e_{0}\Mc_{\ppp}'\,\!^\mathsf{r}$ contributes in the same way to the momentum in cases $\ppp=\eee,\ooo$.  
The momentum is related to matter or antimatter according to the sign of the energy  $\bm p(β,\ppp) ⋅ \e_{0}$. This sign is fixed since the (Lorentz) transformation associated with Eq.~\eqref{generalmomentum} is orthochronous (the sign of a timelike vector cannot change). 

To summarize, our plane wave Ansatz of Eqs.~(\ref{STDgeobispinorEVENup}-\ref{STDgeobispinorODDup}) writes in general
\begin{equation}
	\Mc'_{\ppp} ≐ \sqrt{ρ} e^{ℐ_{\2}β/2}e^{ℐ_{\2}\hat{\bm K} η/2} \Rcℐ_{\ppp} \quad \text{ with } \Rc≐ e^{- \B_{3} (\bm k ⋅ \bm x)}  \label{GeneralAnsatz},
\end{equation}
with $\ppp=\eee,\ooo$ the even or odd grade projections. I introduced $ℐ=1+ℐ_{\3}\B_{3}$ such that $ℐ_{\ppp}$ is either $ℐ_{\eee} = 1$ or $ℐ_{\ooo} = ℐ_{\3}\B_{3}$. Then, we can write in general the spacetime momentum and spin
\begin{eqnarray}
	\bm p(β,\ppp) &=& \Mc_{\ppp}' \e_{0} \Mc_{\ppp}'\,\!^\mathsf{r}     \frac{mc}{ρe^{ℐ_{\2}β}}  =   (ℐ_{\3} \cosh η  - ℐ_{\1} \hat{\bm K}  \sinh η ) \frac{mc}{e^{ℐ_{\2}β}}, \label{spacetimemomentum} \\
	\bm w(β,\ppp) &=& \Mc_{\ppp}' \e_{3} \Mc_{\ppp}'\,\!^\mathsf{r}    \frac{mc}{ρe^{ℐ_{\2}β}} = χ(\ppp)\left[ -(\B_{3} ⋅  \hat{\bm K}) ℐ_{\3} \sinh η   - \B_{3} ℐ_{\1} \cosh^{2}\frac{η}{2} +  \hat{\bm K}  \B_{3}  \hat{\bm K} ℐ_{\1} \sinh^{2} \frac{η}{2}\right] \frac{mc}{e^{ℐ_{\2}β}}, \nonumber \\ \label{spacetimespin}
\end{eqnarray}
where we have defined $χ(\eee)=+1$ and $χ(\ooo)=-1$. Indeed, $(ℐ_{\ppp})\e_{0}(ℐ_{\ppp})^\mathsf{r} = \e_{0}$, but $(ℐ_{\ppp})\e_{3}(ℐ_{\ppp})^\mathsf{r} = χ(\ppp)\e_{3}$. Hence, the spin spacetime vector has a different sign when computed with the even and odd parts of the quadrispinor $\Mc'$.
I used that $\B_{3} ≐ \Rc \B_{3} \Rc^\mathsf{r}$ given that $\Rc$ is a rotation in $\B_{3}$ plane. 
I interpret $ρ$ as the density amplitude, $β$ as a phase describing particle-antiparticle superpositions ($β=0$ for positive energy solution and $β=π$ for negative energy solutions), $\hat{\bm K}$ as a unit space bivector whose dual indicates the mixed plane of boost with rapidity $η$ and, finally, $\Rc $ as a space rotor providing the phase of the fermion's spin. 
We have by construction that $\bm p ⋅ \bm w = 0 $ since
\begin{equation}
	 (\Mc_{\ppp}' \e_{0} \Mc_{\ppp}'\,\!^\mathsf{r} \Mc_{\ppp}' \e_{3} \Mc_{\ppp}'\,\!^\mathsf{r})_{\vide}  = (\Mc_{\ppp}'\,\!^\mathsf{r} \Mc_{\ppp}' \e_{0} \Mc_{\ppp}'\,\!^\mathsf{r} \Mc_{\ppp}' \e_{3} )_{\vide} = ρ^{2}  \e_{0}⋅ \e_{3} = 0 \qquad \text{when } β=0,π.
\end{equation}
Similarly, we have $\bm p^{2} + \bm w^{2} = 0 $. These are two expected properties for the spacetime spin and momentum, justifying Eq.~\eqref{spacetimespin}.

\paragraph{Plane wave solution of the wave equation in the chiral representation}
Starting from the plane wave solution of Eq.~\eqref{GeneralAnsatz},
we look for the corresponding quadrispinor in chiral representation
\begin{eqnarray}
		\Mc &=& \Mc_{\0}+\Mc_{\2}\B_{3}ℐ_{\2} + (\Mc_{\3} + \Mc_{\1}ℐ_{\2}\B_{3}) ℐ_{\3}\B_{3}, \label{RappelChiralQuadrispinor}\\ 
		\Mc &=& \Mc_{\eee}+\Mc_{\ooo},  \qquad (\text{even and odd projection}) \label{Mcevenoddproj}
\end{eqnarray}
and more precisely for the space spinor components $\Mc_{\bbmu}$. To do so in parallel for the even and odd grade parts of a quadrispinor, we define
\begin{eqnarray}
	\lbb(\ppp) &=& \left\{ \begin{array}{l} \0  \quad \text{ for } \ppp = \eee \\ \3 \quad \text{ for } \ppp = \ooo \end{array} \right. \\
	\rr(\ppp) &=& \left\{ \begin{array}{l} \2  \quad \text{ for } \ppp = \eee \\ \1 \quad \text{ for } \ppp = \ooo \end{array} \right.
\end{eqnarray}
that allows to express Eqs.~(\ref{RappelChiralQuadrispinor}--\ref{Mcevenoddproj}) as
\begin{equation}
	\Mc_{\ppp} = (\Mc_{\lbb(\ppp)}+\Mc_{\rr(\ppp)}ℐ_{\2}\B_{3})ℐ_{\ppp}.
	\label{GeneralMcPPPchiral}
\end{equation}
We use the operator $\mathsf{U}_{-}$ on $\Mc'_{\ppp}$ (in standard representation) to switch to chiral representation such that
\begin{eqnarray}
	\Mc_{\ppp} &=& \mathsf{U}_{-} (\Mc_{\ppp}') = \frac{\Mc'_{\ppp}-ℐ_{\1} \Mc'_{\ppp}ℐ_{\3}\B_{3}}{\sqrt{2}}.
\end{eqnarray}
We notice that $\mathsf{U}_{-}$ preserve the grade parity and commutes with right multiplication by $ℐ_{\3}\B_{3}$. Let's use now Eq.~\eqref{GeneralAnsatz} to get
\begin{equation}
	\Mc_{\ppp} = \sqrt{\frac{ρ}{2}} \left[ e^{ℐ_{\2}β/2}e^{ℐ_{\2}\hat{\bm K} η/2} \Rc + ℐ_{\2} e^{-ℐ_{\2}β/2}e^{-ℐ_{\2}\hat{\bm K} η/2}   \Rc \B_{3} \right]ℐ_{\ppp}. \label{McWITHansatz}
\end{equation}
To obtain the space spinor components, we have to separate the terms proportional to $ℐ_{\2}$ from those that are not by using Moivre's formula:
\begin{eqnarray}
	e^{±ℐ_{\2}β/2}e^{±ℐ_{\2}\hat{\bm K} η/2} &=& \left(\cos \frac{β}{2} ± ℐ_{\2} \sin\frac{β}{2} \right) \left(\cosh \frac{η}{2} ± ℐ_{\2}\hat {\bm K} \sinh \frac{η}{2}\right),  \\
	&=& \cos \frac{β}{2} \cosh \frac{η}{2} - \hat{\bm K} \sin \frac{β}{2} \sinh \frac{η}{2}   ± \left( \sin\frac{β}{2} \cosh \frac{η}{2}  +\hat{\bm K} \cos \frac{β}{2} \sinh \frac{η}{2}  \right) ℐ_{\2}, \\
%%%%%%%%%%%%%%%%%%
ℐ_{\2} e^{-ℐ_{\2}β/2}e^{+ℐ_{\2}\hat {\bm K} η/2}&=&
\left(\cos \frac{β}{2} \cosh \frac{η}{2} - \hat{\bm K} \sin \frac{β}{2} \sinh \frac{η}{2}\right)ℐ_{\2} + \left ( \sin\frac{β}{2} \cosh \frac{η}{2}  +\hat{\bm K} \cos \frac{β}{2} \sinh \frac{η}{2}  \right). \nonumber \\ 
\end{eqnarray}
using this into the grade $\ppp$ projection of the quadrispinor $\Mc$ of Eq.~\eqref{McWITHansatz}
\begin{align}
	\Mc_{\ppp} = \sqrt{\frac{ρ}{2}} &\left[ 
	    \left(\cos \frac{β}{2} \cosh \frac{η}{2} - \hat{\bm K} \sin \frac{β}{2} \sinh \frac{η}{2}\right) \Rc  \right. \nonumber \\ 
	&-  \left( \sin\frac{β}{2} \cosh \frac{η}{2}  +\hat{\bm K} \cos \frac{β}{2} \sinh \frac{η}{2}  \right)ℐ_{\2} \Rc\B_{3}^{2}  \nonumber \\
	&+ \left(\cos \frac{β}{2} \cosh \frac{η}{2} - \hat{\bm K} \sin \frac{β}{2} \sinh \frac{η}{2}\right)ℐ_{\2} \Rc   \B_{3} \nonumber  \\
	&+ \left. \left( \sin\frac{β}{2} \cosh \frac{η}{2}  +\hat{\bm K} \cos \frac{β}{2} \sinh \frac{η}{2}  \right) \Rc   \B_{3} \right]ℐ_{\ppp} 
\end{align}
and comparing with Eq.~\eqref{GeneralMcPPPchiral}, we find the space spinor components (in which a factorization arises)
\begin{eqnarray}
	\Mc_{\lbb(\ppp)} &=& \sqrt{\frac{ρ}{2}} \left[ 
	    \left(\cos \frac{β}{2} \cosh \frac{η}{2} - \hat{\bm K} \sin \frac{β}{2} \sinh \frac{η}{2}\right) \Rc  + \left( \sin\frac{β}{2} \cosh \frac{η}{2}  +\hat{\bm K} \cos \frac{β}{2} \sinh \frac{η}{2}  \right) \Rc   \B_{3} \right], \nonumber \\
	    &=& \sqrt{\frac{ρ}{2}} \left[ 
	    \cosh \left(\frac{η}{2} \right) \Rc \left( \cos \frac{β}{2} + \B_{3}\sin \frac{β}{2}  \right) 
	    + \sinh \left( \frac{η}{2} \right)  \hat{\bm K} \Rc \B_{3} \left( \cos \frac{β}{2} + \B_{3}\sin \frac{β}{2}  \right)   \right], \nonumber  \\
	\Mc_{\rr(\ppp)} &=& \sqrt{\frac{ρ}{2}} \left[ 
	    -\left( \sin\frac{β}{2} \cosh \frac{η}{2}  +\hat{\bm K} \cos \frac{β}{2} \sinh \frac{η}{2}  \right) \Rc\B_{3} + \left(\cos \frac{β}{2} \cosh \frac{η}{2} - \hat {\bm K} \sin \frac{β}{2} \sinh \frac{η}{2}\right)\Rc \right], \nonumber \\
	    &=& \sqrt{\frac{ρ}{2}} \left[ 
	    \cosh \left(\frac{η}{2} \right) \Rc\left( \cos \frac{β}{2} - \B_{3}\sin\frac{β}{2}   \right) -\sinh \left(\frac{η}{2}\right) \hat{\bm K}  \Rc \B_{3} \left(\cos \frac{β}{2}  -  \B_{3}  \sin \frac{β}{2} \right)  \right]. \nonumber \\
\end{eqnarray}
We conclude that the spinorial components of the quadrispinor in chiral representation are written as
\begin{eqnarray}
	\label{LEFTspinor}  \Mc_{\lbb(\ppp)} &=& \sqrt{\frac{ρ}{2}} \left[ 
	    \cosh  \left(\frac{η}{2} \right)  + \sinh \left(\frac{η}{2} \right)  \hat{\bm K} \B_{3}    \right]\Rc e^{\B_{3}β/2}, %≐ \Nc_{-}, 
	    \\ 
	\label{RIGHTspinor} \Mc_{\rr(\ppp)} &=& \sqrt{\frac{ρ}{2}} \left[ 
	    \cosh \left(\frac{η}{2} \right) -\sinh \left(\frac{η}{2}\right) \hat{\bm K} \B_{3} \right]\Rc e^{-\B_{3}β/2} . % ≐ \Nc_{+}.
\end{eqnarray}
They describe two pairs of particle/antiparticle (with mass) associated with the quadrispinor of Eq.~\eqref{RappelChiralQuadrispinor} in chiral representation.

From these space spinors, we define the following two null spacetime vectors (i.e., of null square)
\begin{eqnarray}
	\bm u(β,\ppp) &=& ζ\Mc_{\rr(\ppp)} (\e_{0}+\e_{3}) \Mc_{\rr(\ppp)}^\mathsf{r}, \label{nullvector-u}\\
	\bm v(β,\ppp) &=& ζ\Mc_{\lbb(\ppp)} (\e_{0}-\e_{3}) \Mc_{\lbb(\ppp)}^\mathsf{r}, \label{nullvector-v}
\end{eqnarray}
with $ζ=e^{β\B_{3}}=1,-1$ for $β = 0,π$ respectively. Then, the spacetime vectors for linear momentum and spin are
\begin{eqnarray}
	\bm p(β,\ppp) &=& mcρ^{-1}[\bm u(β,\ppp) + \bm v(β,\ppp)],  \label{4momentum}\\
	\bm w(β,\ppp) &=& χ(\ppp)mcρ^{-1}[\bm u(β,\ppp) - \bm v(β,\ppp)]. \label{4spin}  
\end{eqnarray}
In the remainder of this Section, we show that the above formula and Eqs.~(\ref{spacetimemomentum}--\ref{spacetimespin}) agree with each other. Let's first compute $\bm u(0,\ppp)$ and $\bm v (0,\ppp)$ by calculating $\Nc_{χ}\e_{0} \Nc_{χ}^\mathsf{r}$ and $\Nc_{χ}\e_{3} \Nc_{χ}^\mathsf{r} $ 
with 
\begin{equation}
	\Nc_{χ}≐\sqrt{\frac{ρ}{2}} \left[ 
	    \cosh  \left(\frac{η}{2} \right)  -χ \sinh \left(\frac{η}{2} \right)  \hat {\bm K} \B_{3}    \right]\Rc e^{-χ\B_{3}β/2},
\end{equation}
where $χ=-1$ for the left chiral spinorial component and $χ=+1$ for the right chiral one. The first gives
\begin{eqnarray}
\Nc_{χ}\e_{0} \Nc_{χ}^\mathsf{r} &=& \frac{ρ}{2} \left[ 
	    \cosh  \left(\frac{η}{2} \right)  -χ \sinh \left(\frac{η}{2} \right)  \hat {\bm K} \B_{3}    \right]\e_{0}\left[ 
	    \cosh  \left(\frac{η}{2} \right)  -χ \sinh \left(\frac{η}{2} \right) \B_{3}   \hat {\bm K}   \right] ,    \\
	    &=&  \frac{ρ}{2} \left[ 
	     \cosh (η)\e_{0}  - χ \sinh (η)  (\hat {\bm K} ⋅ \B_{3} ) \e_{0}  \right]  ,
\end{eqnarray}
and similarly 
for the second
\begin{eqnarray}
	\Nc_{χ}\e_{3} \Nc_{χ}^\mathsf{r} &=& \frac{ρ}{2} \left[ 
	    \cosh  \left(\frac{η}{2} \right)  -χ \sinh \left(\frac{η}{2} \right)  \hat{\bm K} \B_{3}    \right]\e_{3}\left[ 
	    \cosh  \left(\frac{η}{2} \right)  -χ \sinh \left(\frac{η}{2} \right) \B_{3}   \hat{\bm K}   \right],     \\
	    &=& \frac{ρ}{2} \left[ 
	     \cosh^{2}  \left(\frac{η}{2} \right) \e_{3} - \sinh^{2} \left(\frac{η}{2} \right)  \hat {\bm K} \e_{3} \hat {\bm K} - χℐ_{\1}\hat{\bm K} \sinh (η)\right] .
\end{eqnarray}
We find the null vectors
\begin{eqnarray}
	\bm u(0,\ppp) &=& \frac{ρ}{2} \left[ 
	     \cosh (η)\e_{0}  -  \sinh (η)  (\hat {\bm K} ⋅ \B_{3} ) \e_{0}  
	     +\cosh^{2}  \left(\frac{η}{2} \right) \e_{3} 
	     - \sinh^{2} \left(\frac{η}{2} \right)  \hat {\bm K} \e_{3} \hat {\bm K} 
	     - ℐ_{\1}\hat {\bm K} \sinh (η)\right] \\
	\bm v(0,\ppp) &=& \frac{ρ}{2} \left[ 
	     \cosh (η)\e_{0}  +  \sinh (η)  (\hat {\bm K} ⋅ \B_{3} ) \e_{0}  
	     -\cosh^{2}  \left(\frac{η}{2} \right) \e_{3} 
	     + \sinh^{2} \left(\frac{η}{2} \right)  \hat {\bm K} \e_{3} \hat {\bm K} 
	     - ℐ_{\1}\hat {\bm K} \sinh (η)\right]
\end{eqnarray}
and the corresponding spacetime vectors for linear momentum and spin
\begin{eqnarray}
	\bm p(0,\ppp) &=&  mc\left[ 
	     \cosh (η)\e_{0}  - ℐ_{\1}\hat {\bm K} \sinh (η)\right]\\
	\bm w(0,\ppp) &=& χ(\ppp) mc \left[ 
	       -\sinh (η)  (\hat {\bm K} ⋅ \B_{3} ) \e_{0}  + 
	     \cosh^{2}  \left(\frac{η}{2} \right) \e_{3} - \sinh^{2} \left(\frac{η}{2} \right)  \hat {\bm K} \e_{3} \hat {\bm K} \right] \label{4spin-intermediate}
\end{eqnarray}
respectively, with opposite signs in the case $β=π$, in agreement with Eqs.~(\ref{spacetimemomentum}-\ref{spacetimespin}). 

We remark that multiplying the space spinors of Eqs.~(\ref{LEFTspinor}--\ref{RIGHTspinor}) by $±\B_{i}$ with $i=1,2$ results (following exactly the same calculations) in changing the sign of the space vectors of linear momentum and spin (no change on components $\e_{0}$). However, changing the sign of $\hat{\bm K}$ changes the helicity of the spinor $\Rc$ in Eq.~\eqref{GeneralAnsatz}. Accordingly, this flips the sign of the spatial part of the linear momentum without changing the spin vector. Hence, if they were aligned before, they become anti-aligned after this change (and vice versa).  

\paragraph{Spacetime vectors of spin and momentum}
From Eqs.~\eqref{cosh} and \eqref{sinh}, we find
\begin{equation}
	\bm p(0,\ppp) = \frac{ε}{c}\e_{0} - ℐ_{\1} \hat {\bm K} |\bm P| = \frac{ε}{c}\e_{0} - ℐ_{\1} \frac{ℏ \bm K}{|ℏ\bm K|} |\bm P| = \frac{ε}{c}\e_{0}+ \vec p.
\end{equation}
For consistency, we consider $\hat{\bm K}$ as the unit bivector perpendicular to the (space) wave vector $\vec k = -\bm K ℐ_{\1}$ and momentum $\vec p = ℏ \vec k = -\bm P ℐ_{\1}$, such that $ℏ\bm K = |\bm P|\hat{\bm K} = \bm P$ is the (space) bivector momentum of the particle. Similarly, the parameter $η>0$ is the rapidity for the boost between the observer frame and the center of mass frame of the particle-antiparticle pair.

The projection of the vector $\e_{3}$ onto the space vector $\vec p$ is $(\vec p ⋅ \e_{3})\vec p / |\bm P|^{2}$ while its rejection (i.e., the projection in plane $\hat {\bm K}$) is $\e_{3}-(\vec p ⋅ \e_{3})\vec p/|\bm P|^{2} $. Since the projection commutes with $ \hat {\bm K}$ whereas the rejection anticommutes, we find that the last term appearing in the spin spacetime vector is directed by
\begin{eqnarray}
	\hat {\bm K}\e_{3}\hat {\bm K} &=& \hat {\bm K}\left[ \e_{3}-\frac{(\vec p ⋅ \e_{3})\vec p}{|\bm P|^{2}} \right]\hat {\bm K} + \hat {\bm K}\frac{(\vec p ⋅ \e_{3})\vec p}{|\bm P|^{2}} \hat {\bm K},   \\
	&=&  \e_{3}-\frac{2(\vec p ⋅ \e_{3})\vec p}{|\bm P|^{2}}.
\end{eqnarray}
Using this in Eq.~\eqref{4spin-intermediate}, we find for the spacetime vector of spin-momentum
\begin{eqnarray}
	\bm w(0,\ppp) &=& χ(\ppp) mc \left[ 
	       -\sinh (η)  (\hat {\bm K} ⋅ \B_{3} ) \e_{0}  + 
	      \e_{3} -2\sinh^{2} \left(\frac{η}{2} \right) \frac{(\vec p ⋅ \e_{3})\vec p}{|\bm P|^{2}} \right], \\
	\bm w(0,\ppp) &=& χ(\ppp) \left[ 
	       -  (\vec p ⋅ \e_{3} ) \e_{0}  + 
	     mc  \e_{3} -\frac{c(\vec p ⋅ \e_{3})\vec p}{ε+mc^{2}} \right], \label{Pauli-Lubanski-vector}
\end{eqnarray}
where we have used Eqs.~(\ref{sinhhalfeta},\ref{sinh}) and $|\bm P|^{2}c^{2} = (ε-mc^{2})(ε+mc^{2})$ to get the second equality. Eq.~\eqref{Pauli-Lubanski-vector} provides the general spacetime vector of Pauli-Lubanski for a massive particle that verifies $\bm p ⋅ \bm w = 0$, $\bm p^{2}+\bm w^{2}=0$ and $\bm w^{2} = -m^{2}c^{2}$. Therefore, we recover the conventional interpretation in geometric algebra that $\e_{3}$ is the spin vector of the particle-antiparticle pair in their center of mass frame (when $\B_{3}$ is chosen as a complex number in Eq.~\eqref{WEappendix}).

In the massless limit, the spacetime vectors of spin and linear momentum become, as expected, null vectors
\begin{eqnarray}
	\bm p(0,\ppp) &\underset{m \rightarrow 0}{=}& |\vec p|\e_{0}+ |\vec p|\e_{3}, \\
	\bm w(0,\ppp) &\underset{m \rightarrow 0}{=}& χ(\ppp)\left(-|\vec p|\e_{0} + |\vec p| \e_{3}\right)
\end{eqnarray}

\paragraph{From spacetime vectors of spin and momentum to space spinors}
We solve the inverse problem of finding the two spinorial components compatible with some spacetime vectors of spin and momentum for a particle--antiparticle pair. The vectors $\u$ and $\v$ are null vectors directly determined from the space spinors $\Mc_{\lbb}$ and $\Mc_{\rr}$ (such as $\Mc_{\0}$ and $\Mc_{\2}$) in Eqs.~(\ref{nullvector-u}--\ref{nullvector-v}). Of the $8$ real degrees of freedom available in the two space spinors, only $6$ are used to define the spin and the momentum of a massive particle. The remaining two parameters have nothing to do with kinematics and encode the probability amplitude and the coefficients of the linear combination between the positive and negative energy solutions.

From the kinematics (spacetime momentum $\bm p $, spacetime spin $\bm w$) we get the null vectors $\u = \Mc_{\2}(\e_{0}+\e_{3})\Mc_{\2}^\mathsf{r}$ and $\v = \Mc_{\0}(\e_{0}-\e_{3})\Mc_{\0}^\mathsf{r}$ that each has 3 independent real degrees of freedom. From these null spacetime vectors, we determine below the constraints on the space spinors $\Mc_{\0}$ and $\Mc_{\2}$ parameterized as follows:
\begin{equation}
	\Mc_{\bbmu} = ∑_{μ} ν_{\bbmu}^{μ} \B_{μ}.
\end{equation}
The component of each space spinor is expressed using angles $α_{\bbmu},β_{\bbmu},η_{\bbmu}$ and amplitude $ν_{\bbmu}$ as 
\begin{eqnarray}
	ν_{\bbmu}^{0}&=&  ν_{\bbmu} \sin α_{\bbmu} \cos η_{\bbmu}, \label{reparam0}\\
	ν_{\bbmu}^{1}&=&  ν_{\bbmu} \sin β_{\bbmu} \sin η_{\bbmu}, \\
	ν_{\bbmu}^{2}&=&  ν_{\bbmu} \cos β_{\bbmu} \sin η_{\bbmu}, \\
	ν_{\bbmu}^{3}&=&  ν_{\bbmu} \cos α_{\bbmu} \cos η_{\bbmu}, \label{reparam3}
\end{eqnarray}
such that
\begin{eqnarray}
	{(ν_{\bbmu}^{0})}^{2} + {(ν_{\bbmu}^{3})}^{2} &=& {(ν_{\bbmu})}^{2}\cos^{2} η_{\bbmu}, \\
	{(ν_{\bbmu}^{1})}^{2}+{(ν_{\bbmu}^{2})}^{2} &=&  {(ν_{\bbmu})}^{2}\sin^{2} η_{\bbmu}. 
\end{eqnarray}

The direct computation of the components $ u_{μ}$ of the spacetime vector $\bm u$ gives, in terms of the angles and the amplitude:
\begin{align}
	u_{0} &= ∑_{μ=0}^{3}{(ν_{\2}^{μ})}^{2} = (ν_{\2})^{2}, \\
	u_{1} &= 2 \left(ν_{\2}^{0}ν_{\2}^{2}+ν_{\2}^{1}ν_{\2}^{3}\right) =  2 (ν_{\2})^{2}( \sin α_{\2} \cos β_{\2}  +  \sin β_{\2} \cos α_{\2}  )\cos η_{\2}\sin η_{\2} = (ν_{\2})^{2} \sin( α_{\2}+β_{\2}) \sin 2η_{\2},\\
	u_{2} &= 2 \left(ν_{\2}^{2}ν_{\2}^{3}-ν_{\2}^{0}ν_{\2}^{1}\right) = 2 (ν_{\2})^{2} ( \cos β_{\2}  \cos α_{\2}  -\sin α_{\2}  \sin β_{\2} )\cos η_{\2}\sin η_{\2} = (ν_{\2})^{2} \cos( α_{\2}+β_{\2}) \sin 2η_{\2}, \\
	u_{3} &= {(ν_{\2}^{0})}^{2}-{ν(_{\2}^{1})}^{2}-{(ν_{\2}^{2})}^{2}+{(ν_{\2}^{3})}^{2} = {(ν_{\2})}^{2}\cos 2 η_{\2} .
\end{align}
Solving the above equations for $α_{\2},β_{\2},η_{\2},ν_{\2}$, the space spinor components of $\Mc_{\2}$ are given by Eqs.~(\ref{reparam0}--\ref{reparam3}) with angles and amplitude
\begin{eqnarray}
	ν_{\2} &=& ±\sqrt{u_{0}}, \\
	α_{\2}+β_{\2} &=& \arctan \frac{u_{1}}{u_{2}}, \\
	η_{\2} &=& \frac{1}{2} \arccos \frac{u_{3}}{u_{0}}.
\end{eqnarray}
This means that, in our parametrization of $\Mc_{\2}$, the sum $α_{\2}+β_{\2}$ is fixed, but not the difference. 

Similarly, the direct computation of the components $ v_{μ}$ of spacetime vector $\bm v$ gives 
\begin{align}
	v_{0} &= ∑_{μ}{(ν_{\0}^{μ})}^{2}  = {(ν_{\0})}^{2},\\
	v_{1} &= -2 \left(ν_{\0}^{0}ν_{\0}^{2}+ν_{\0}^{1}ν_{\0}^{3}\right), \\
	v_{2} &= -2 \left(ν_{\0}^{2}ν_{\0}^{3}-ν_{\0}^{0}ν_{\0}^{1}\right), \\
	v_{3} &= -{(ν_{\0}^{0})}^{2}+{ν(_{\0}^{1})}^{2}+{(ν_{\0}^{2})}^{2}-{(ν_{\0}^{3})}^{2}.
\end{align}
We conclude for the angles $α_{\0},β_{\0},η_{\0}$ and the amplitude $ν_{\0}$:
\begin{eqnarray}
	ν_{\0} &=& ±\sqrt{v_{0}}, \\
	α_{\0}+β_{\0} &=& \arctan \frac{v_{1}}{v_{2}}, \\
	η_{\0} &=& \frac{1}{2} \arccos \frac{-v_{3}}{v_{0}}.
\end{eqnarray}
Therefore, one can find the two space spinors associated with a particle- antiparticle pair with given spin and linear momentum up to a phase in each spinor. This phase is the difference $α_{\bbmu}-β_{\bbmu}$, and its value shall be related to the amplitude of the wave function and/or to the angle describing the superposition of matter and antimatter in a general solution (since those are not present in the spin-momentum vectors).

\bibliographystyle{unsrt}
\bibliography{Ma_base_de_papier}

\end{document}